\documentclass[longauth]{aa}

\usepackage{graphicx}
\usepackage{natbib}
\usepackage{scalerel}

\usepackage[table]{xcolor}

\usepackage{subcaption} 
\usepackage[font=small]{caption} 
\usepackage[labelfont=bf]{caption} 

\usepackage{txfonts}
\usepackage[pdfencoding=auto,psdextra]{hyperref}
\hypersetup{
    colorlinks=true,
    linkcolor=blue,
    filecolor=magenta,      
    urlcolor=blue,
    citecolor=blue
}
\makeatletter
\renewcommand*\aa@pageof{, page \thepage{} of \pageref*{LastPage}}
\makeatother

\usepackage[utf8]{inputenc}

\usepackage[switch, modulo]{lineno}

\usepackage{euclid}

\begin{document}
   \title{\Euclid\ preparation. Exploring the properties of proto-clusters in the Simulated
   Euclid Wide Survey}

\newcommand{\orcid}[1]{} 
\author{Euclid Collaboration: H.~B\"ohringer\orcid{0000-0001-8241-4204}\thanks{\email{hxb@mpe.mpg.de}}\inst{\ref{aff1},\ref{aff2},\ref{aff3}}
\and G.~Chon\inst{\ref{aff2},\ref{aff3}}
\and O.~Cucciati\orcid{0000-0002-9336-7551}\inst{\ref{aff4}}
\and H.~Dannerbauer\orcid{0000-0001-7147-3575}\inst{\ref{aff5}}
\and M.~Bolzonella\orcid{0000-0003-3278-4607}\inst{\ref{aff4}}
\and G.~De~Lucia\orcid{0000-0002-6220-9104}\inst{\ref{aff6}}
\and A.~Cappi\inst{\ref{aff4},\ref{aff7}}
\and L.~Moscardini\orcid{0000-0002-3473-6716}\inst{\ref{aff8},\ref{aff4},\ref{aff9}}
\and C.~Giocoli\orcid{0000-0002-9590-7961}\inst{\ref{aff4},\ref{aff10}}
\and G.~Castignani\orcid{0000-0001-6831-0687}\inst{\ref{aff4}}
\and N.~A.~Hatch\orcid{0000-0001-5600-0534}\inst{\ref{aff11}}
\and S.~Andreon\orcid{0000-0002-2041-8784}\inst{\ref{aff12}}
\and E.~Ba\~nados\orcid{0000-0002-2931-7824}\inst{\ref{aff13}}
\and S.~Ettori\orcid{0000-0003-4117-8617}\inst{\ref{aff4},\ref{aff14}}
\and F.~Fontanot\orcid{0000-0003-4744-0188}\inst{\ref{aff6},\ref{aff15}}
\and H.~Gully\orcid{0000-0001-7616-7278}\inst{\ref{aff11}}
\and M.~Hirschmann\orcid{0000-0002-3301-3321}\inst{\ref{aff16},\ref{aff6}}
\and M.~Maturi\orcid{0000-0002-3517-2422}\inst{\ref{aff17},\ref{aff18}}
\and S.~Mei\orcid{0000-0002-2849-559X}\inst{\ref{aff19}}
\and L.~Pozzetti\orcid{0000-0001-7085-0412}\inst{\ref{aff4}}
\and T.~Schlenker\orcid{0009-0000-6916-1038}\inst{\ref{aff17}}
\and M.~Spinelli\orcid{0000-0003-0148-3254}\inst{\ref{aff7},\ref{aff6},\ref{aff20}}
\and N.~Aghanim\orcid{0000-0002-6688-8992}\inst{\ref{aff21}}
\and B.~Altieri\orcid{0000-0003-3936-0284}\inst{\ref{aff22}}
\and N.~Auricchio\orcid{0000-0003-4444-8651}\inst{\ref{aff4}}
\and C.~Baccigalupi\orcid{0000-0002-8211-1630}\inst{\ref{aff15},\ref{aff6},\ref{aff23},\ref{aff24}}
\and M.~Baldi\orcid{0000-0003-4145-1943}\inst{\ref{aff25},\ref{aff4},\ref{aff9}}
\and S.~Bardelli\orcid{0000-0002-8900-0298}\inst{\ref{aff4}}
\and C.~Bodendorf\inst{\ref{aff1}}
\and D.~Bonino\orcid{0000-0002-3336-9977}\inst{\ref{aff26}}
\and E.~Branchini\orcid{0000-0002-0808-6908}\inst{\ref{aff27},\ref{aff28},\ref{aff12}}
\and M.~Brescia\orcid{0000-0001-9506-5680}\inst{\ref{aff29},\ref{aff30},\ref{aff31}}
\and J.~Brinchmann\orcid{0000-0003-4359-8797}\inst{\ref{aff32},\ref{aff33}}
\and S.~Camera\orcid{0000-0003-3399-3574}\inst{\ref{aff34},\ref{aff35},\ref{aff26}}
\and V.~Capobianco\orcid{0000-0002-3309-7692}\inst{\ref{aff26}}
\and C.~Carbone\orcid{0000-0003-0125-3563}\inst{\ref{aff36}}
\and J.~Carretero\orcid{0000-0002-3130-0204}\inst{\ref{aff37},\ref{aff38}}
\and S.~Casas\orcid{0000-0002-4751-5138}\inst{\ref{aff39}}
\and F.~J.~Castander\orcid{0000-0001-7316-4573}\inst{\ref{aff40},\ref{aff41}}
\and M.~Castellano\orcid{0000-0001-9875-8263}\inst{\ref{aff42}}
\and S.~Cavuoti\orcid{0000-0002-3787-4196}\inst{\ref{aff30},\ref{aff31}}
\and A.~Cimatti\inst{\ref{aff43}}
\and C.~Colodro-Conde\inst{\ref{aff44}}
\and G.~Congedo\orcid{0000-0003-2508-0046}\inst{\ref{aff45}}
\and C.~J.~Conselice\orcid{0000-0003-1949-7638}\inst{\ref{aff46}}
\and L.~Conversi\orcid{0000-0002-6710-8476}\inst{\ref{aff47},\ref{aff22}}
\and Y.~Copin\orcid{0000-0002-5317-7518}\inst{\ref{aff48}}
\and F.~Courbin\orcid{0000-0003-0758-6510}\inst{\ref{aff49}}
\and H.~M.~Courtois\orcid{0000-0003-0509-1776}\inst{\ref{aff50}}
\and A.~Da~Silva\orcid{0000-0002-6385-1609}\inst{\ref{aff51},\ref{aff52}}
\and H.~Degaudenzi\orcid{0000-0002-5887-6799}\inst{\ref{aff53}}
\and A.~M.~Di~Giorgio\orcid{0000-0002-4767-2360}\inst{\ref{aff54}}
\and J.~Dinis\orcid{0000-0001-5075-1601}\inst{\ref{aff51},\ref{aff52}}
\and M.~Douspis\orcid{0000-0003-4203-3954}\inst{\ref{aff21}}
\and F.~Dubath\orcid{0000-0002-6533-2810}\inst{\ref{aff53}}
\and C.~A.~J.~Duncan\inst{\ref{aff46}}
\and X.~Dupac\inst{\ref{aff22}}
\and S.~Dusini\orcid{0000-0002-1128-0664}\inst{\ref{aff55}}
\and M.~Farina\orcid{0000-0002-3089-7846}\inst{\ref{aff54}}
\and S.~Farrens\orcid{0000-0002-9594-9387}\inst{\ref{aff56}}
\and F.~Faustini\orcid{0000-0001-6274-5145}\inst{\ref{aff57},\ref{aff42}}
\and P.~Fosalba\orcid{0000-0002-1510-5214}\inst{\ref{aff41},\ref{aff58}}
\and M.~Frailis\orcid{0000-0002-7400-2135}\inst{\ref{aff6}}
\and E.~Franceschi\orcid{0000-0002-0585-6591}\inst{\ref{aff4}}
\and M.~Fumana\orcid{0000-0001-6787-5950}\inst{\ref{aff36}}
\and S.~Galeotta\orcid{0000-0002-3748-5115}\inst{\ref{aff6}}
\and B.~Gillis\orcid{0000-0002-4478-1270}\inst{\ref{aff45}}
\and P.~G\'omez-Alvarez\orcid{0000-0002-8594-5358}\inst{\ref{aff59},\ref{aff22}}
\and A.~Grazian\orcid{0000-0002-5688-0663}\inst{\ref{aff60}}
\and F.~Grupp\inst{\ref{aff1},\ref{aff61}}
\and S.~V.~H.~Haugan\orcid{0000-0001-9648-7260}\inst{\ref{aff62}}
\and W.~Holmes\inst{\ref{aff63}}
\and F.~Hormuth\inst{\ref{aff64}}
\and A.~Hornstrup\orcid{0000-0002-3363-0936}\inst{\ref{aff65},\ref{aff66}}
\and P.~Hudelot\inst{\ref{aff67}}
\and K.~Jahnke\orcid{0000-0003-3804-2137}\inst{\ref{aff13}}
\and M.~Jhabvala\inst{\ref{aff68}}
\and B.~Joachimi\orcid{0000-0001-7494-1303}\inst{\ref{aff69}}
\and E.~Keih\"anen\orcid{0000-0003-1804-7715}\inst{\ref{aff70}}
\and S.~Kermiche\orcid{0000-0002-0302-5735}\inst{\ref{aff71}}
\and A.~Kiessling\orcid{0000-0002-2590-1273}\inst{\ref{aff63}}
\and M.~Kilbinger\orcid{0000-0001-9513-7138}\inst{\ref{aff56}}
\and B.~Kubik\orcid{0009-0006-5823-4880}\inst{\ref{aff48}}
\and M.~K\"ummel\orcid{0000-0003-2791-2117}\inst{\ref{aff61}}
\and M.~Kunz\orcid{0000-0002-3052-7394}\inst{\ref{aff72}}
\and H.~Kurki-Suonio\orcid{0000-0002-4618-3063}\inst{\ref{aff73},\ref{aff74}}
\and S.~Ligori\orcid{0000-0003-4172-4606}\inst{\ref{aff26}}
\and P.~B.~Lilje\orcid{0000-0003-4324-7794}\inst{\ref{aff62}}
\and V.~Lindholm\orcid{0000-0003-2317-5471}\inst{\ref{aff73},\ref{aff74}}
\and I.~Lloro\inst{\ref{aff75}}
\and G.~Mainetti\orcid{0000-0003-2384-2377}\inst{\ref{aff76}}
\and D.~Maino\inst{\ref{aff77},\ref{aff36},\ref{aff78}}
\and E.~Maiorano\orcid{0000-0003-2593-4355}\inst{\ref{aff4}}
\and O.~Mansutti\orcid{0000-0001-5758-4658}\inst{\ref{aff6}}
\and O.~Marggraf\orcid{0000-0001-7242-3852}\inst{\ref{aff79}}
\and K.~Markovic\orcid{0000-0001-6764-073X}\inst{\ref{aff63}}
\and M.~Martinelli\orcid{0000-0002-6943-7732}\inst{\ref{aff42},\ref{aff80}}
\and N.~Martinet\orcid{0000-0003-2786-7790}\inst{\ref{aff81}}
\and F.~Marulli\orcid{0000-0002-8850-0303}\inst{\ref{aff8},\ref{aff4},\ref{aff9}}
\and R.~Massey\orcid{0000-0002-6085-3780}\inst{\ref{aff82}}
\and S.~Maurogordato\inst{\ref{aff7}}
\and E.~Medinaceli\orcid{0000-0002-4040-7783}\inst{\ref{aff4}}
\and Y.~Mellier\inst{\ref{aff83},\ref{aff67}}
\and M.~Meneghetti\orcid{0000-0003-1225-7084}\inst{\ref{aff4},\ref{aff9}}
\and G.~Meylan\inst{\ref{aff49}}
\and M.~Moresco\orcid{0000-0002-7616-7136}\inst{\ref{aff8},\ref{aff4}}
\and S.-M.~Niemi\inst{\ref{aff84}}
\and C.~Padilla\orcid{0000-0001-7951-0166}\inst{\ref{aff85}}
\and S.~Paltani\orcid{0000-0002-8108-9179}\inst{\ref{aff53}}
\and F.~Pasian\orcid{0000-0002-4869-3227}\inst{\ref{aff6}}
\and K.~Pedersen\inst{\ref{aff86}}
\and V.~Pettorino\inst{\ref{aff84}}
\and S.~Pires\orcid{0000-0002-0249-2104}\inst{\ref{aff56}}
\and G.~Polenta\orcid{0000-0003-4067-9196}\inst{\ref{aff57}}
\and M.~Poncet\inst{\ref{aff87}}
\and L.~A.~Popa\inst{\ref{aff88}}
\and F.~Raison\orcid{0000-0002-7819-6918}\inst{\ref{aff1}}
\and R.~Rebolo\inst{\ref{aff44},\ref{aff89}}
\and A.~Renzi\orcid{0000-0001-9856-1970}\inst{\ref{aff90},\ref{aff55}}
\and J.~Rhodes\orcid{0000-0002-4485-8549}\inst{\ref{aff63}}
\and G.~Riccio\inst{\ref{aff30}}
\and E.~Romelli\orcid{0000-0003-3069-9222}\inst{\ref{aff6}}
\and M.~Roncarelli\orcid{0000-0001-9587-7822}\inst{\ref{aff4}}
\and E.~Rossetti\orcid{0000-0003-0238-4047}\inst{\ref{aff25}}
\and R.~Saglia\orcid{0000-0003-0378-7032}\inst{\ref{aff61},\ref{aff1}}
\and Z.~Sakr\orcid{0000-0002-4823-3757}\inst{\ref{aff17},\ref{aff91},\ref{aff92}}
\and A.~G.~S\'anchez\orcid{0000-0003-1198-831X}\inst{\ref{aff1}}
\and D.~Sapone\orcid{0000-0001-7089-4503}\inst{\ref{aff93}}
\and B.~Sartoris\orcid{0000-0003-1337-5269}\inst{\ref{aff61},\ref{aff6}}
\and M.~Schirmer\orcid{0000-0003-2568-9994}\inst{\ref{aff13}}
\and P.~Schneider\orcid{0000-0001-8561-2679}\inst{\ref{aff79}}
\and M.~Scodeggio\inst{\ref{aff36}}
\and A.~Secroun\orcid{0000-0003-0505-3710}\inst{\ref{aff71}}
\and G.~Seidel\orcid{0000-0003-2907-353X}\inst{\ref{aff13}}
\and S.~Serrano\orcid{0000-0002-0211-2861}\inst{\ref{aff41},\ref{aff94},\ref{aff40}}
\and C.~Sirignano\orcid{0000-0002-0995-7146}\inst{\ref{aff90},\ref{aff55}}
\and G.~Sirri\orcid{0000-0003-2626-2853}\inst{\ref{aff9}}
\and L.~Stanco\orcid{0000-0002-9706-5104}\inst{\ref{aff55}}
\and J.~Steinwagner\inst{\ref{aff1}}
\and P.~Tallada-Cresp\'{i}\orcid{0000-0002-1336-8328}\inst{\ref{aff37},\ref{aff38}}
\and A.~N.~Taylor\inst{\ref{aff45}}
\and I.~Tereno\inst{\ref{aff51},\ref{aff95}}
\and R.~Toledo-Moreo\orcid{0000-0002-2997-4859}\inst{\ref{aff96}}
\and F.~Torradeflot\orcid{0000-0003-1160-1517}\inst{\ref{aff38},\ref{aff37}}
\and I.~Tutusaus\orcid{0000-0002-3199-0399}\inst{\ref{aff91}}
\and T.~Vassallo\orcid{0000-0001-6512-6358}\inst{\ref{aff61},\ref{aff6}}
\and G.~Verdoes~Kleijn\orcid{0000-0001-5803-2580}\inst{\ref{aff97}}
\and A.~Veropalumbo\orcid{0000-0003-2387-1194}\inst{\ref{aff12},\ref{aff28},\ref{aff98}}
\and Y.~Wang\orcid{0000-0002-4749-2984}\inst{\ref{aff99}}
\and J.~Weller\orcid{0000-0002-8282-2010}\inst{\ref{aff61},\ref{aff1}}
\and G.~Zamorani\orcid{0000-0002-2318-301X}\inst{\ref{aff4}}
\and E.~Zucca\orcid{0000-0002-5845-8132}\inst{\ref{aff4}}
\and E.~Bozzo\orcid{0000-0002-8201-1525}\inst{\ref{aff53}}
\and C.~Burigana\orcid{0000-0002-3005-5796}\inst{\ref{aff100},\ref{aff14}}
\and M.~Calabrese\orcid{0000-0002-2637-2422}\inst{\ref{aff101},\ref{aff36}}
\and D.~Di~Ferdinando\inst{\ref{aff9}}
\and J.~A.~Escartin~Vigo\inst{\ref{aff1}}
\and F.~Finelli\orcid{0000-0002-6694-3269}\inst{\ref{aff4},\ref{aff14}}
\and J.~Gracia-Carpio\inst{\ref{aff1}}
\and S.~Matthew\orcid{0000-0001-8448-1697}\inst{\ref{aff45}}
\and N.~Mauri\orcid{0000-0001-8196-1548}\inst{\ref{aff43},\ref{aff9}}
\and M.~P\"ontinen\orcid{0000-0001-5442-2530}\inst{\ref{aff73}}
\and C.~Porciani\orcid{0000-0002-7797-2508}\inst{\ref{aff79}}
\and V.~Scottez\inst{\ref{aff83},\ref{aff102}}
\and M.~Tenti\orcid{0000-0002-4254-5901}\inst{\ref{aff9}}
\and M.~Viel\orcid{0000-0002-2642-5707}\inst{\ref{aff15},\ref{aff6},\ref{aff24},\ref{aff23},\ref{aff103}}
\and M.~Wiesmann\orcid{0009-0000-8199-5860}\inst{\ref{aff62}}
\and Y.~Akrami\orcid{0000-0002-2407-7956}\inst{\ref{aff104},\ref{aff105}}
\and V.~Allevato\orcid{0000-0001-7232-5152}\inst{\ref{aff30}}
\and S.~Alvi\orcid{0000-0001-5779-8568}\inst{\ref{aff106}}
\and S.~Anselmi\orcid{0000-0002-3579-9583}\inst{\ref{aff55},\ref{aff90},\ref{aff107}}
\and M.~Archidiacono\orcid{0000-0003-4952-9012}\inst{\ref{aff77},\ref{aff78}}
\and F.~Atrio-Barandela\orcid{0000-0002-2130-2513}\inst{\ref{aff108}}
\and A.~Balaguera-Antolinez\orcid{0000-0001-5028-3035}\inst{\ref{aff44},\ref{aff89}}
\and M.~Ballardini\orcid{0000-0003-4481-3559}\inst{\ref{aff106},\ref{aff4},\ref{aff109}}
\and A.~Blanchard\orcid{0000-0001-8555-9003}\inst{\ref{aff91}}
\and L.~Blot\orcid{0000-0002-9622-7167}\inst{\ref{aff110},\ref{aff107}}
\and S.~Borgani\orcid{0000-0001-6151-6439}\inst{\ref{aff111},\ref{aff15},\ref{aff6},\ref{aff23}}
\and S.~Bruton\orcid{0000-0002-6503-5218}\inst{\ref{aff112}}
\and R.~Cabanac\orcid{0000-0001-6679-2600}\inst{\ref{aff91}}
\and A.~Calabro\orcid{0000-0003-2536-1614}\inst{\ref{aff42}}
\and F.~Caro\inst{\ref{aff42}}
\and C.~S.~Carvalho\inst{\ref{aff95}}
\and T.~Castro\orcid{0000-0002-6292-3228}\inst{\ref{aff6},\ref{aff23},\ref{aff15},\ref{aff103}}
\and K.~C.~Chambers\orcid{0000-0001-6965-7789}\inst{\ref{aff113}}
\and S.~Contarini\orcid{0000-0002-9843-723X}\inst{\ref{aff1}}
\and A.~R.~Cooray\orcid{0000-0002-3892-0190}\inst{\ref{aff114}}
\and M.~Costanzi\orcid{0000-0001-8158-1449}\inst{\ref{aff111},\ref{aff6},\ref{aff15}}
\and B.~De~Caro\inst{\ref{aff36}}
\and G.~Desprez\inst{\ref{aff115}}
\and A.~D\'iaz-S\'anchez\orcid{0000-0003-0748-4768}\inst{\ref{aff116}}
\and S.~Di~Domizio\orcid{0000-0003-2863-5895}\inst{\ref{aff27},\ref{aff28}}
\and H.~Dole\orcid{0000-0002-9767-3839}\inst{\ref{aff21}}
\and S.~Escoffier\orcid{0000-0002-2847-7498}\inst{\ref{aff71}}
\and A.~G.~Ferrari\orcid{0009-0005-5266-4110}\inst{\ref{aff43},\ref{aff9}}
\and P.~G.~Ferreira\orcid{0000-0002-3021-2851}\inst{\ref{aff117}}
\and I.~Ferrero\orcid{0000-0002-1295-1132}\inst{\ref{aff62}}
\and A.~Fontana\orcid{0000-0003-3820-2823}\inst{\ref{aff42}}
\and F.~Fornari\orcid{0000-0003-2979-6738}\inst{\ref{aff14}}
\and L.~Gabarra\orcid{0000-0002-8486-8856}\inst{\ref{aff117}}
\and K.~Ganga\orcid{0000-0001-8159-8208}\inst{\ref{aff19}}
\and J.~Garc\'ia-Bellido\orcid{0000-0002-9370-8360}\inst{\ref{aff104}}
\and T.~Gasparetto\orcid{0000-0002-7913-4866}\inst{\ref{aff6}}
\and V.~Gautard\inst{\ref{aff118}}
\and E.~Gaztanaga\orcid{0000-0001-9632-0815}\inst{\ref{aff40},\ref{aff41},\ref{aff119}}
\and F.~Giacomini\orcid{0000-0002-3129-2814}\inst{\ref{aff9}}
\and F.~Gianotti\orcid{0000-0003-4666-119X}\inst{\ref{aff4}}
\and A.~H.~Gonzalez\orcid{0000-0002-0933-8601}\inst{\ref{aff120}}
\and G.~Gozaliasl\orcid{0000-0002-0236-919X}\inst{\ref{aff121},\ref{aff73}}
\and C.~M.~Gutierrez\orcid{0000-0001-7854-783X}\inst{\ref{aff122}}
\and A.~Hall\orcid{0000-0002-3139-8651}\inst{\ref{aff45}}
\and W.~G.~Hartley\inst{\ref{aff53}}
\and H.~Hildebrandt\orcid{0000-0002-9814-3338}\inst{\ref{aff123}}
\and J.~Hjorth\orcid{0000-0002-4571-2306}\inst{\ref{aff124}}
\and A.~Jimenez~Mu\~noz\orcid{0009-0004-5252-185X}\inst{\ref{aff125}}
\and J.~J.~E.~Kajava\orcid{0000-0002-3010-8333}\inst{\ref{aff126},\ref{aff127}}
\and V.~Kansal\orcid{0000-0002-4008-6078}\inst{\ref{aff128},\ref{aff129}}
\and D.~Karagiannis\orcid{0000-0002-4927-0816}\inst{\ref{aff130},\ref{aff20}}
\and C.~C.~Kirkpatrick\inst{\ref{aff70}}
\and L.~Legrand\orcid{0000-0003-0610-5252}\inst{\ref{aff131}}
\and J.~Lesgourgues\orcid{0000-0001-7627-353X}\inst{\ref{aff39}}
\and T.~I.~Liaudat\orcid{0000-0002-9104-314X}\inst{\ref{aff132}}
\and A.~Loureiro\orcid{0000-0002-4371-0876}\inst{\ref{aff133},\ref{aff134}}
\and J.~Macias-Perez\orcid{0000-0002-5385-2763}\inst{\ref{aff125}}
\and G.~Maggio\orcid{0000-0003-4020-4836}\inst{\ref{aff6}}
\and M.~Magliocchetti\orcid{0000-0001-9158-4838}\inst{\ref{aff54}}
\and C.~Mancini\orcid{0000-0002-4297-0561}\inst{\ref{aff36}}
\and F.~Mannucci\orcid{0000-0002-4803-2381}\inst{\ref{aff135}}
\and R.~Maoli\orcid{0000-0002-6065-3025}\inst{\ref{aff136},\ref{aff42}}
\and C.~J.~A.~P.~Martins\orcid{0000-0002-4886-9261}\inst{\ref{aff137},\ref{aff32}}
\and L.~Maurin\orcid{0000-0002-8406-0857}\inst{\ref{aff21}}
\and R.~B.~Metcalf\orcid{0000-0003-3167-2574}\inst{\ref{aff8},\ref{aff4}}
\and M.~Miluzio\inst{\ref{aff22},\ref{aff138}}
\and P.~Monaco\orcid{0000-0003-2083-7564}\inst{\ref{aff111},\ref{aff6},\ref{aff23},\ref{aff15}}
\and A.~Montoro\orcid{0000-0003-4730-8590}\inst{\ref{aff40},\ref{aff41}}
\and A.~Mora\orcid{0000-0002-1922-8529}\inst{\ref{aff139}}
\and C.~Moretti\orcid{0000-0003-3314-8936}\inst{\ref{aff24},\ref{aff103},\ref{aff6},\ref{aff15},\ref{aff23}}
\and G.~Morgante\inst{\ref{aff4}}
\and Nicholas~A.~Walton\orcid{0000-0003-3983-8778}\inst{\ref{aff140}}
\and L.~Patrizii\inst{\ref{aff9}}
\and V.~Popa\orcid{0000-0002-9118-8330}\inst{\ref{aff88}}
\and D.~Potter\orcid{0000-0002-0757-5195}\inst{\ref{aff141}}
\and I.~Risso\orcid{0000-0003-2525-7761}\inst{\ref{aff98}}
\and P.-F.~Rocci\inst{\ref{aff21}}
\and M.~Sahl\'en\orcid{0000-0003-0973-4804}\inst{\ref{aff142}}
\and A.~Schneider\orcid{0000-0001-7055-8104}\inst{\ref{aff141}}
\and M.~Schultheis\inst{\ref{aff7}}
\and M.~Sereno\orcid{0000-0003-0302-0325}\inst{\ref{aff4},\ref{aff9}}
\and F.~Shankar\orcid{0000-0001-8973-5051}\inst{\ref{aff143}}
\and P.~Simon\inst{\ref{aff79}}
\and A.~Spurio~Mancini\orcid{0000-0001-5698-0990}\inst{\ref{aff144},\ref{aff145}}
\and J.~Stadel\orcid{0000-0001-7565-8622}\inst{\ref{aff141}}
\and S.~A.~Stanford\orcid{0000-0003-0122-0841}\inst{\ref{aff146}}
\and K.~Tanidis\inst{\ref{aff117}}
\and C.~Tao\orcid{0000-0001-7961-8177}\inst{\ref{aff71}}
\and G.~Testera\inst{\ref{aff28}}
\and R.~Teyssier\orcid{0000-0001-7689-0933}\inst{\ref{aff147}}
\and S.~Toft\orcid{0000-0003-3631-7176}\inst{\ref{aff148},\ref{aff149}}
\and S.~Tosi\orcid{0000-0002-7275-9193}\inst{\ref{aff27},\ref{aff28}}
\and A.~Troja\orcid{0000-0003-0239-4595}\inst{\ref{aff90},\ref{aff55}}
\and M.~Tucci\inst{\ref{aff53}}
\and C.~Valieri\inst{\ref{aff9}}
\and J.~Valiviita\orcid{0000-0001-6225-3693}\inst{\ref{aff73},\ref{aff74}}
\and D.~Vergani\orcid{0000-0003-0898-2216}\inst{\ref{aff4}}
\and G.~Verza\orcid{0000-0002-1886-8348}\inst{\ref{aff150},\ref{aff151}}}
										   
\institute{Max Planck Institute for Extraterrestrial Physics, Giessenbachstr. 1, 85748 Garching, Germany\label{aff1}
\and
Ludwig-Maximilians-University, Schellingstrasse 4, 80799 Munich, Germany\label{aff2}
\and
Max-Planck-Institut f\"ur Physik, Boltzmannstr. 8, 85748 Garching, Germany\label{aff3}
\and
INAF-Osservatorio di Astrofisica e Scienza dello Spazio di Bologna, Via Piero Gobetti 93/3, 40129 Bologna, Italy\label{aff4}
\and
Instituto de Astrof\'isica de Canarias (IAC); Departamento de Astrof\'isica, Universidad de La Laguna (ULL), 38200, La Laguna, Tenerife, Spain\label{aff5}
\and
INAF-Osservatorio Astronomico di Trieste, Via G. B. Tiepolo 11, 34143 Trieste, Italy\label{aff6}
\and
Universit\'e C\^{o}te d'Azur, Observatoire de la C\^{o}te d'Azur, CNRS, Laboratoire Lagrange, Bd de l'Observatoire, CS 34229, 06304 Nice cedex 4, France\label{aff7}
\and
Dipartimento di Fisica e Astronomia "Augusto Righi" - Alma Mater Studiorum Universit\`a di Bologna, via Piero Gobetti 93/2, 40129 Bologna, Italy\label{aff8}
\and
INFN-Sezione di Bologna, Viale Berti Pichat 6/2, 40127 Bologna, Italy\label{aff9}
\and
Istituto Nazionale di Fisica Nucleare, Sezione di Bologna, Via Irnerio 46, 40126 Bologna, Italy\label{aff10}
\and
School of Physics and Astronomy, University of Nottingham, University Park, Nottingham NG7 2RD, UK\label{aff11}
\and
INAF-Osservatorio Astronomico di Brera, Via Brera 28, 20122 Milano, Italy\label{aff12}
\and
Max-Planck-Institut f\"ur Astronomie, K\"onigstuhl 17, 69117 Heidelberg, Germany\label{aff13}
\and
INFN-Bologna, Via Irnerio 46, 40126 Bologna, Italy\label{aff14}
\and
IFPU, Institute for Fundamental Physics of the Universe, via Beirut 2, 34151 Trieste, Italy\label{aff15}
\and
Institute of Physics, Laboratory for Galaxy Evolution, Ecole Polytechnique F\'ed\'erale de Lausanne, Observatoire de Sauverny, CH-1290 Versoix, Switzerland\label{aff16}
\and
Institut f\"ur Theoretische Physik, University of Heidelberg, Philosophenweg 16, 69120 Heidelberg, Germany\label{aff17}
\and
Zentrum f\"ur Astronomie, Universit\"at Heidelberg, Philosophenweg 12, 69120 Heidelberg, Germany\label{aff18}
\and
Universit\'e Paris Cit\'e, CNRS, Astroparticule et Cosmologie, 75013 Paris, France\label{aff19}
\and
Department of Physics and Astronomy, University of the Western Cape, Bellville, Cape Town, 7535, South Africa\label{aff20}
\and
Universit\'e Paris-Saclay, CNRS, Institut d'astrophysique spatiale, 91405, Orsay, France\label{aff21}
\and
ESAC/ESA, Camino Bajo del Castillo, s/n., Urb. Villafranca del Castillo, 28692 Villanueva de la Ca\~nada, Madrid, Spain\label{aff22}
\and
INFN, Sezione di Trieste, Via Valerio 2, 34127 Trieste TS, Italy\label{aff23}
\and
SISSA, International School for Advanced Studies, Via Bonomea 265, 34136 Trieste TS, Italy\label{aff24}
\and
Dipartimento di Fisica e Astronomia, Universit\`a di Bologna, Via Gobetti 93/2, 40129 Bologna, Italy\label{aff25}
\and
INAF-Osservatorio Astrofisico di Torino, Via Osservatorio 20, 10025 Pino Torinese (TO), Italy\label{aff26}
\and
Dipartimento di Fisica, Universit\`a di Genova, Via Dodecaneso 33, 16146, Genova, Italy\label{aff27}
\and
INFN-Sezione di Genova, Via Dodecaneso 33, 16146, Genova, Italy\label{aff28}
\and
Department of Physics "E. Pancini", University Federico II, Via Cinthia 6, 80126, Napoli, Italy\label{aff29}
\and
INAF-Osservatorio Astronomico di Capodimonte, Via Moiariello 16, 80131 Napoli, Italy\label{aff30}
\and
INFN section of Naples, Via Cinthia 6, 80126, Napoli, Italy\label{aff31}
\and
Instituto de Astrof\'isica e Ci\^encias do Espa\c{c}o, Universidade do Porto, CAUP, Rua das Estrelas, PT4150-762 Porto, Portugal\label{aff32}
\and
Faculdade de Ci\^encias da Universidade do Porto, Rua do Campo de Alegre, 4150-007 Porto, Portugal\label{aff33}
\and
Dipartimento di Fisica, Universit\`a degli Studi di Torino, Via P. Giuria 1, 10125 Torino, Italy\label{aff34}
\and
INFN-Sezione di Torino, Via P. Giuria 1, 10125 Torino, Italy\label{aff35}
\and
INAF-IASF Milano, Via Alfonso Corti 12, 20133 Milano, Italy\label{aff36}
\and
Centro de Investigaciones Energ\'eticas, Medioambientales y Tecnol\'ogicas (CIEMAT), Avenida Complutense 40, 28040 Madrid, Spain\label{aff37}
\and
Port d'Informaci\'{o} Cient\'{i}fica, Campus UAB, C. Albareda s/n, 08193 Bellaterra (Barcelona), Spain\label{aff38}
\and
Institute for Theoretical Particle Physics and Cosmology (TTK), RWTH Aachen University, 52056 Aachen, Germany\label{aff39}
\and
Institute of Space Sciences (ICE, CSIC), Campus UAB, Carrer de Can Magrans, s/n, 08193 Barcelona, Spain\label{aff40}
\and
Institut d'Estudis Espacials de Catalunya (IEEC),  Edifici RDIT, Campus UPC, 08860 Castelldefels, Barcelona, Spain\label{aff41}
\and
INAF-Osservatorio Astronomico di Roma, Via Frascati 33, 00078 Monteporzio Catone, Italy\label{aff42}
\and
Dipartimento di Fisica e Astronomia "Augusto Righi" - Alma Mater Studiorum Universit\`a di Bologna, Viale Berti Pichat 6/2, 40127 Bologna, Italy\label{aff43}
\and
Instituto de Astrof\'isica de Canarias, Calle V\'ia L\'actea s/n, 38204, San Crist\'obal de La Laguna, Tenerife, Spain\label{aff44}
\and
Institute for Astronomy, University of Edinburgh, Royal Observatory, Blackford Hill, Edinburgh EH9 3HJ, UK\label{aff45}
\and
Jodrell Bank Centre for Astrophysics, Department of Physics and Astronomy, University of Manchester, Oxford Road, Manchester M13 9PL, UK\label{aff46}
\and
European Space Agency/ESRIN, Largo Galileo Galilei 1, 00044 Frascati, Roma, Italy\label{aff47}
\and
Universit\'e Claude Bernard Lyon 1, CNRS/IN2P3, IP2I Lyon, UMR 5822, Villeurbanne, F-69100, France\label{aff48}
\and
Institute of Physics, Laboratory of Astrophysics, Ecole Polytechnique F\'ed\'erale de Lausanne (EPFL), Observatoire de Sauverny, 1290 Versoix, Switzerland\label{aff49}
\and
UCB Lyon 1, CNRS/IN2P3, IUF, IP2I Lyon, 4 rue Enrico Fermi, 69622 Villeurbanne, France\label{aff50}
\and
Departamento de F\'isica, Faculdade de Ci\^encias, Universidade de Lisboa, Edif\'icio C8, Campo Grande, PT1749-016 Lisboa, Portugal\label{aff51}
\and
Instituto de Astrof\'isica e Ci\^encias do Espa\c{c}o, Faculdade de Ci\^encias, Universidade de Lisboa, Campo Grande, 1749-016 Lisboa, Portugal\label{aff52}
\and
Department of Astronomy, University of Geneva, ch. d'Ecogia 16, 1290 Versoix, Switzerland\label{aff53}
\and
INAF-Istituto di Astrofisica e Planetologia Spaziali, via del Fosso del Cavaliere, 100, 00100 Roma, Italy\label{aff54}
\and
INFN-Padova, Via Marzolo 8, 35131 Padova, Italy\label{aff55}
\and
Universit\'e Paris-Saclay, Universit\'e Paris Cit\'e, CEA, CNRS, AIM, 91191, Gif-sur-Yvette, France\label{aff56}
\and
Space Science Data Center, Italian Space Agency, via del Politecnico snc, 00133 Roma, Italy\label{aff57}
\and
Institut de Ciencies de l'Espai (IEEC-CSIC), Campus UAB, Carrer de Can Magrans, s/n Cerdanyola del Vall\'es, 08193 Barcelona, Spain\label{aff58}
\and
FRACTAL S.L.N.E., calle Tulip\'an 2, Portal 13 1A, 28231, Las Rozas de Madrid, Spain\label{aff59}
\and
INAF-Osservatorio Astronomico di Padova, Via dell'Osservatorio 5, 35122 Padova, Italy\label{aff60}
\and
Universit\"ats-Sternwarte M\"unchen, Fakult\"at f\"ur Physik, Ludwig-Maximilians-Universit\"at M\"unchen, Scheinerstrasse 1, 81679 M\"unchen, Germany\label{aff61}
\and
Institute of Theoretical Astrophysics, University of Oslo, P.O. Box 1029 Blindern, 0315 Oslo, Norway\label{aff62}
\and
Jet Propulsion Laboratory, California Institute of Technology, 4800 Oak Grove Drive, Pasadena, CA, 91109, USA\label{aff63}
\and
Felix Hormuth Engineering, Goethestr. 17, 69181 Leimen, Germany\label{aff64}
\and
Technical University of Denmark, Elektrovej 327, 2800 Kgs. Lyngby, Denmark\label{aff65}
\and
Cosmic Dawn Center (DAWN), Denmark\label{aff66}
\and
Institut d'Astrophysique de Paris, UMR 7095, CNRS, and Sorbonne Universit\'e, 98 bis boulevard Arago, 75014 Paris, France\label{aff67}
\and
NASA Goddard Space Flight Center, Greenbelt, MD 20771, USA\label{aff68}
\and
Department of Physics and Astronomy, University College London, Gower Street, London WC1E 6BT, UK\label{aff69}
\and
Department of Physics and Helsinki Institute of Physics, Gustaf H\"allstr\"omin katu 2, 00014 University of Helsinki, Finland\label{aff70}
\and
Aix-Marseille Universit\'e, CNRS/IN2P3, CPPM, Marseille, France\label{aff71}
\and
Universit\'e de Gen\`eve, D\'epartement de Physique Th\'eorique and Centre for Astroparticle Physics, 24 quai Ernest-Ansermet, CH-1211 Gen\`eve 4, Switzerland\label{aff72}
\and
Department of Physics, P.O. Box 64, 00014 University of Helsinki, Finland\label{aff73}
\and
Helsinki Institute of Physics, Gustaf H{\"a}llstr{\"o}min katu 2, University of Helsinki, Helsinki, Finland\label{aff74}
\and
NOVA optical infrared instrumentation group at ASTRON, Oude Hoogeveensedijk 4, 7991PD, Dwingeloo, The Netherlands\label{aff75}
\and
Centre de Calcul de l'IN2P3/CNRS, 21 avenue Pierre de Coubertin 69627 Villeurbanne Cedex, France\label{aff76}
\and
Dipartimento di Fisica "Aldo Pontremoli", Universit\`a degli Studi di Milano, Via Celoria 16, 20133 Milano, Italy\label{aff77}
\and
INFN-Sezione di Milano, Via Celoria 16, 20133 Milano, Italy\label{aff78}
\and
Universit\"at Bonn, Argelander-Institut f\"ur Astronomie, Auf dem H\"ugel 71, 53121 Bonn, Germany\label{aff79}
\and
INFN-Sezione di Roma, Piazzale Aldo Moro, 2 - c/o Dipartimento di Fisica, Edificio G. Marconi, 00185 Roma, Italy\label{aff80}
\and
Aix-Marseille Universit\'e, CNRS, CNES, LAM, Marseille, France\label{aff81}
\and
Department of Physics, Institute for Computational Cosmology, Durham University, South Road, DH1 3LE, UK\label{aff82}
\and
Institut d'Astrophysique de Paris, 98bis Boulevard Arago, 75014, Paris, France\label{aff83}
\and
European Space Agency/ESTEC, Keplerlaan 1, 2201 AZ Noordwijk, The Netherlands\label{aff84}
\and
Institut de F\'{i}sica d'Altes Energies (IFAE), The Barcelona Institute of Science and Technology, Campus UAB, 08193 Bellaterra (Barcelona), Spain\label{aff85}
\and
Department of Physics and Astronomy, University of Aarhus, Ny Munkegade 120, DK-8000 Aarhus C, Denmark\label{aff86}
\and
Centre National d'Etudes Spatiales -- Centre spatial de Toulouse, 18 avenue Edouard Belin, 31401 Toulouse Cedex 9, France\label{aff87}
\and
Institute of Space Science, Str. Atomistilor, nr. 409 M\u{a}gurele, Ilfov, 077125, Romania\label{aff88}
\and
Departamento de Astrof\'isica, Universidad de La Laguna, 38206, La Laguna, Tenerife, Spain\label{aff89}
\and
Dipartimento di Fisica e Astronomia "G. Galilei", Universit\`a di Padova, Via Marzolo 8, 35131 Padova, Italy\label{aff90}
\and
Institut de Recherche en Astrophysique et Plan\'etologie (IRAP), Universit\'e de Toulouse, CNRS, UPS, CNES, 14 Av. Edouard Belin, 31400 Toulouse, France\label{aff91}
\and
Universit\'e St Joseph; Faculty of Sciences, Beirut, Lebanon\label{aff92}
\and
Departamento de F\'isica, FCFM, Universidad de Chile, Blanco Encalada 2008, Santiago, Chile\label{aff93}
\and
Satlantis, University Science Park, Sede Bld 48940, Leioa-Bilbao, Spain\label{aff94}
\and
Instituto de Astrof\'isica e Ci\^encias do Espa\c{c}o, Faculdade de Ci\^encias, Universidade de Lisboa, Tapada da Ajuda, 1349-018 Lisboa, Portugal\label{aff95}
\and
Universidad Polit\'ecnica de Cartagena, Departamento de Electr\'onica y Tecnolog\'ia de Computadoras,  Plaza del Hospital 1, 30202 Cartagena, Spain\label{aff96}
\and
Kapteyn Astronomical Institute, University of Groningen, PO Box 800, 9700 AV Groningen, The Netherlands\label{aff97}
\and
Dipartimento di Fisica, Universit\`a degli studi di Genova, and INFN-Sezione di Genova, via Dodecaneso 33, 16146, Genova, Italy\label{aff98}
\and
Infrared Processing and Analysis Center, California Institute of Technology, Pasadena, CA 91125, USA\label{aff99}
\and
INAF, Istituto di Radioastronomia, Via Piero Gobetti 101, 40129 Bologna, Italy\label{aff100}
\and
Astronomical Observatory of the Autonomous Region of the Aosta Valley (OAVdA), Loc. Lignan 39, I-11020, Nus (Aosta Valley), Italy\label{aff101}
\and
Junia, EPA department, 41 Bd Vauban, 59800 Lille, France\label{aff102}
\and
ICSC - Centro Nazionale di Ricerca in High Performance Computing, Big Data e Quantum Computing, Via Magnanelli 2, Bologna, Italy\label{aff103}
\and
Instituto de F\'isica Te\'orica UAM-CSIC, Campus de Cantoblanco, 28049 Madrid, Spain\label{aff104}
\and
CERCA/ISO, Department of Physics, Case Western Reserve University, 10900 Euclid Avenue, Cleveland, OH 44106, USA\label{aff105}
\and
Dipartimento di Fisica e Scienze della Terra, Universit\`a degli Studi di Ferrara, Via Giuseppe Saragat 1, 44122 Ferrara, Italy\label{aff106}
\and
Laboratoire Univers et Th\'eorie, Observatoire de Paris, Universit\'e PSL, Universit\'e Paris Cit\'e, CNRS, 92190 Meudon, France\label{aff107}
\and
Departamento de F{\'\i}sica Fundamental. Universidad de Salamanca. Plaza de la Merced s/n. 37008 Salamanca, Spain\label{aff108}
\and
Istituto Nazionale di Fisica Nucleare, Sezione di Ferrara, Via Giuseppe Saragat 1, 44122 Ferrara, Italy\label{aff109}
\and
Kavli Institute for the Physics and Mathematics of the Universe (WPI), University of Tokyo, Kashiwa, Chiba 277-8583, Japan\label{aff110}
\and
Dipartimento di Fisica - Sezione di Astronomia, Universit\`a di Trieste, Via Tiepolo 11, 34131 Trieste, Italy\label{aff111}
\and
Minnesota Institute for Astrophysics, University of Minnesota, 116 Church St SE, Minneapolis, MN 55455, USA\label{aff112}
\and
Institute for Astronomy, University of Hawaii, 2680 Woodlawn Drive, Honolulu, HI 96822, USA\label{aff113}
\and
Department of Physics \& Astronomy, University of California Irvine, Irvine CA 92697, USA\label{aff114}
\and
Department of Astronomy \& Physics and Institute for Computational Astrophysics, Saint Mary's University, 923 Robie Street, Halifax, Nova Scotia, B3H 3C3, Canada\label{aff115}
\and
Departamento F\'isica Aplicada, Universidad Polit\'ecnica de Cartagena, Campus Muralla del Mar, 30202 Cartagena, Murcia, Spain\label{aff116}
\and
Department of Physics, Oxford University, Keble Road, Oxford OX1 3RH, UK\label{aff117}
\and
CEA Saclay, DFR/IRFU, Service d'Astrophysique, Bat. 709, 91191 Gif-sur-Yvette, France\label{aff118}
\and
Institute of Cosmology and Gravitation, University of Portsmouth, Portsmouth PO1 3FX, UK\label{aff119}
\and
Department of Astronomy, University of Florida, Bryant Space Science Center, Gainesville, FL 32611, USA\label{aff120}
\and
Department of Computer Science, Aalto University, PO Box 15400, Espoo, FI-00 076, Finland\label{aff121}
\and
Instituto de Astrof\'\i sica de Canarias, c/ Via Lactea s/n, La Laguna E-38200, Spain. Departamento de Astrof\'\i sica de la Universidad de La Laguna, Avda. Francisco Sanchez, La Laguna, E-38200, Spain\label{aff122}
\and
Ruhr University Bochum, Faculty of Physics and Astronomy, Astronomical Institute (AIRUB), German Centre for Cosmological Lensing (GCCL), 44780 Bochum, Germany\label{aff123}
\and
DARK, Niels Bohr Institute, University of Copenhagen, Jagtvej 155, 2200 Copenhagen, Denmark\label{aff124}
\and
Univ. Grenoble Alpes, CNRS, Grenoble INP, LPSC-IN2P3, 53, Avenue des Martyrs, 38000, Grenoble, France\label{aff125}
\and
Department of Physics and Astronomy, Vesilinnantie 5, 20014 University of Turku, Finland\label{aff126}
\and
Serco for European Space Agency (ESA), Camino bajo del Castillo, s/n, Urbanizacion Villafranca del Castillo, Villanueva de la Ca\~nada, 28692 Madrid, Spain\label{aff127}
\and
ARC Centre of Excellence for Dark Matter Particle Physics, Melbourne, Australia\label{aff128}
\and
Centre for Astrophysics \& Supercomputing, Swinburne University of Technology,  Hawthorn, Victoria 3122, Australia\label{aff129}
\and
School of Physics and Astronomy, Queen Mary University of London, Mile End Road, London E1 4NS, UK\label{aff130}
\and
ICTP South American Institute for Fundamental Research, Instituto de F\'{\i}sica Te\'orica, Universidade Estadual Paulista, S\~ao Paulo, Brazil\label{aff131}
\and
IRFU, CEA, Universit\'e Paris-Saclay 91191 Gif-sur-Yvette Cedex, France\label{aff132}
\and
Oskar Klein Centre for Cosmoparticle Physics, Department of Physics, Stockholm University, Stockholm, SE-106 91, Sweden\label{aff133}
\and
Astrophysics Group, Blackett Laboratory, Imperial College London, London SW7 2AZ, UK\label{aff134}
\and
INAF-Osservatorio Astrofisico di Arcetri, Largo E. Fermi 5, 50125, Firenze, Italy\label{aff135}
\and
Dipartimento di Fisica, Sapienza Universit\`a di Roma, Piazzale Aldo Moro 2, 00185 Roma, Italy\label{aff136}
\and
Centro de Astrof\'{\i}sica da Universidade do Porto, Rua das Estrelas, 4150-762 Porto, Portugal\label{aff137}
\and
HE Space for European Space Agency (ESA), Camino bajo del Castillo, s/n, Urbanizacion Villafranca del Castillo, Villanueva de la Ca\~nada, 28692 Madrid, Spain\label{aff138}
\and
Aurora Technology for European Space Agency (ESA), Camino bajo del Castillo, s/n, Urbanizacion Villafranca del Castillo, Villanueva de la Ca\~nada, 28692 Madrid, Spain\label{aff139}
\and
Institute of Astronomy, University of Cambridge, Madingley Road, Cambridge CB3 0HA, UK\label{aff140}
\and
Department of Astrophysics, University of Zurich, Winterthurerstrasse 190, 8057 Zurich, Switzerland\label{aff141}
\and
Theoretical astrophysics, Department of Physics and Astronomy, Uppsala University, Box 515, 751 20 Uppsala, Sweden\label{aff142}
\and
School of Physics \& Astronomy, University of Southampton, Highfield Campus, Southampton SO17 1BJ, UK\label{aff143}
\and
Department of Physics, Royal Holloway, University of London, TW20 0EX, UK\label{aff144}
\and
Mullard Space Science Laboratory, University College London, Holmbury St Mary, Dorking, Surrey RH5 6NT, UK\label{aff145}
\and
Department of Physics and Astronomy, University of California, Davis, CA 95616, USA\label{aff146}
\and
Department of Astrophysical Sciences, Peyton Hall, Princeton University, Princeton, NJ 08544, USA\label{aff147}
\and
Cosmic Dawn Center (DAWN)\label{aff148}
\and
Niels Bohr Institute, University of Copenhagen, Jagtvej 128, 2200 Copenhagen, Denmark\label{aff149}
\and
Center for Cosmology and Particle Physics, Department of Physics, New York University, New York, NY 10003, USA\label{aff150}
\and
Center for Computational Astrophysics, Flatiron Institute, 162 5th Avenue, 10010, New York, NY, USA\label{aff151}}

 \date{Received  July 28, 2024}
%
%
\abstract{Galaxy proto-clusters are receiving an increased interest since most of the processes shaping the structure of clusters of galaxies and their galaxy population are happening at early stages of their formation. 
The Euclid Survey will provide a unique opportunity to discover a large number of proto-clusters over a large fraction of the sky ( 14\,500 deg$^2$). 
In this paper, we explore the expected observational properties of proto-clusters in the Euclid Wide Survey by means of theoretical models and simulations. 
We provide an overview of the predicted proto-cluster extent, galaxy density profiles, mass-richness relations, abundance, and sky-filling as a function of redshift.
Useful analytical approximations for the functions of these properties are provided. 
The focus is on the redshift range $z= 1.5$ to $4$. 
We discuss in particular the density contrast with which proto-clusters can be observed against the background in the galaxy distribution if photometric galaxy redshifts are used as supplied by the ESA \Euclid mission together with the ground-based photometric surveys.
We show that the obtainable detection significance is sufficient to find
large numbers of interesting proto-cluster candidates. 
For quantitative studies, additional spectroscopic follow-up is required to confirm the proto-clusters and establish their richness.
}
%
%
%
%
   \titlerunning{\Euclid\/: Euclid proto-cluster properties}
   \authorrunning{H. B\"ohringer et al.}
   
   \maketitle
%
%
\section{\label{sc:Intro}Introduction}

The interest in galaxy proto-clusters has strongly increased in the recent past, thanks to 
observational capabilities of new survey instruments, which
revealed how some of the most essential processes shaping the present-day
galaxy population in clusters happened already in the very early stages of
cluster formation.
The desire to obtain direct observational evidence of these processes at high redshifts has motivated many recent observational studies of proto-clusters (e.g. \citealt{Ove2016,Alb2022}). 

Galaxy proto-clusters have been found in different ways. 
On the one hand, as serendipitous detections in systematic surveys: (1) in photometric surveys often conducted to find distant galaxies. 
One of the first of these discoveries is the proto-cluster in the SSA22 field as an overdensity of Ly-break galaxies at redshift $z$\,$\sim$\,$3$ \citep{Ste1998}. 
Other such detections include overdensities 
of Ly$\alpha$-emitters (\citealt{Shi2003} -- Subaru Deep Field, \citealt{Ouc2005} - COSMOS Survey, \citealt{Hig2019} -- SUBARU HSC Survey), i-band drop outs (\citealt{Tos2012} -- Subaru Deep Field), and multi-band photometric redshifts (\citealt{Chi2014} -- COSMOS Survey). 
They were also found (2) in spectroscopic surveys, for example, in the VIMOS Ultra Deep Survey \citep{Cuc2014,Cuc2018,Lem2014, Har2019}, and (3) as concentrations of sub-mm sources in the Planck Survey \citep{Pla2015,Flo2016,Cal2023}, by the South Pole Telescope \citep{vie2010,Mil2018}, by the Herschel Space Observatory \citep{Cle2014,Gre2018},  and in other deep survey fields \citep{Dad2009,Dan2014,Cas2015,Ote2018,Gom2019,Wan2021}.
Recently also the detection of proto-clusters as overdensities of passive galaxies was reported \citep{Str2015,Mcc2022,Ito2023}, where the objects of \citet{Str2015} are not easily classified into either clusters or proto-clusters.
On the other hand, interesting proto-cluster systems were discovered around particular objects marking dense regions of the Universe, so-called signposts for proto-clusters, like (4) radio galaxies, which have been used for quite some time to search for dense environments \citep{Lef1996, Pen2000,Kur2000,Kur2004,Ven2002,Ven2004,Ven2007,Mil2008,Kui2011,Hat2011a,Hat2011b,Gal2012,Wyl2013,Koy2013,Cas2014a,Cas2014b}, 
(5) AGN (active galactic nuclei) playing a similar role as radio galaxies \citep{Djo2003,Hen2015,Gar2017,Gar2019,Gar2022}, 
(6) Ly-{$\alpha$} blobs \citep{Cal2023}, and last but not least (7) by absorption in the light of background objects  \citep{Fra1996,Ste1998,Hen2015}.

While these observed systems span a range of properties (not all of these objects may end up in one galaxy cluster), one looks for a unifying description. 
The general idea is, hereby,  to call a proto-cluster those structures which are expected to evolve into a galaxy cluster by redshift $z=0$ (e.g. \citealt{Ste1998,Ove2016}), whereby theoretical modelling or simulations are used to connect the observations to present day cluster properties. 
In \citet{Ste1998}, in one of the earliest of such studies, the structure evolution model of a spherical top-hat overdensity was used to relate an observed overdensity to the expectation for a galaxy cluster at the present day.
Another approach is to use $N$-body simulations to trace the evolution of $z=0$ clusters back to the observations' redshifts and provide, in this way, relations between present-day cluster masses and the properties of their precursors at high redshift.
\citet{Chi2013} provide results from such a study and present correlations between the proto-cluster overdensity and the expected cluster mass at $z=0$, which has been applied with some success to several proto-cluster observations (e.g. \citealt{Cuc2014}). Also, \citet{Con2016} studied proto-cluster sizes in simulations. However, they used boxes instead of spheres, which makes a comparison to other work more difficult.

The ESA-\Euclid mission \citep{Scaramella-EP1,Cropper2024,Jahnke2024,Hormuth2024,Mellier2024} with its deep near-infrared and high-angular-resolution visual band survey, together with the auxiliary ground-based optical survey data, will provide a unique opportunity to search for proto-clusters over a large region of the sky. This will not only increase the number of known proto-clusters and improve the statistics on their properties, but also turn up rare and massive systems, which can only be found in large survey volumes.
The main part of the Euclid Survey is the Wide Survey of the sky outside the Galactic band with an area of about 14\,00\,deg$^2$ over six years.
It will reach estimated limiting AB magnitudes (5\,$\sigma$ for point-like sources) of about 26.2 in the visual band , \IE \footnote{The visual band of the \Euclid VIS instrument covers a wavelength range from 500 to 900 nm.}, and 24.5 for the near-infrared bands \YE, \JE, \HE \citep{Scaramella-EP1} .
This paper provides studies of proto-cluster properties and their appearance in the Euclid Wide Survey by means of simulations in order to explore the prospects for the search of proto-clusters in the \Euclid sky. 
Euclid will enormously increase the number of known
proto-clusters at high redshifts and thus provide the base for precise statistical studies on proto-cluster structure, early galaxy evolution in dense environments,
and the origin of present day cluster properties. But it will also provide
the large survey volume to find the most interesting objects, the precursors of the most massive galaxy clusters.

In this paper we define a proto-cluster as a matter and galaxy concentration at higher redshift that is bound to develop into a galaxy cluster by redshift zero with a mass larger than $10^{14}$\,\si{\solarmass} inside $r_{200}$.\footnote{The radius $r_{200}$ is the radius inside which the mean matter overdensity of the cluster is 200 times the critical density of the Universe at the cluster redshift.} 
We focus mainly on the redshift range $z = 1.5$ to $4$. 
At higher redshifts, the galaxy density in the Euclid Survey is too sparse to effectively characterise proto-clusters, while at lower redshifts, galaxy clusters are already abundant.
In the following, we explore the observational features of such proto-clusters using analytical models and cosmological simulations. We also estimate their abundance as a function of redshift. 

The paper is organised in the following way. 
In Section 2, we describe the formalism of proto-cluster evolution with a top-hat model, while in Section 3 we give information on the cosmological simulations used to explore proto-cluster properties. 
The following sections are focused on different proto-cluster properties, such as sizes  (Sect. 4), galaxy density profiles (Sect. 5), projected contrast in observations (Sects. 6 and 7), and the mass-richness relation (Sect. 8). 
Section 9 discusses the expected proto-cluster abundances and their sky-filling factors
(how much of the sky is covered by proto-clusters in projection).
Discussions of the findings are provided in Section 10 and in Section 11 we present our summary and conclusions. 
Unless stated otherwise, we use a cosmological model with 
$h_{100}= 0.7 = H_0 / 100\,{\rm km}\,{\rm s}^{-1}\,{\rm Mpc}^{-1}$, $\Omega{\rm _m} = 0.3$, and a flat metric, which is referred to as ``reference model'' in the following.

\section{\label{sc:model}Proto-cluster model}

As a basic characterisation of proto-clusters, we explore their overdensity evolution in this section.
For galaxy clusters and their formation, a simple, general concept that can be expressed with analytic formulas has helped us very much in guiding our thoughts, the so-called ``Press--Schechter Model'' \citep{Pre1974} and its extensions (e.g. \citealt{Bon1991,She1999}). 
It is in its original form based on the collapse model of a homogeneous overdense sphere and the first-order statistics of density peaks in the large-scale matter distribution. 
It provides the most essential information to characterise the galaxy cluster population, such as the mass function, their characteristic sizes, and their number density evolution with time. 
Its precision for the prediction of number counts is usually better than a factor of two for low redshifts and not extremely high masses  $(\lesssim 10^{15}$ \si{\solarmass}) \citep{Bon1991}. 
This is not enough for precise cosmological modelling. 
But this concept has also provided the frame in which more precise analytical models have been devised, which have been calibrated with $N$-body simulations (e.g. \citealt{Jen2001,Evr2002,Tin2010,despali16,Cas2021}).
In recent work, we applied this concept also successfully to those superclusters, which are expected to collapse in the future -- calling them ``superstes-clusters'' \citep{Cho2015}. 
These superclusters have the same relation to future galaxy clusters as high-redshift proto-clusters to galaxy clusters of the present day. 

Therefore, it is well justified to apply this concept analogously to our proto-cluster project. 
The adopted model describes the evolution of a homogeneous top-hat spherical overdensity in a $\Lambda$CDM universe. 
The calculation of the evolution is based on Birkhoff's theorem, that a homogeneous sphere in a homogeneous background universe evolves like a universe with the local parameters as cosmological parameters.
After finding the proper initial conditions for a collapse at redshift zero, we follow the evolution of the overdensity by integrating the Friedmann equations, including a $\Lambda$ term starting at high redshift to $z = 0$. 
In the calculation, we determine the overdensity with respect to the mean matter density and the evolution of the radius of the overdense region with respect to $r_{200}$ at $z = 0$. 

\begin{table}[htbp!]
\caption{Cosmological parameters used in the different models,
assuming a flat metric}
\label{table:cosmpars}
\[
    \begin{array}{llll}
        \hline
        \noalign{\smallskip}
        {\rm model}   & \Omega{\rm _m} &  h_{100} & \sigma_8   \\ 
        \noalign{\smallskip}
        \hline
        \noalign{\smallskip}
        {\Planck}     & 0.311 &  0.677  &  0.816 \\
        {\rm Millennium} & 0.25 &  0.730  &  0.90 \\
        {\rm reference}  & 0.30 &  0.700  &   -    \\
        {\rm {\sf REFLEX}~clusters} & 0.29 & 0.700^* & 0.77 \\
        \noalign{\smallskip}
        \hline
        \noalign{\smallskip}
    \end{array}
\]

$^*$ The Hubble parameter was present in this study. 
Its influence on the results is small and discussed in~\citet{Boe2014}.
The {\sf REFLEX} cluster model is only used in Sects. 9 and 10.
\end{table}

We define a matter overdensity ratio as the ratio of the mean proto-cluster density, $\bar\rho_{\rm pc}(z)$, to the background density, $R_{\rm ov-DM}(z) = \bar\rho_{\rm pc}(z) / \rho_{\rm m}(z)$,  where $\rho_{\rm m}(z)$ is the mean matter density at the given redshift. 
We have calculated $R_{\rm ov-DM}(z)$ by means of the spherical collapse model for three different sets of cosmological parameters: for the reference model defined in the introduction, the cosmology resulting from the Planck Survey  \citep{Pla2016}, and the cosmology used in the Millennium Simulations \citep{Spr2005}. 
Table~\ref{table:cosmpars} gives these three sets of cosmological parameters. 
We also list the cosmological parameters inferred from the present-day cluster population in the {\sf REFLEX} cluster survey \citep{Boe2014}.
The cluster mass function used later in this study \citep{Boe2017} is based on these results. 

\begin{figure}[htbp!]
   \includegraphics[width=\columnwidth]{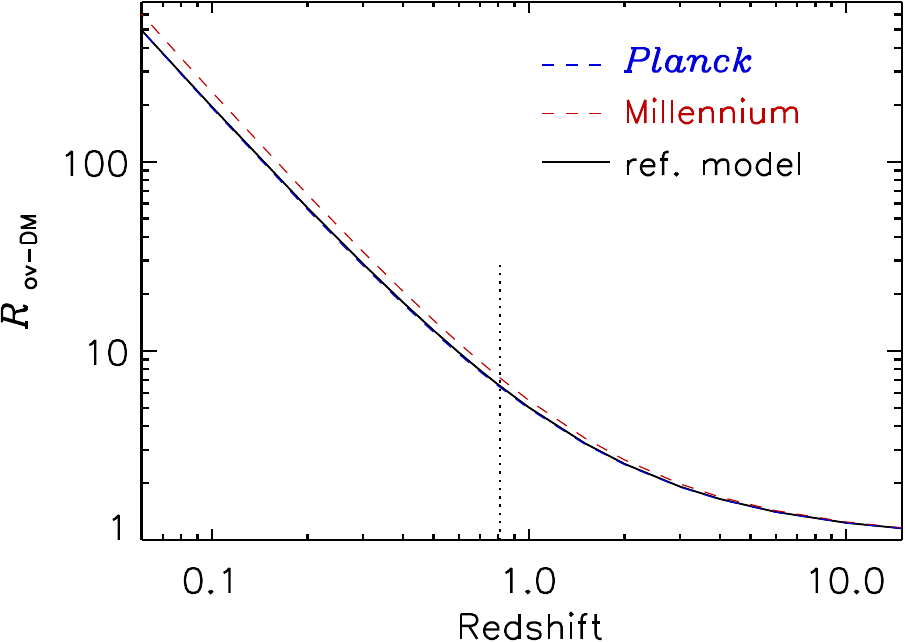}
    \caption{Evolution of the overdensity ratio, $R_{\rm ov-DM}(z)$, with redshift for an object that collapses at $z = 0$. 
The results for three cosmological models: the reference model, Planck cosmology, and Millennium cosmology are shown. The first two models can hardly be distinguished in the plot. 
The vertical line indicates the redshift of the turn-around point. 
}
\label{fig:fig1}
\end{figure}

The resulting evolution of $R_{\rm ov-DM}(z)$ is shown in Fig.~\ref{fig:fig1}. 
The curves for the reference model and the {\it Planck} cosmology can hardly be distinguished in the figure, but the Millennium result is slightly different. The difference is mostly due to the choice of $\Omega_{\rm m}$ and $\sigma_8$, for which an older preference was used in the Millennium Simulations. The difference in the cosmological models does not significantly change the observables discussed in the following.
We also indicate in the figure the turn-around redshift, at which the overdensity stops expanding and starts to collapse. 

As reference we provide a few numerical values of $R_{\rm ov-DM}(z)$:
3.24, 2.52, 1.91, 1.64 for $z = 1.5$, 2, 3, 4, respectively.
The overdensities are small for the time before the turn-around, which occurs at $ z \sim 0.8$ in the reference model.
For an object that collapses at $z = 0$, the overdensity ratio at turn-around is about 6.6. After the turn-around, the overdensity ratio increases rapidly.

\section{\label{sc:sims}Cosmological simulations}

Three sets of simulations in the form of lightcones are used for the following study. The MAMBO, GAEA H, and GAEA  F lightcones are derived from the Millennium Simulations \citep{Spr2005,Boy2009}. The size of the lightcones and the cosmological parameters on which the simulations were based are given in Table  \ref{table:tab3}.
Importantly, these lightcones contain information about the merger trees, such that for each galaxy, one can find out if it is included in a larger dark matter halo, a group or cluster of galaxies, at redshift zero. The mass of this final host halo is known, and this parameter is used throught this paper. These three simulations were the only ones available to us where this information necessary for our work was included. 
For the two GAEA light cones different semi-analytical models were used to determine the properties of the galaxy population.
Some of the following analysis required special information only available for the MAMBO light cone at the time
of writing. In this case the results are shown only for this simulation. One of the critical differences in the semi-analytic modelling of the galaxy population which can effect the observables derived from the light cones is the stellar mass function used, as shown by \citet{Fu2024}. 

The MAMBO simulation based on empirical relations, developed by M. Bolzonella, L. Pozzetti, and G. Girelli, uses the halo and sub-halo positions as well as dark matter masses of one of the 24 lightcones built by \citet{Henr2015}. Under the assumption that each subhalo is hosting one galaxy, the stellar mass was derived from the stellar-to-halo mass relation \citep{Gir2020}, and the physical and observed properties (e.g. star formation rate, dust attenuation, gas metallicity, morphology, emission lines, broadband rest-frame and observed fluxes) from empirical relations implemented in a modified version of the Empirical Galaxy Generator code (EGG; \citealt{Sch2017}).
In \citet{Henr2015}, the Millennium Simulations have been re-scaled to a cosmology consistent with the {\it Planck} results (\citealt{Pla2016};  see Table~\ref{table:cosmpars}). 

For the GAEA lightcones, we take advantage of lightcones built from predictions based on the Galaxy Evolution and Assembly ({\sc GAEA}) theoretical model. 
This is coupled with dark matter merger trees extracted from the Millennium Simulation \citep{Spr2005}, which makes it possible to track the evolution of each model galaxy both to higher and lower redshifts (in terms of their progenitors and descendants). 
This approach allows us to study the later evolution of systems that are identified as proto-cluster regions and characterise the $z=0$ descendant mass distribution.
GAEA follows the evolution of galaxies across different cosmic epochs and environments by means of a coupled system of differential equations, each of them describing a single physical process responsible for the exchange of mass and energy among the different baryonic components.  
The individual prescriptions can be of empirical, analytical, or theoretical derivation and can involve the definition of free parameters that are usually calibrated against a selected set of observational constraints. 

In this paper, we consider two different GAEA realisations based on the model versions published in \citet{Hir2016} and \citet{Fon2020}. 
Hereafter, we will refer to these realisations as GAEA-H and GAEA-F, respectively. 
Both model runs include a detailed treatment for non-instantaneous chemical enrichment \citep{DeLu2014} and a prescription for stellar feedback partly based on hydro-simulations. 
These prescriptions allow us to reproduce the evolution of the galaxy stellar mass function as well as cosmic star formation rate up to $z\lesssim7$ \citep{Fon2017}, as well the evolution of the mass-metallicity relations and their secondary dependencies \citep{DeLu2020, Fon2021}. 
GAEA-F also includes an improved treatment of cold gas accretion onto supermassive black holes, an explicit treatment for AGN-driven winds \citep{Fon2020}, and an updated tracing of the angular momentum exchanges between different galactic components, which we use to model
galaxy structural properties \citep{Xie2020}. 
While retaining all successes of the previous model, this update also reproduces the properties of the AGN populations up to $z \sim 4$.

For each model galaxy, GAEA predicts the expected broadband photometry in the {\it H}-band and in the \Euclid visual band, in addition to a number of physical properties and other photometric bands. 
These model outputs have been used to construct two independent lightcones (each for each model realisation) using the algorithm described in \citet{Zol2017}. Each lightcone covers an aperture of \ang{5.27} diameter and includes all model galaxies from $z=0$ to $z=4$ down to $\HE=25$.

In the simulations, galaxies keep their identity through time and can be traced through the merger trees. Therefore, galaxies that are found to belong to a galaxy cluster at redshift zero with a mass $M_{200}\ge 10^{14}$ \si{\solarmass} inside $r_{200}$ can be identified and labelled as members of a proto-cluster at higher redshift in the lightcone. The ensemble of the member galaxies defines the proto-cluster.

\begin{table}
    \caption{Parameters for the lightcones and the underlying cosmological simulations.}
    \label{table:tab3}    
  \[
         \begin{array}{llllll}
            \hline
            \noalign{\smallskip}
{\rm data~set}&{\rm radius}&{\rm Sky~area}& h_{100} & \Omega_{\rm m}& \sigma_8 \\
              & {\rm deg} & {\rm deg}^2 &  &   \\ 
            \noalign{\smallskip}
            \hline
            \noalign{\smallskip}
{\rm MAMBO}  & 1.0   &    3.14 &  0.673 &  0.311 & 0.816 \\
{\rm GAEA~H} & 2.635 &  21.813 &  0.73  &  0.25  & 0.9   \\
{\rm GAEA~F} & 2.635 &  21.813 &  0.73  &  0.25  & 0.9   \\
            \noalign{\smallskip}
            \hline
            \noalign{\smallskip}
         \end{array}
      \]
   \end{table}

For our studies, we selected galaxies with the following magnitude limits, $m_{\HE} \le 24.25$ for the near-infrared {\it H}-band and $m_{\IE} \le 25.10$ for the visual band of \Euclid.  
The lightcone databases provide observed and randomly perturbed magnitudes according to the expected measurement errors. Here, we use the unperturbed magnitudes.
The limits correspond to an expected S/N\,=\,5 \Euclid will reach in the Wide Survey for extended sources. 
For each galaxy, redshifts calculated from the simulations with and without peculiar motions are available.  We use the values without peculiar motion. Also, the photometric redshift was derived for each galaxy with the SED fitting code \texttt{Phosphoros} developed in the Euclid Collaboration (Paltani et al. in preparation), taking into account the photometric noise expected in \Euclid bands complemented by the ground-based ones that will be available at the time of the Data Release 3 southern hemisphere\footnote{The photometric data used for the determination of the photometric redshifts include the \Euclid bands: \IE, \YE, \JE, \HE~ and the LSST bands: {\it u, g, r, i, z.}} \citep{Scaramella-EP1}. 
The lightcone databases contain the entire probability distribution of the derived photometric redshift. Here, we use only the median values.


\section{\label{sc:extent}Extent of proto-clusters}

In this section, we present estimates of the proto-cluster radius, $r_{\rm pc}$. We first derive a theoretical estimate for the radius evolution of a spherical overdensity in Sect. 4.1 and compare the results to simulations in Sect. 4.2.

\subsection{Theoretical estimate}

For $r_{\rm pc}$, we take the radius of the overdensity, which evolves into a galaxy cluster with radius $r_{200}$ in the spherical collapse model described in Sect. 2. 
The proto-cluster radius, $r_{\rm pc}$ in comoving units, $r_{\rm com}$, can then be determined from the overdensity ratio through the relation: 

\begin{equation}
 r_{\rm com} = r_{200}~ \left( \frac{\Omega_{\rm m}~  R_{\rm ov-DM}}{200} \right) ^{-1/3} ,
\end{equation}
and the physical radius of the proto-cluster is given by 
$r_{\rm phys} = r_{\rm com}/(1+z)$. 

\begin{figure}[htbp!]
\includegraphics[width=\columnwidth]{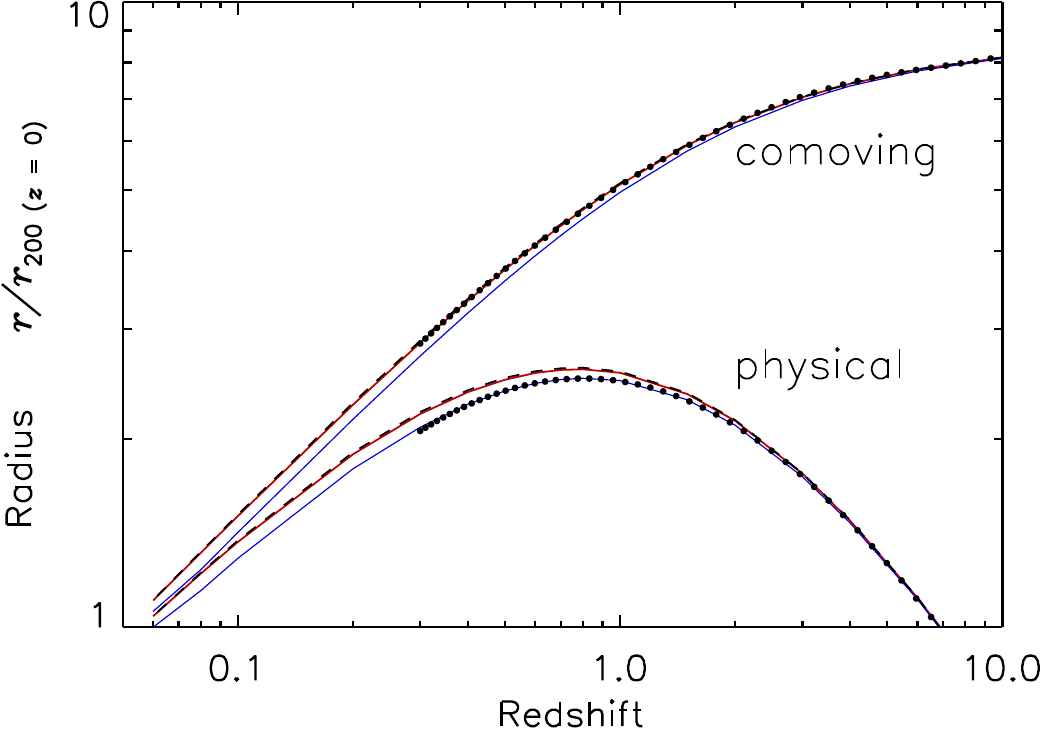}
    \caption{Evolution of the proto-cluster radius in comoving and physical units scaled to $r_{200}$ at z=0.
Results for three cosmological models are shown by red lines (reference model), blue lines (Millennium cosmology) and dashed black lines (Planck cosmology).  The over-plotted black dots show the parameterised approximation of $r_{\rm com}$  (see Eq. 2) for the reference cosmology and $r_{\rm phys}$ for the Millennium cosmology.
}
\label{fig:radius}
\end{figure}

Figure~\ref{fig:radius} shows the results of the calculations for the three cosmological models listed in Table~\ref{table:cosmpars}.
The results for the reference and {\it Planck} cosmology are practically identical, while those for the Millennium cosmology are different by a few per cent. 
\citet{Chi2017} published similar calculations based on simulations, which are in good agreement with our results. Similarly, \citet{Mul2015} used simulations to assess the evolution of the proto-cluster radius, defined as enclosing 90\% of the stellar mass ending up in the $z = 0$ cluster and found similar results. In their Fig.~2, we see that the physical radius shows a similar function of time with a value $\sim 2.5$ times higher at $z = 1$ than at $z=0$ and a factor of $\sim 1.2$ higher at $z = 5$.

For further practical work, we derived numerical fits to the results for $r_{\rm com}$. 
The following approximation, valid for the redshift range $z = 0.5$ to $8$, provides an accuracy better than 1\%, with parameters listed in Table~\ref{table:tab4},

\begin{equation}
r_{\rm com}/r_{200} = A~z^{0.7} + B~z^{0.95} + C~z^{1.1} + D ~~.
\end{equation}

We use this relation for the reference cosmology in the further work in this paper.

   \begin{table}
      \caption{Parameters describing the evolution of the proto-cluster 
               radius with redshift (see Eq. 2), only valid in the redshift range $z = 0.5$ to 8.}
  \[
         \begin{array}{lrrrr}
            \hline
            \noalign{\smallskip}
{\rm model}   & A &  B & C & D  \\ 
            \noalign{\smallskip}
            \hline
            \noalign{\smallskip}
{\rm reference}  & 19.50 & -22.75  & 9.05  & -0.71 \\
{\it Planck}     & 19.54 & -22.83  & 9.08  & -0.68 \\
{\rm Millennium} & 19.38 & -22.43  & 8.88  & -0.88 \\
            \noalign{\smallskip}
            \hline
            \noalign{\smallskip}
         \end{array}
      \]
\label{table:tab4}
   \end{table}

\begin{figure}[h]
   \includegraphics[width=\columnwidth]{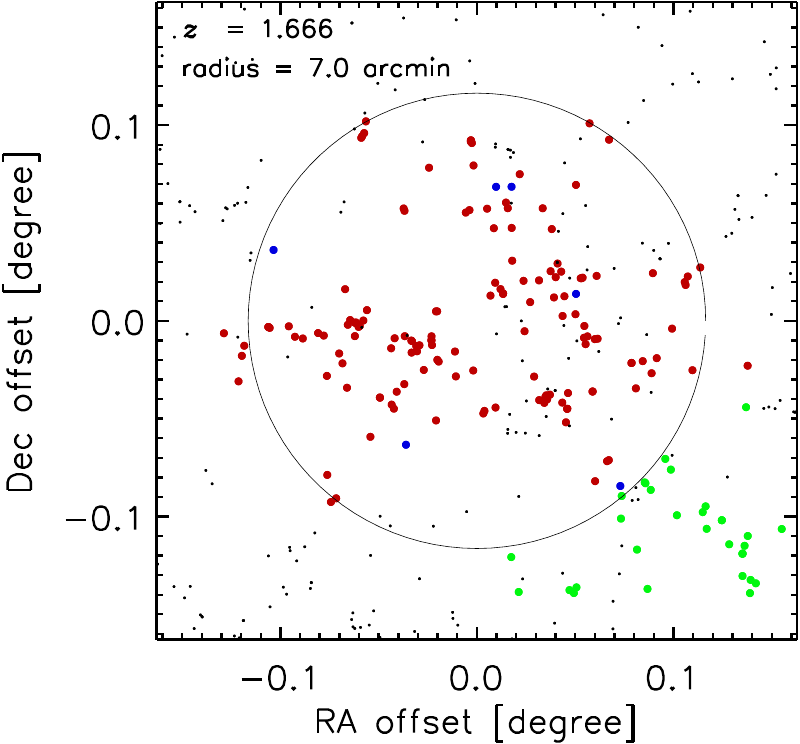}
   \includegraphics[width=\columnwidth]{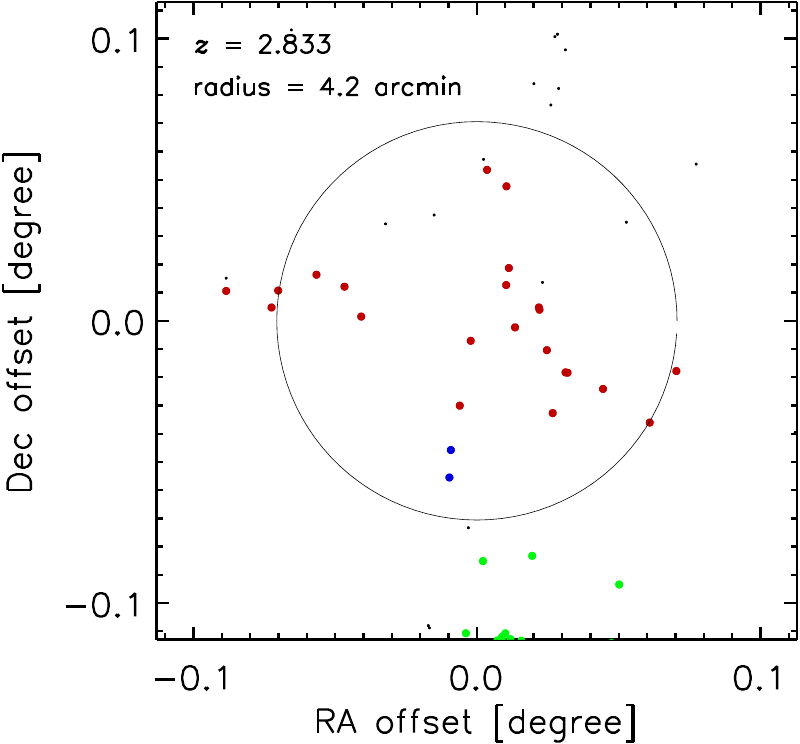}
\caption{Examples of two proto-clusters at redshifts 1.666 and 2.8335 with 148 and 22 members, respectively. 
The proto-cluster members are shown as red-filled circles, while non-members inside the proto-cluster radius are shown as
blue circles. Other non-members outside the proto-cluster radius in three dimensions are shown as small black dots, and members of other proto-clusters are shown as full green circles.  The large black circle indicates the estimated proto-cluster radius, whose
size is indicated in the top left corner of the plot.
}
\label{fig:fig3}
\end{figure}

   \begin{table}
      \caption{Completeness and contamination of the members inside the 
          estimated proto-cluster radius, $r_{\rm pc}$ and $1.2\,r_{\rm pc}$, 
                for different redshift shells for the MAMBO and GAEA samples}
      \[
         \begin{array}{rrrrr}
            \hline
            \noalign{\smallskip}
{\rm redshift}&{\rm complet.}& {\rm contam. } & {\rm complet.}& {\rm contam.} \\
              &  r \le r_{\rm pc} &  r \le r_{\rm pc}& r \le1.2\,r_{\rm pc} & r \le 1.2\, r_{\rm pc}  \\
            \noalign{\smallskip}
            \hline
            \noalign{\smallskip}
{\rm GAEA~H}  &&&&\\
             \noalign{\smallskip}
            \hline
            \noalign{\smallskip}           
 1.5 - 4 & 0.74 &  0.13 & 0.88 & 0.21 \\
 1.5 - 2 & 0.70 &  0.17 & 0.84 & 0.25 \\
   2 - 3 & 0.76 &  0.12 & 0.89 & 0.20 \\
   3 - 4 & 0.80 &  0.08 & 0.91 & 0.16 \\
               \noalign{\smallskip}
            \hline
            \noalign{\smallskip}
{\rm GAEA~F}  &&&&\\           
            \noalign{\smallskip}
            \hline
            \noalign{\smallskip}
 1.5 - 4 & 0.75 & 0.12 & 0.88 & 0.21 \\
 1.5 - 2 & 0.70 & 0.16 & 0.84 & 0.25 \\
   2 - 3 & 0.76 & 0.11 & 0.90 & 0.20 \\
   3 - 4 & 0.81 & 0.07 & 0.90 & 0.15 \\
            \noalign{\smallskip}
            \hline
            \noalign{\smallskip}
{\rm MAMBO}  &&&&\\ 
            \noalign{\smallskip}
            \hline
            \noalign{\smallskip}
 1.5 - 4 &  0.78 & 0.043 & 0.91 & 0.11 \\
 1.5 - 2 &  0.76 & 0.053 & 0.90 & 0.12 \\
   2 - 3 &  0.79 & 0.034 & 0.91 & 0.11 \\
   3 - 4 &  0.81 & 0.038 & 0.92 & 0.11 \\
            \noalign{\smallskip}
            \hline
            \noalign{\smallskip}            
         \end{array}
      \]
\label{table:tab5}
   \end{table}

\subsection{Comparison to simulations} 

We can compare the theoretical predictions for the proto-cluster radius with the distribution of the cluster member galaxies in the MAMBO and GAEA lightcones. 
Here and in the following, we use only proto-clusters that are fully contained in the field of view of the lightcone and in addition reject a small number of objects that are artifacts originating from common problems with the lightcone construction.

Figure~\ref{fig:fig3} shows two proto-clusters from the MAMBO simulation, where galaxies inside and outside the proto-cluster radius (in three dimensions) are marked as members and non-members. As the centre of the proto-clusters, we chose the barycentre. 
As proto-cluster members, we take those galaxies which are members of the descendent cluster at redshift zero inside $r_{200}$.
All galaxies in a redshift slice of $\Delta =$ five times the proto-cluster radius are shown in projection on the sky. 
We note that the majority of the proto-cluster members are located inside the estimated proto-cluster radius, and a few non-members, marked in blue in the figure, are found inside. These non-members are found close to the cluster boundary and are not bound into the cluster during the following collapse of the system. Some of the non-members are seen inside the circle in projection in the plots as black dots, but they are outside the proto-cluster spheres. Members of neighbouring proto-clusters in the same redshift interval are shown as green symbols.

\begin{figure}[h]
   \includegraphics[width=\columnwidth]{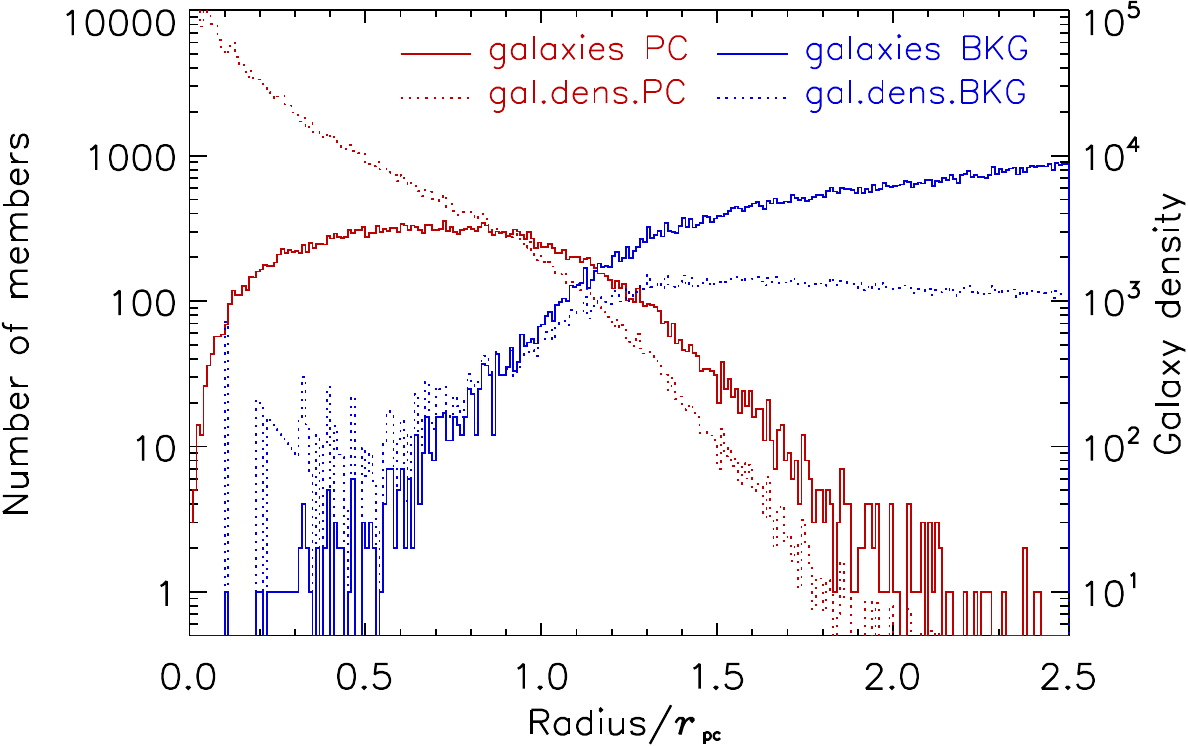}
   \includegraphics[width=\columnwidth]{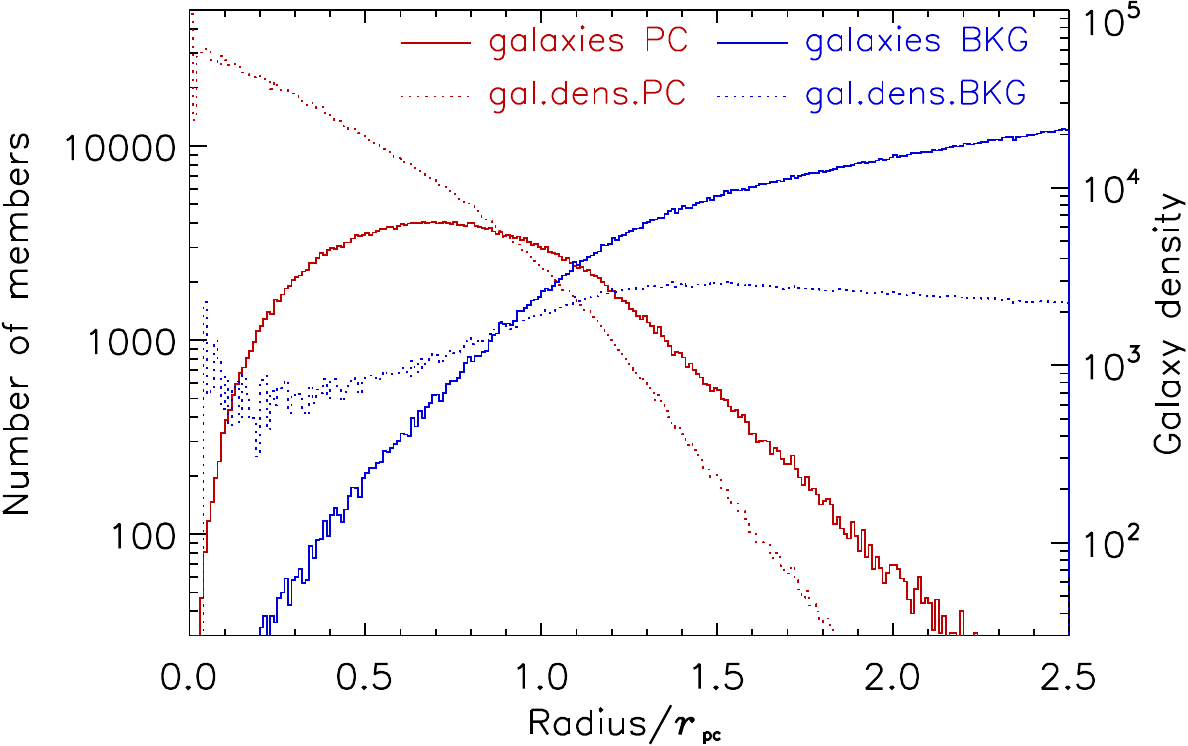} 
   \includegraphics[width=\columnwidth]{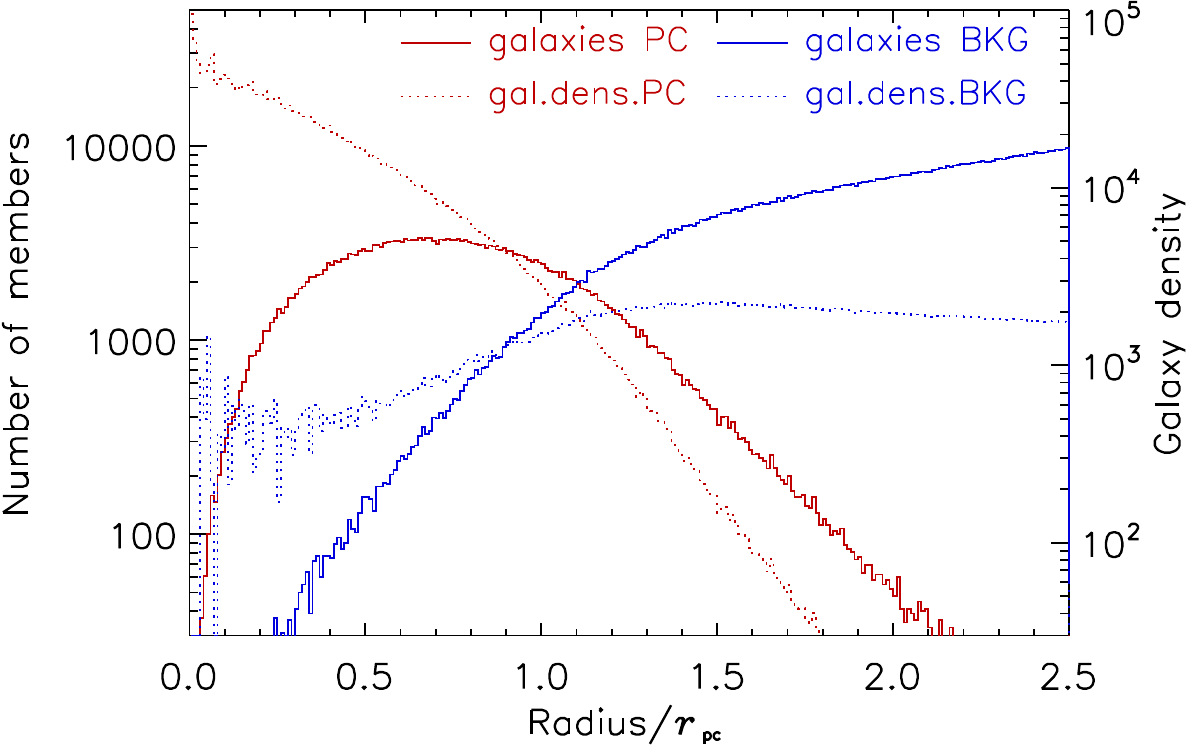}
\caption{Three-dimensional radial distribution of the member galaxies in all MAMBO (top), GAEA-H (middle), and GAEA-F (bottom) proto-clusters ($z = 1.5$ to $4$), where the radius is in units of the estimated proto-cluster radius, $r_{\rm pc}$. The red lines show the numbers and densities of the proto-cluster members, and the blue lines those of the non-member galaxies. The solid lines show the galaxy number in shells (left $Y$-axis), while the dotted lines show the galaxy density in the shells (right $Y$-axis). 
}
\label{fig4}
\end{figure}

Figure~\ref{fig4} shows the three-dimensional radial distribution of the galaxy number and number density of the members of MAMBO, GAEA-H, and GAEA-F proto-clusters compared to the radial distribution of non-members. The differential distribution in spherical shells is shown.
For the proto-cluster centres, we used the barycentre, the mean of the mass distribution of all proto-cluster member galaxies. 
The distance of the galaxies to the centre in all proto-clusters was scaled to the estimated proto-cluster radius, $r_{\rm pc} = r_{\rm phys}~ r_{200}$. 
The radius, $r_{\rm phys}$, has been defined in the sentence connected to Equation~1. 
For the positions, the true locations of the galaxies were used without redshift space distortions due to galaxy peculiar motions.

The distributions of the galaxies in all three simulation data sets look very similar. The contamination of non-members inside the proto-cluster radius in the GAEA lightcones is higher than that in the MAMBO lightcone.
About 80\% of the members are located inside $r_{\rm pc}$ with a contamination of $\sim 5\%$ (MAMBO), $\sim 10\%$ (GAEA) non-members in three dimensions. About 90\% of the members are located inside $1.2\,r_{\rm pc}$ with a contamination of $\sim 10\%$ (MAMBO) and $\sim 20\%$ (GAEA) . 
The theoretically estimated proto-cluster radius provides thus a good orientation for the expected size of proto-clusters. 
Results for splitting the proto-cluster sample up into three redshift shells are given in Table~\ref{table:tab5}, where the completeness is defined as the fraction of member-galaxies contained inside the aperture radius compared to the total number of member galaxies. 
We note little variation with redshift.

\section{\label{sc:radialprofile}Radial galaxy density profiles}

\begin{figure}[h]
   \includegraphics[width=\columnwidth]{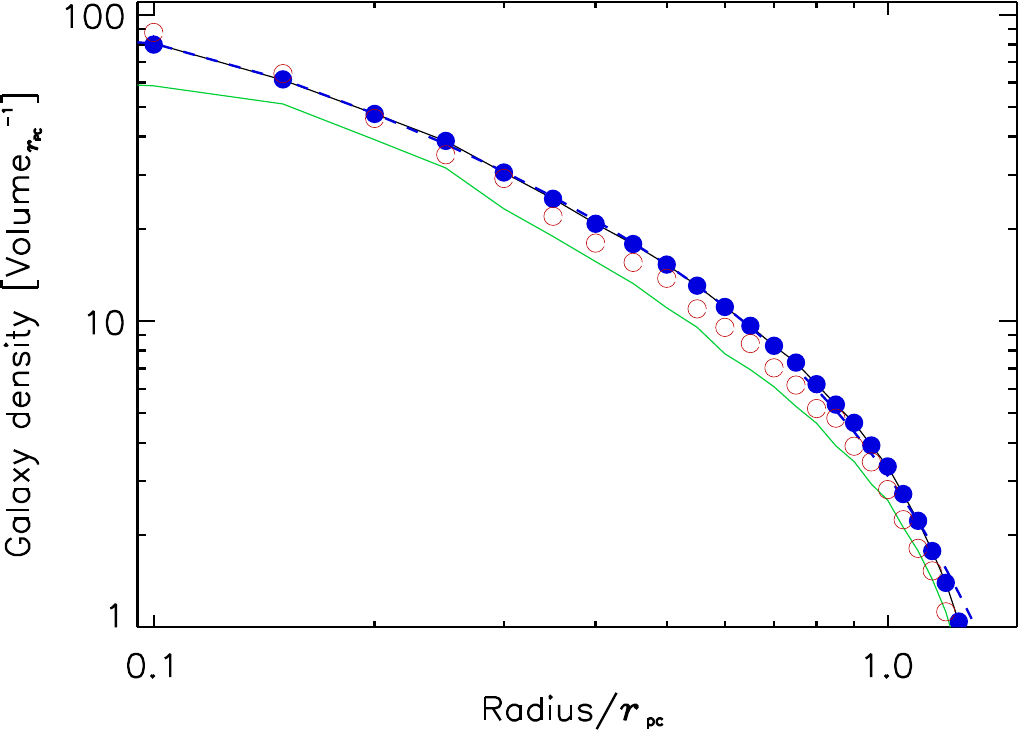}
\caption{Mean three-dimensional radial density profile of proto-clusters.
The black solid line and blue points show the mean profile for all proto-clusters in the GAEA F lightcone in the redshift range $z = 1.5$ to 4, while the green curve shows that for proto-clusters with $z = 1.5$ to 2. The blue dashed curve shows the model fit.
The red open symbols show the density profile for the clusters in the MAMBO 
lightcone ($z$ = 1.5 to 4) for comparison. The galaxy density is scaled to the volume of the proto-cluster, $volume_{r_{\rm PC}} = (4/3) \pi r_{\rm PC}^3$.}
\label{fig:fig5}
\end{figure}

We used the GAEA and MAMBO simulations to study the typical radial density profiles of proto-clusters. Here and throughout the paper, we use the barycentre of the galaxy population as the proto-cluster centre.
In Fig.\,\ref{fig:fig5}  we show the mean
three-dimensional proto-cluster profile for the GAEA-F and MAMBO samples
for the redshift range $z$=1.5 to 4. In addition we show with data for GAEA-F in the range $z$=1.5 to 2, that there is no substantial change with redshift. 
The profiles were generated involving 6533 and 786 proto-clusters
from the two simulations, respectively. 
Since we used radii scaled to the proto-cluster radius, $r_{\rm pc}$, we also determined the mean densities as scaled parameters in units of $ 4/3 \pi r_{\rm pc}^3$. 
The mean profile for GAEA-H has the same shape but about 10\% higher normalisation since this sample has relatively more proto-cluster members. 

We fitted the resulting profile with a model including a power-law profile with a core and an additional inner and outer slope: 

\begin{equation}
n(r) = A~\left(\frac{r}{r_{\rm c}}\right)^{-\alpha}~ \left[1 + \left(\frac{r}{r_{\rm c}}\right)^2 \right]^{-\beta +\frac{\alpha}{2}}  \left[1 + \left(\frac{r}{r_{\rm s}}\right)^2 \right]^{-\gamma +\beta +\frac{\alpha}{2}} ~.
\end{equation}

The resulting fit parameters for the three-dimensional case are shown in Table~\ref{tab_prof}.
We also determined the projected two-dimensional profiles, which look similar with a flatter slope. 
The profiles were fitted with the same relation, and the results are also shown in Table~\ref{tab_prof}. 
In this case, $r_{\rm c}$ and $r_{\rm s}$ are projected radii.
Figure~\ref{fig:fig5} shows that the fit provides reasonable approximations. Actually, a fit with a function including only the first four parameters, with the first two terms on the right hand side of Eq. (3), provides a good approximation up to $r \sim r_{\rm PC}$, but leaves a small shallow tail beyond. To remove this particular feature with an additional outer slope requires large values for the core radius, $r_s$, and the slope parameter. 

   \begin{table}
      \caption{Fit parameters for the proto-cluster profiles in three and two dimensions 
       for the formula in Eq.~(3). The radii, $r_{\rm c}$ and $r_{\rm s}$, are in units of $r_{\rm pc}$.}
      \[
         \begin{array}{lrrrrrr}
            \hline
            \noalign{\smallskip}
{\rm sample}&  {\rm A}& r_{\rm c} & \alpha & \beta & r_{\rm s} & \gamma \\
           \noalign{\smallskip}
 \hline
           \noalign{\smallskip}
{\rm 3~dim.}  &&&&&&\\
           \noalign{\smallskip}
 \hline
           \noalign{\smallskip}
{\rm GAEA~H} &117.7& 0.1413 & 0.2163 & 0.635 & 11.35 & 115.9\\
{\rm GAEA~F} &187.9& 0.0625 & -0.989 & 0.530 & 15.74 & 286.6\\
{\rm MAMBO}  &55.32& 0.2371 & 0.6368 & 0.790 & 16.71 & 198.3\\
          \noalign{\smallskip}
 \hline
           \noalign{\smallskip}
{\rm 2~dim.}  &&&&&&\\
           \noalign{\smallskip}
 \hline
           \noalign{\smallskip}
{\rm GAEA~H} &34.79& 1.004  & 0.2121 & 2.98 & 242.2 & 1050 \\
{\rm GAEA~F} &59.09& 0.0968 & -0.110 & 0.254 & 15.61 & 364.0\\
{\rm MAMBO}  &127.4& 0.0209 & -2.940 & 0.279 & 9.475 & 128.8\\
           \noalign{\smallskip}
            \hline
            \noalign{\smallskip}
         \end{array}
      \]
\label{tab_prof}
   \end{table}

\begin{table}
    \caption{Definition of the five different categories of
  proto-cluster density profiles}
\[
    \begin{array}{llrrr}
    \hline
    \noalign{\smallskip}
{\rm category}&{\rm condition}& N_{\rm pc} & {\rm percent} \\
    \noalign{\smallskip}
\hline
    \noalign{\smallskip}
 1 &{\rm D1 > D2 > D3} & 542 & 69 \\
 2 &{\rm D1 > D3 > D2}  &  33 &  4.2 \\
 3 &{\rm D2 > D1 > D3}  &  67 &  8.5 \\
 4 &{\rm D2 > D3 > D1}  & 127 & 16.1 \\
 5 & {\rm D3 > D1~ and~ D2          } &  17 &  2.2 \\ 
    \noalign{\smallskip}
    \hline
    \noalign{\smallskip}
    \end{array}
\]
{{\bf Notes:} D1, D2, and D3 are the mean galaxy densities in the three radial intervals, 0 to 1/3, 1/3 to 2/3, and 2/3 to 1 $r_{\rm pc}$, respectively. The last two columns give the number of proto-clusters and the percentage in each category.}
\label{table:tab6}
\end{table}

One interesting result from the average proto-cluster profile to keep in mind is that about half of the member galaxies reside inside
$0.5~ r_{\rm pc}$. That implies that the density inside an aperture of $0.5~ r_{\rm pc}$ is about 3 times higher than the density in the annulus
at $r = 0.5$ to $1~r_{\rm pc}$.

Inspection of individual profiles shows that there is a large variety of profile shapes.  
Figure~\ref{fig:fig3} and in the Appendix Fig.~\ref{fig52} show a selection of proto-clusters with some emphasis on systems with substructure.
To devise a cluster detection method that works with assumptions on the proto-cluster shapes, as, for example, a matched filter algorithm, one needs an overview of the variation of the proto-cluster structures. 
Any treatment that regards azimuthally symmetric shapes in first order would rely on knowledge of profiles. 

To get such an overview, we dissected the three-dimensional profiles out to $r_{\rm pc}$ into three equal radial intervals and classified the proto-cluster profiles into five categories according to the density ratios of the different regions as listed in Table~\ref{table:tab6}. Category 1 and 2 have a central peak, 3 and 4 have the maximum in the middle, and 5 has the maximum near $r_{\rm pc}$.
Table~\ref{table:tab6} also shows the number of proto-clusters which fall into each category. 

To display the variation of the profiles, we have determined the mean profile for each category with a resolution of 8 bins out to $1.6~ r_{\rm pc}$. 
Note that this provides a higher radial resolution of 5 bins inside $r_{\rm pc}$ than the three radial bins used for the categorisation. 
This allows us to show the profiles in more detail.
Not to let those proto-clusters with the largest number of galaxy members completely dominate the results, we use a weighting with a factor of $n_{\rm gal}^{-1/2}$ per proto-cluster in averaging the profiles, where $n_{\rm gal}$ is the total number of members in the simulations.
Due to the scaling with proto-cluster radius and this weighing scheme, the resulting densities loose the normal physical units, and we show relative values. 
These mean profiles for categories 1 to 5 are shown in Figs.~\ref{fig50} and \ref{fig51} in the Appendix.
We note first of all that 73.2\% of the proto-clusters have a high central density inside $0.2~ r_{\rm pc}$ (much higher than in the other radial bins). 
4.2\% of these proto-clusters have a higher density in the third than in the second bin. 
But as shown in 
Fig.~\ref{fig50} they have mostly very compact cores with few galaxies outside. 
The 24.6\% of category 3 and 4 have the highest density in the middle radial region.
Two examples of such clusters are shown in 
the upper two panels of Fig.~\ref{fig52}.

Only 17 proto-clusters have the highest density in the region from $2/3\,r_{\rm pc}$ to $1\,r_{\rm pc}$.
An example is shown in the bottom 
panel of Fig.~\ref{fig52}. 
Among these cases with high density in the outer annuli we find binary and multiple systems, which will nevertheless collapse into a single cluster by $z = 0$.
Thus, the majority of the proto-clusters will show a significantly higher contrast of the central region compared to the overall system. 
It also implies that the central region will collapse earlier than redshift zero for most proto-clusters to form a smaller galaxy group or cluster first.

\section{\label{sc:densitycontrast}Density Contrast in the projected galaxy distribution}

\subsection{Density contrast with respect to the global sky background}

To explore the efficiency with which proto-clusters can be detected in the Euclid Wide Survey, we study the projected galaxy density contrast of proto-clusters against the galaxy background in this section. 
We make use of the photometric redshifts that have been modelled in the simulations. To find the majority of the members of a proto-cluster we have to consider a selection window for the redshifts wide enough to cover the uncertainties of the photometric redshifts.
Using the MAMBO lightcone, 
we illustrate in Fig. 6 the median photometric errors corresponding to
including 50\% -- 90\% of all galaxies at a given redshift, which we define as ``completeness''.
These photometric errors constitute the redshift windows used to detect PC.
The half window size is shown as photo-z error parameter as a function of redshift in the Figure. 
We will designate the redshift interval for a 50\% completeness limit as $\Delta_{z~(50\%)}$. 
These calculations have been performed for all galaxies in the lightcone, including non-members, for better statistics in 132 redshift bins. 
In each bin, the galaxies were sorted by $\Delta z$, the deviation of the photometric redshift from the true one. The maximum $\Delta z$ of the 50\% smallest values yields then, for example, $\Delta_{z~(50\%)}$. 
This completeness limit is identical to the median.
We used this value in most of the following examples because it provides a detection efficiency close to the maximum, as shown in Sect. 6.2 and 7.
Also shown is the official \Euclid requirement for the redshift accuracy, $\Delta z = 0.05 ~(1+z)$.
This requirement is close to the median curve. At redshifts between z = 1.5 and 2.2, it is worse than the requirement. 
With increasing redshift, it gets better and falls below the requirement at $z = 3$ to $4$. At these higher redshifts, the photometric bands bracket the Ly-break better.

\begin{figure}[h]
   \includegraphics[width=\columnwidth]{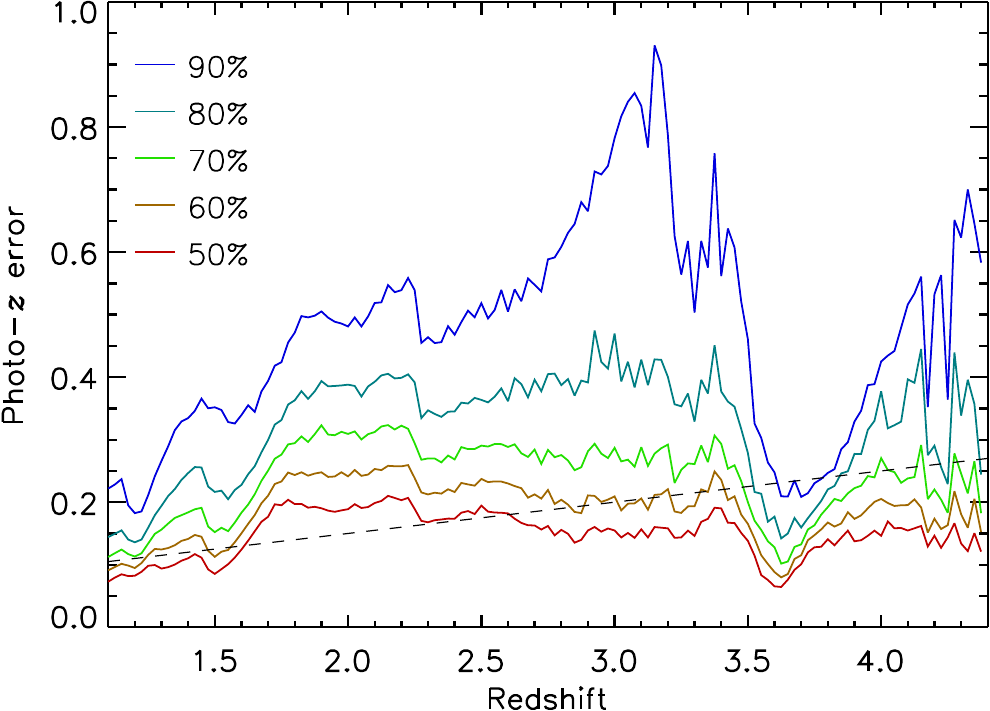}
\caption{Photo-$z$ error as a function of redshift for different completeness limits. The parameter photo-$z$ error describes the half-width of the redshift window necessary to contain 50\%, 60\%, 70\%, 80\%, and 90\% of all galaxies at a certain redshift.
The dashed line shows the official requirement for the photometric accuracy.
}
\label{fig:photozerror}
\end{figure}

\begin{figure}[htpb!]
\centering
\includegraphics[width=.9\columnwidth]{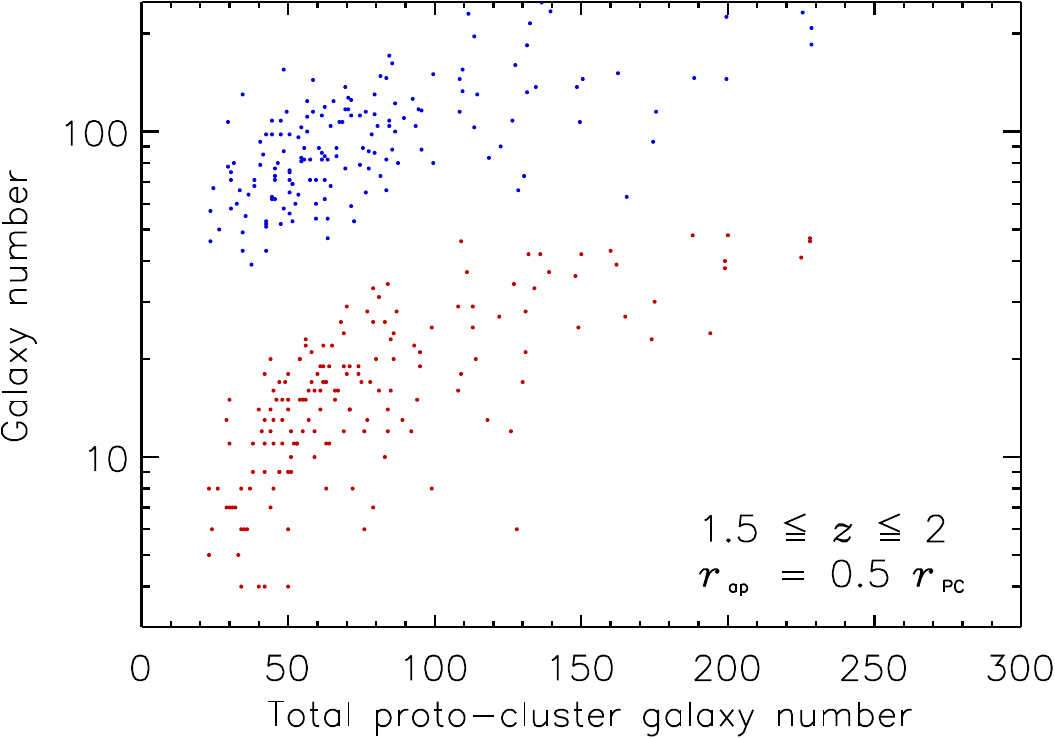}
\includegraphics[width=.9\columnwidth]{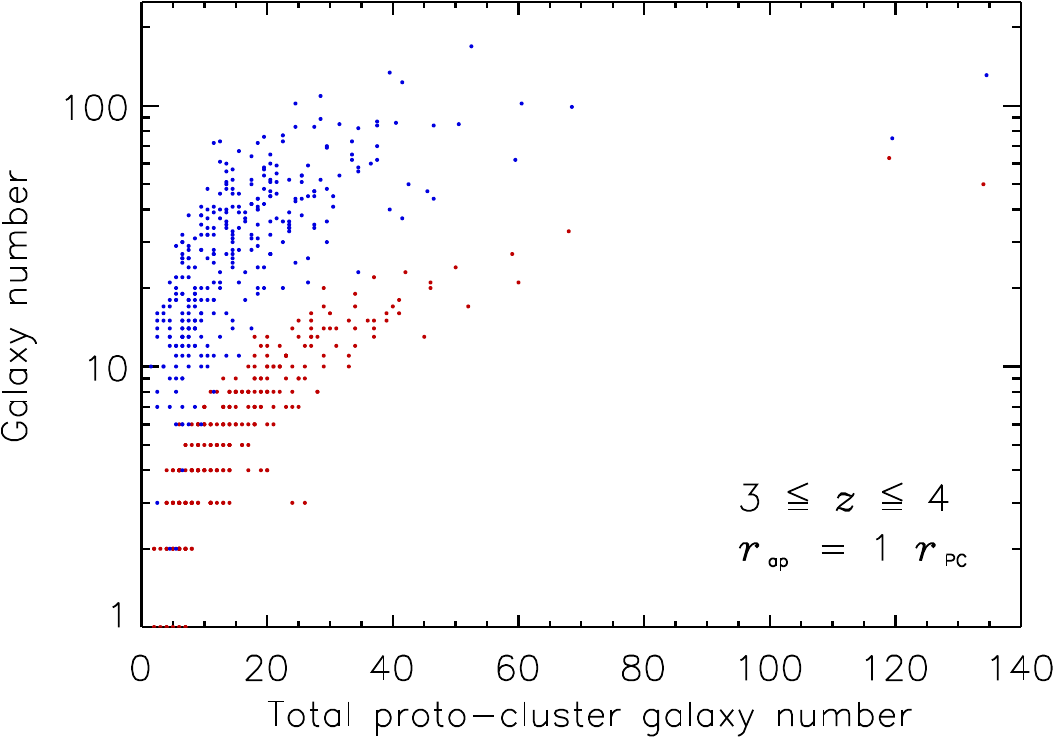}
\caption{
Galaxy counts in proto-clusters (red) in comparison to the background (blue) as a function of the total galaxy number in the proto-clusters.
The top row is taken in the redshift range $z=1.5$ to $2$ with an aperture of $0.5~r_{\rm pc}$ and the bottom row in the range $z=3$ to $4$ inside an aperture of $r_{\rm pc}$. For the galaxy selection a redshift interval corresponding to $\Delta_{z~(50\%)}$ was used.
}
\label{fig:galcts}
\end{figure}

Galaxy counts and densities in proto-clusters and background in the MAMBO lightcone (limited by the redshift range corresponding to given redshift uncertainties) were calculated for different aperture sizes (with respect to $r_{\rm pc}$). In practice, we counted all known proto-cluster members inside the aperture radius and all other galaxies as background in the cylinders defined by the redshift uncertainty, $\Delta{z~_{(X\%)}}$, and the aperture area.
This was done for different redshift regions and different photometric uncertainty limits.
Figure~\ref{fig:galcts} shows two examples of the galaxy counts in proto-clusters and background for apertures, $r_{\rm ap} = 0.5~ r_{\rm pc}$ for $z = 1.5$ to $2$ and $r_{\rm ap} = r_{\rm pc}$ for $z = 3$ to $4$ using a redshift window defined by $\Delta_{z~(50\%)}$. Here and in the following we designate all galaxies outside the proto-clusters either in front or in the back as background.
We note that the background galaxies always outnumber the proto-cluster galaxies. 
We show the data as a function of the total galaxy number of the proto-cluster, which is a good proxy for the proto-cluster mass \citep{Ryk2014}. 
Of course, the number of recovered proto-cluster galaxies increases with the total galaxy number, but also, the background increases somewhat due to the increase of $r_{\rm pc}$. 
Statistically, the number of detected galaxies with large aperture radius ($r \sim 2.5~r_{\rm pc}$) scatters around 50\% of the total number by construction for a $\Delta_{z~(50\%)}$ limit.

\begin{figure}[htpb!]
\centering
\includegraphics[width=.9\columnwidth]{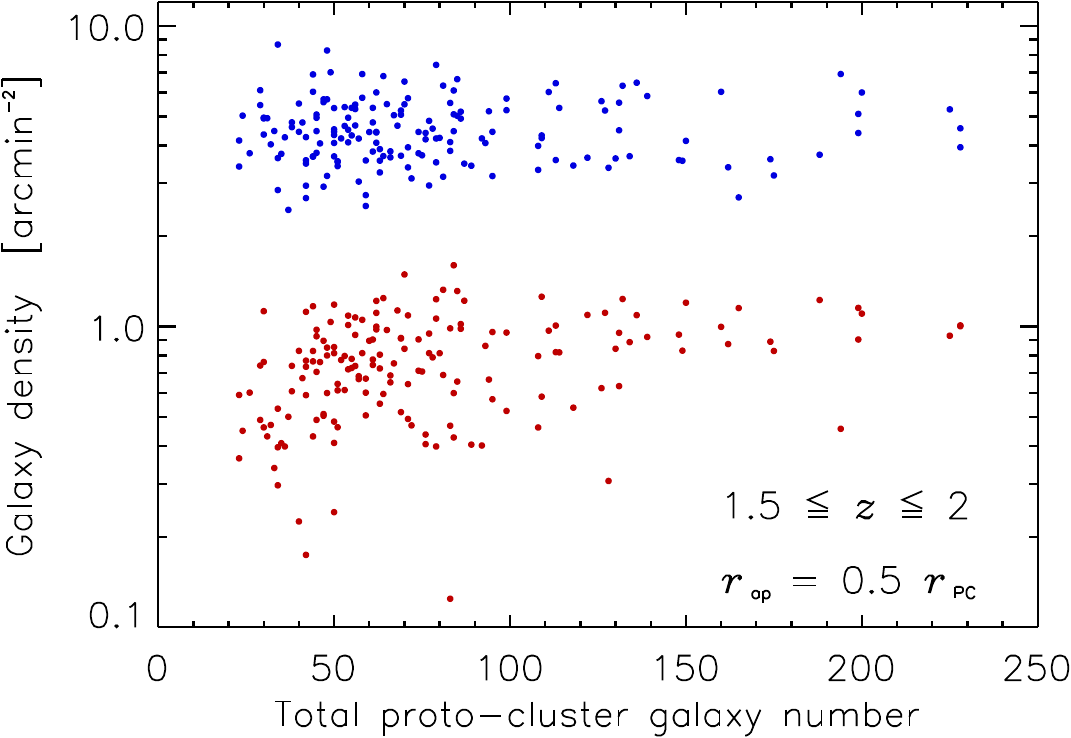}
\includegraphics[width=.9\columnwidth]{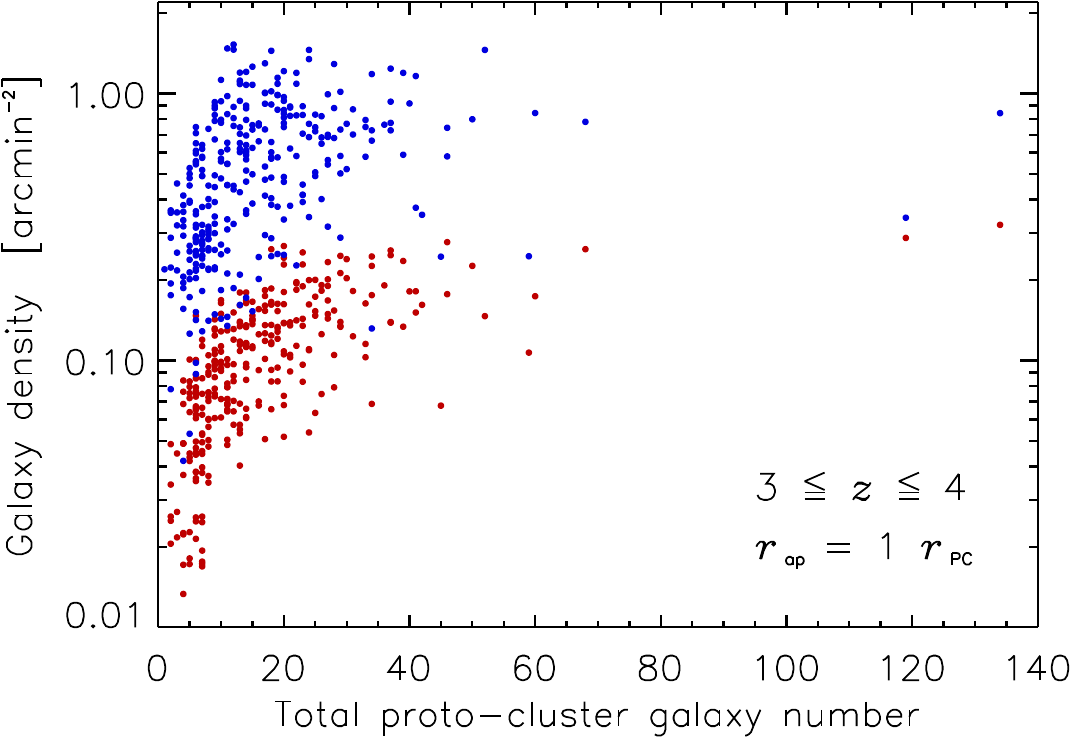}

\caption{
Galaxy densities in proto-clusters (red) in comparison to the background (blue) as a function of the total galaxy number in the proto-clusters.
The top row is taken in the redshift range $z =1.5$ to $2$ with an aperture of $0.5~r_{\rm pc}$ and the bottom row in the range $z=3$ to $4$ inside an aperture of $r_{\rm pc}$.
For the galaxy selection a redshift interval corresponding to $\Delta_{z~(50\%)}$ was used.
}
\label{fig:galdens}
\end{figure}

\begin{figure}[h]
   \includegraphics[width=\columnwidth]{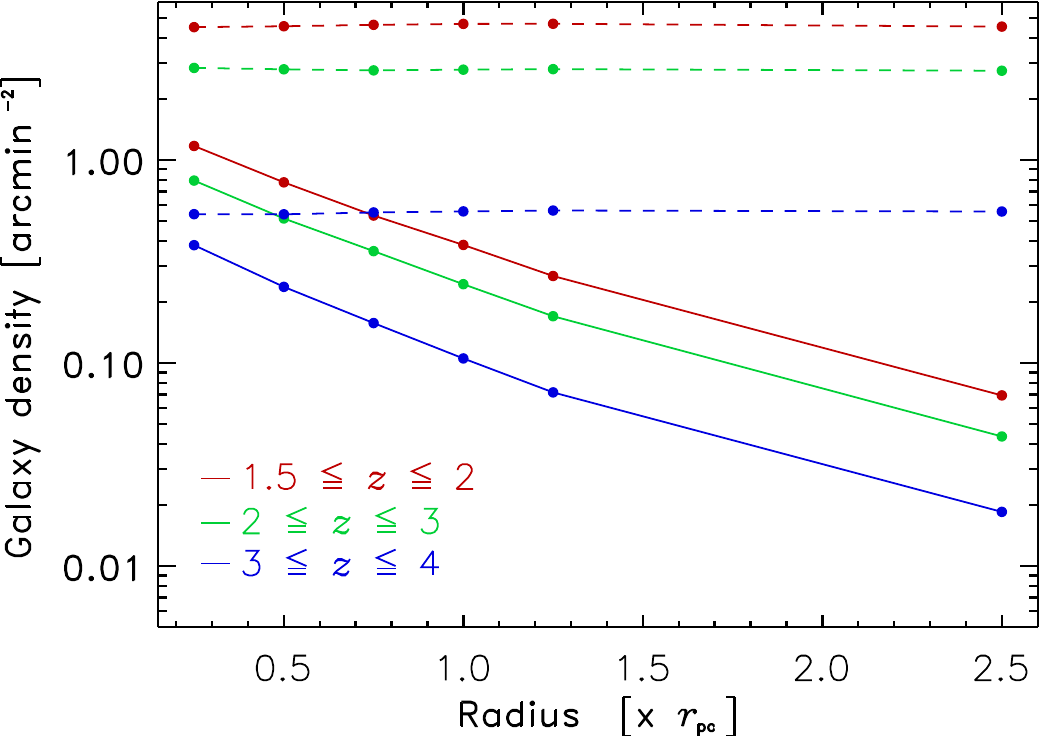}
\caption{Mean projected galaxy densities in proto-clusters (solid lines) as a function of aperture radius in three redshift intervals. 
The dashed curves give the galaxy densities in the background. The results were derived for $\Delta_{z~(50\%)}$.
}
\label{fig:fig18}
\end{figure}

The results for the densities of the detected galaxies are shown in Fig.~\ref{fig:galdens}.
Again, the projected densities of the background are always larger than the proto-cluster densities. Figure~\ref{fig:fig18} summarises the results for the density statistics as a function of aperture radius and redshift for a 50\% completeness and \Euclid magnitude limit.
Here we note that the background densities appear constant with changing aperture radius as expected. 
The densities for the proto-clusters, however, decrease with increasing aperture radius due to the decreasing density profiles.

\begin{figure}[h]
   \includegraphics[width=\columnwidth]{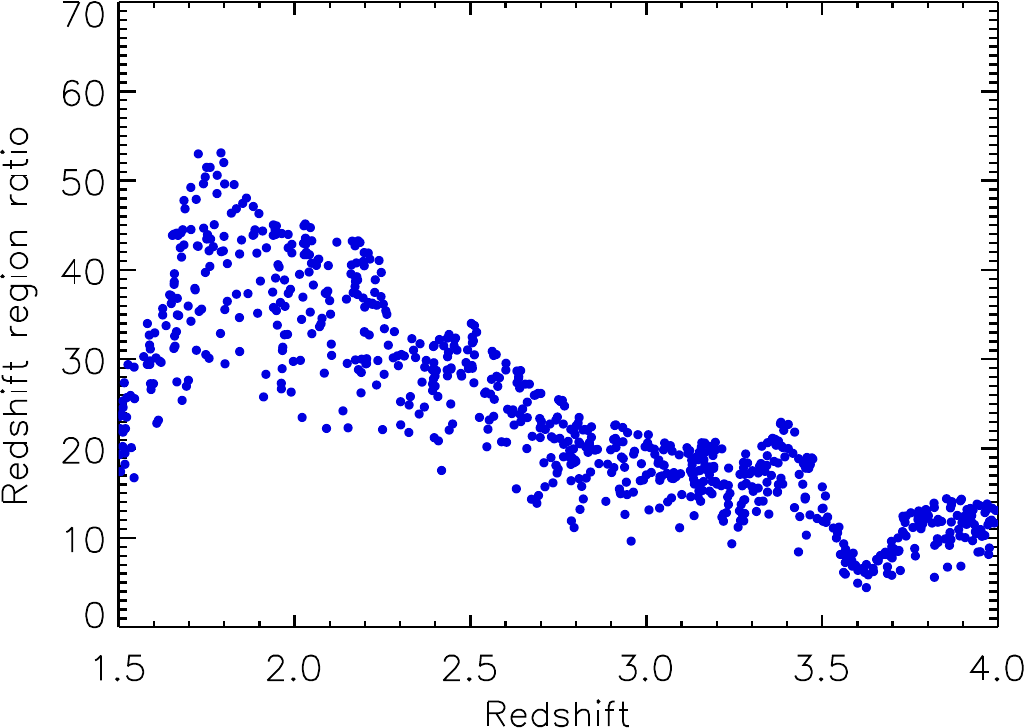}
\caption{Ratio between the line-of-sight interval given by the photometric redshift uncertainty and the line-of-sight covered by proto-clusters,
$r(\Delta_{z~(50\%)})/r_{pc})$.
The ratios for all MAMBO proto-clusters as a function of redshift are shown.
}
\label{fig19}
\end{figure}

The fact that the projected background density is often very much larger than the proto-cluster density is a challenge for the reliable detection of the proto-clusters. 
Thus, before proceeding further, we study the reason for this situation in more detail. 
We have shown in Sect. 2 that the proto-cluster overdensities are not very large since we capture the proto-clusters before turn-around. 
This overdensity has to be compared with the line-of-sight ratio across the proto-cluster and the redshift range defined by the photometric redshift accuracy. 
In projection, we sample all the galaxies in the line-of-sight of the proto-cluster, which have a redshift inside the uncertainty limits of the photometric redshifts. Since the line-of-sight distance interval corresponding to the redshift uncertainties is much larger than the diameter of the proto-clusters, more background galaxies are sampled than proto-cluster members by the photometric redshift selection, in spite of the moderate galaxy overdensity in proto-clusters. To illustrate this, we show in Fig.~\ref{fig19} the ratio between the distance interval corresponding to the photometric redshift uncertainty and the proto-cluster diameter as a function of redshift.
Here, we have used a redshift interval of $\Delta_{z~(50\%)}$; with a smaller completeness, the volume ratio would be smaller, but we would also sample fewer member galaxies.
We clearly see that the volume from which the background galaxies are sampled is much larger than the proto-cluster volume (which has been simply approximated here by a cylinder).
This large volume ratio is the consequence of using only broadband photometry for the redshift estimates. 

\begin{figure}[h]
   \includegraphics[width=\columnwidth]{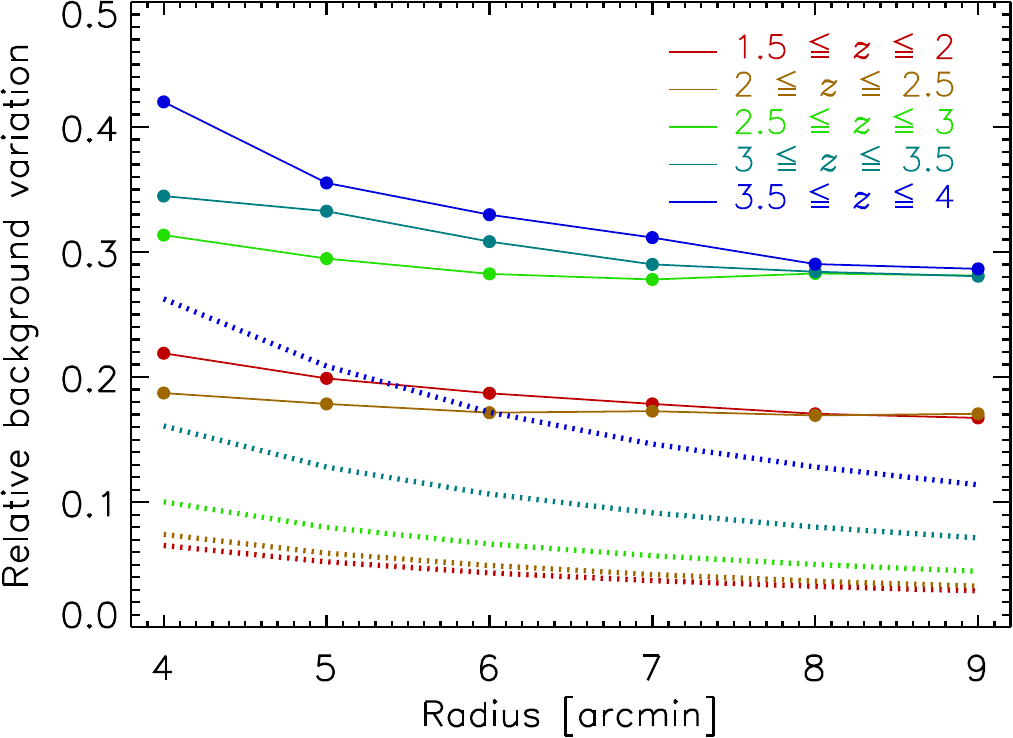}
\caption{Relative root mean square (RMS) deviations of the 
projected background galaxy density fluctuations (solid lines)
as a function of aperture radius for five different redshift 
intervals.
The dashed lines show the estimated Poisson errors for comparison.
}
\label{fig20}
\end{figure}

\subsection{Significance of the density contrast}

An additional problem complicates the detection of proto-clusters. 
While it is sometimes assumed naively that one detects the proto-clusters in a distinct and smooth background field, which is characterised by Poissonian density fluctuations, we are facing the situation that we have to detect the proto-clusters against a background characterised by the presence of cosmic large-scale structure. Since the aperture sizes for detecting proto-clusters sample the background at relatively small scales,  where the cosmic large-scale structure is well in the non-linear regime, we have to cope with background fluctuations larger than the Poisson noise for the relevant galaxy counts.
We illustrate this by means of the MAMBO simulation below.

The rms of the background density fluctuations was determined in different aperture radii in five redshift ranges ($z = 1.5$ to $2$, $2$ to $2.5$, $2.5$ to $3$, $3$ to $3.5$, and $3.5$ to $4$). 
The background densities were evaluated at the positions of the proto-clusters, while the proto-cluster galaxies were excluded from the background density calculation. 
In principle, one could have also used random positions, but in our approach, we include the small effect that a tiny part of the line-of-sight is occupied by the proto-cluster and not by the background.
This time the aperture radius is kept fixed for a chosen aperture value and not scaled with the proto-cluster radius. 
We show the results in Fig.~\ref{fig20} for six aperture radii (4, 5, 6, 7, 8, and 9 arcmin) and compare them with Poisson errors. 
While the Poisson errors decrease steadily with the aperture radius, as expected, we note hardly any decrease in the variance, except for the first aperture radius bin, where shot noise still plays an important role. 
The variance decreases with aperture radius not because of the improving count statistics but because of the decreasing variance of the large-scale structure with scale. 
This decrease is comparatively slow and hardly noticeable in the relevant scale range.
For this reason, the significance of detection above the background cannot easily be improved by increasing the aperture size, as it would be for Poisson errors.

To characterise the significance of detection of the proto-clusters we take the ratio of the detected counts to the rms of the background fluctuations. We do not include the shot noise of the proto-cluster counts, which would be important in the calculation of the error of the detected signal. We look here just at the detection significance. 
We determine the detection significance for every proto-cluster by using the actually detectable counts in the simulations, and for the background, we take the rms of the background counts determined from all proto-cluster positions in the same redshift bin and for the same aperture radius.

\begin{figure}[htbp!]
   \includegraphics[width=\columnwidth]{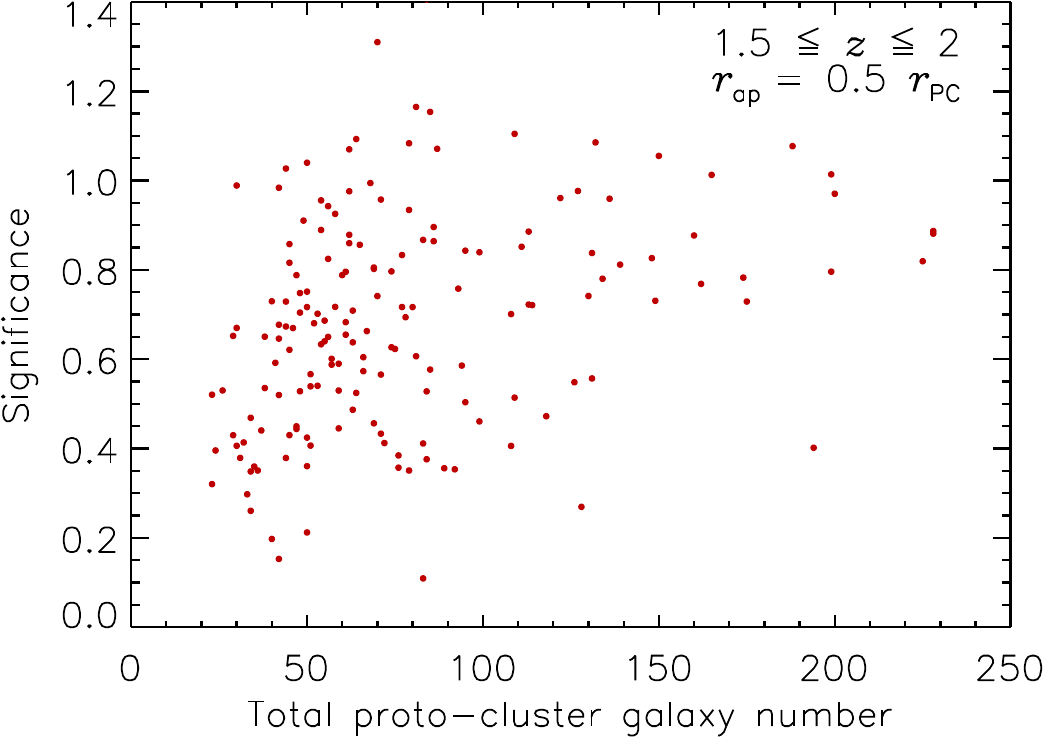}
   \includegraphics[width=\columnwidth]{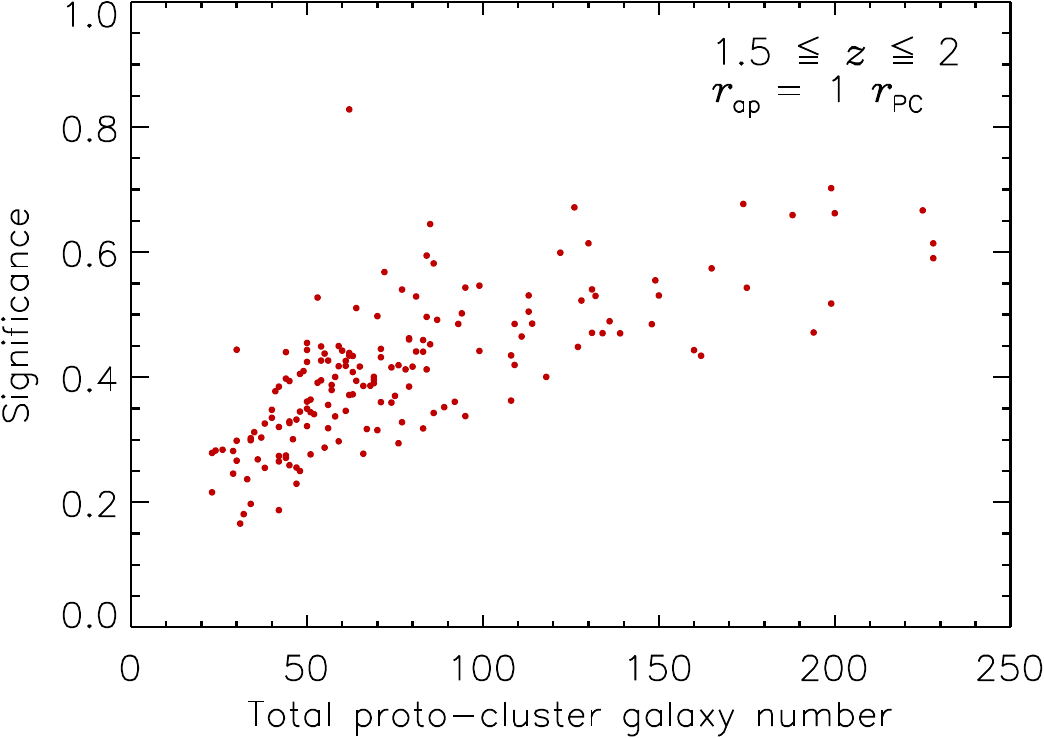}
    \caption{Detection significance of proto-clusters from the MAMBO simulation in the redshift interval $z = 1.5$ to $2$.
    The panels show results for aperture radii of $0.5$ (top) and $1~ r_{\rm pc}$ (bottom).
    For the galaxy selection a redshift interval corresponding to $\Delta_{z~(50\%)}$ was used.
    }
\label{fig21}
\end{figure}

Examples of the results are shown in Fig.~\ref{fig21} for the redshift range $z = 1.5$ to $2$ for aperture radii of $r_{\rm ap} = 0.5$ and $1.0~ r_{\rm pc}$. 
We show the detection significance as a function of the total galaxy number belonging to the proto-clusters in the simulation. 

The values for the significance are small. There is a correlation of the significance with the richness of the proto-cluster, which is more pronounced for the better number statistics with the larger aperture.

\begin{figure}[h]
   \includegraphics[width=\columnwidth]{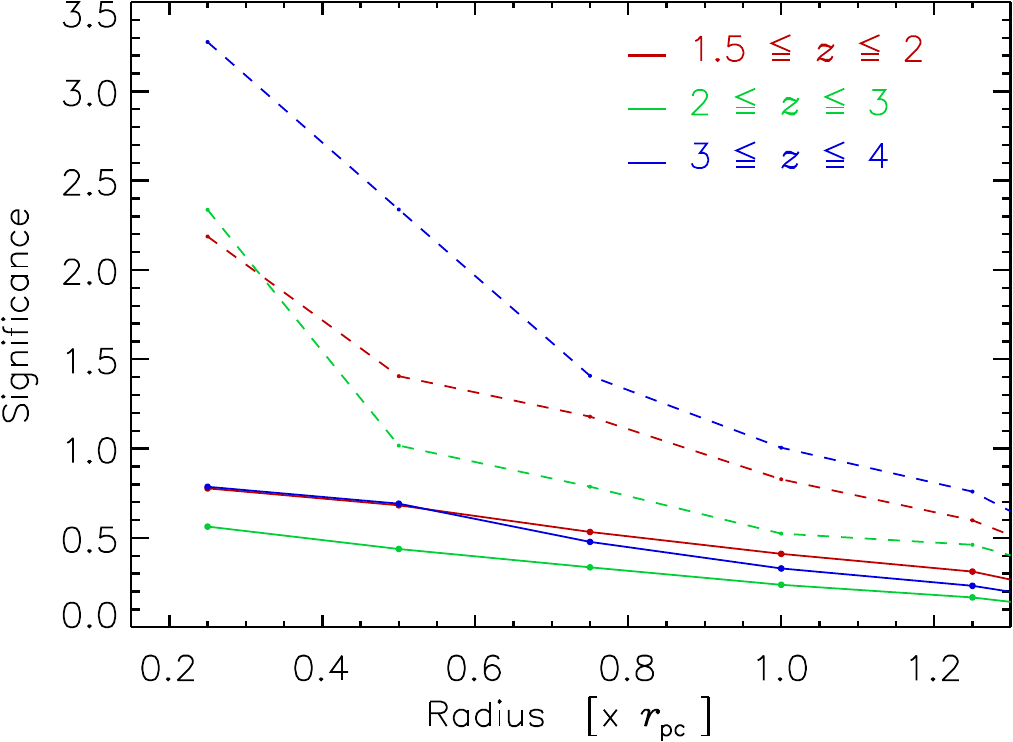}
   \includegraphics[width=\columnwidth]{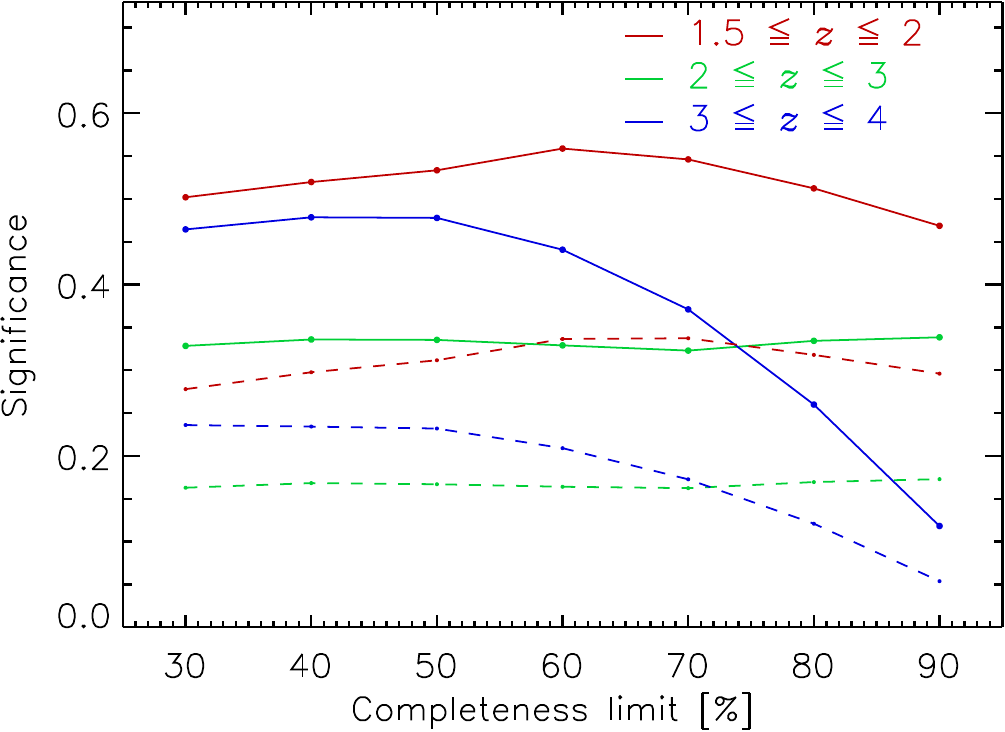}   
\caption{{\bf Top:} Detection significance of proto-clusters as a function of aperture radius for three different redshift intervals.
The solid curves show the mean significance, while the dashed curves show the maxima.
A redshift interval of $\Delta_{z~(50\%)}$ was used. 
{\bf Bottom:} Mean detection significance of proto-clusters as a function of the completeness limit, $\Delta{z~(X\%)}$, in the range 30 to 90\% for three redshift intervals. The solid lines show the results for an aperture radius of 0.5 $r_{\rm pc}$ and dashed lines for 1 $r_{\rm pc}$. 
}
\label{fig23}
\end{figure}

The upper panel of Fig.~\ref{fig23} summarises the significance studied for a $\Delta_{z~(50\%)}$  limit. In the plot, we show both the mean and maximum values for the significance of each parameter selection.
The maximum values for the smallest aperture radius are relatively high due to the large Poisson noise for small counts. 
We note that for the highest redshift bin, the maximum values of the significance is always highest for the same reason, but for the mean values, the highest redshift bin is not more significant than the lowest one. 

We also explored if we can increase the significance by using a higher or lower completeness limit so that we can sample more galaxies for each proto-cluster or deal with a smaller redshift range for the background. 
The lower panel of Fig.~\ref{fig23} shows how the significance changes with the completeness limit for two apertures (0.5 and 1 $r_{\rm pc}$) and three redshift intervals. We note that the 50\% or 60\% completeness limits provide the maximum in all cases. 
Otherwise, the curves are relatively flat. 
Only for the highest redshift bin there is a notable decrease 
of the significance with increasing completeness
because it is closer to the detection limit for the galaxies.
The main conclusion is, however, that there is not much room for improvement in the detection efficiency with a different choice of the completeness limit. 

Thus overall, a strategy which uses $\Delta_{z~(50\%)}$ and apertures smaller than $1~r_{\rm pc}$ (for example $0.5~r_{\rm pc}$) provides a close-to-optimal solution.
While at decreasing radii the signal-to-noise gets better, the statistics gets worse, and therefore a radius of around $0.5~r_{\rm pc}$ gives a good compromise.
For most of the following analysis we will thus use this selection.

An improvement can, of course, be expected from using a probability-based detection algorithm, for example, a matched filter technique. 
The present study provides a useful guideline for such a method since it shows which proto-cluster region contributes most to the detection signal.

\section {Detection significance with local background assessment}

\begin{figure}[h]
   \includegraphics[width=\columnwidth]{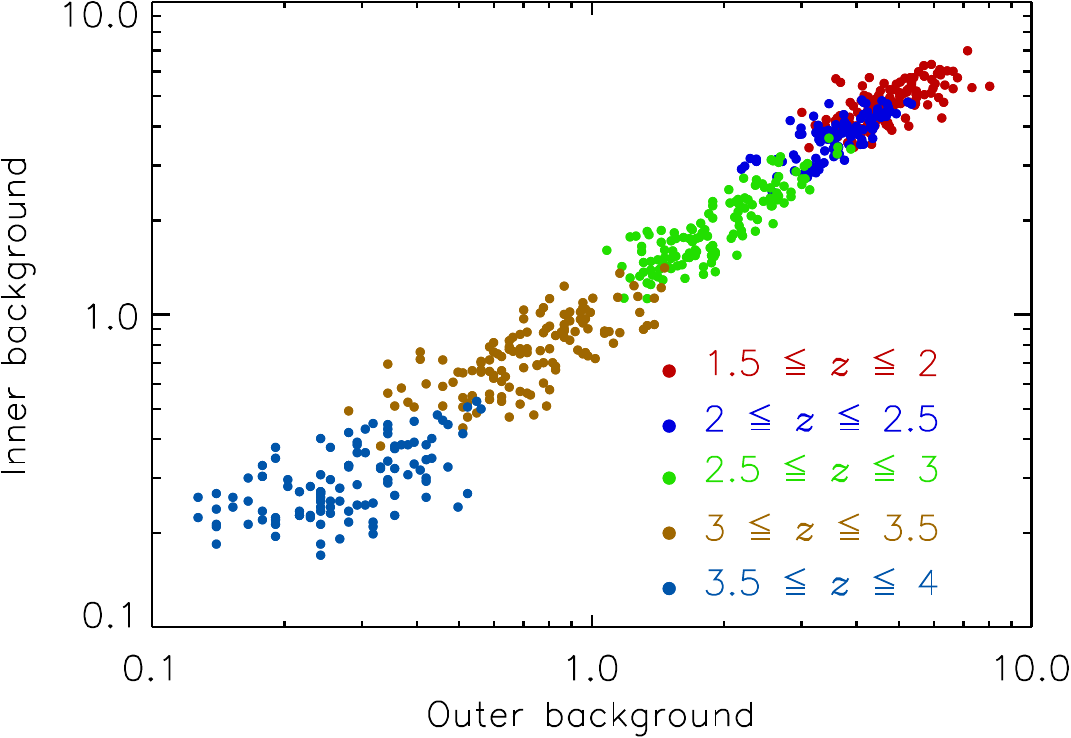}
\caption{Correlation of the galaxy background density for proto-clusters in an aperture radius of 5 arcmin and the background taken from an annulus of 9 to 13 arcmin radius from the cluster centre around each proto-cluster. The colours mark different redshift intervals from $z = 1.5$ to 4.0 with a width of $\Delta z = 0.5$.
}\label{fig26a}
\end{figure}

In this section we explore whether the detection significance can be improved with a local background assessment. 
In the previous section, we have used the global background variance for the significance calculation. Now, we test the behaviour of the background if the background is taken for each proto-cluster from an annulus around it
and its variance determined from the data.
Due to the spatial correlation of the galaxy density in the large-scale structure, we can expect that there is a correlation between the background galaxy density inside the proto-cluster aperture and that of the surrounding annulus.
This correlation is shown for the simulated MAMBO proto-clusters in Fig.~\ref{fig26a}.
Since there is a clear correlation, the estimated background variance can be reduced compared to that considered in the previous section if we add information about the local background.

Now, we estimate the background in an annulus around the detection aperture with radii of 9 and 13 arcmin. 
The radii of proto-clusters for masses of $10^{14}$ and $10^{15}$ \si{\solarmass} for different redshifts can be found in Table~\ref{tab8}. 
For $z = 1.5$ we find $r_{\rm pc} \sim 4.4$ to $9.6$ arcmin and for $z = 4$ we get $r_{\rm pc} \sim 3.4$ to $7.4$ arcmin for this mass range.
The background region is thus outside the proto-clusters. 
The correlation of the background in the aperture and the annulus, shown in 
Fig.~\ref{fig26a}, helps us in the following way. A measurement of the background
in the annulus can be used normalise the background in the aperture region. The residual aperture backround will then be smaller than the variance without this normalisation.
The relevant residual background variance is then the variance of the ratio of the galaxy density inside the aperture to the galaxy density in the background annulus. It is the scatter of the relation shown in Fig.~\ref{fig26a}.

\begin{figure}[h]
   \includegraphics[width=\columnwidth]{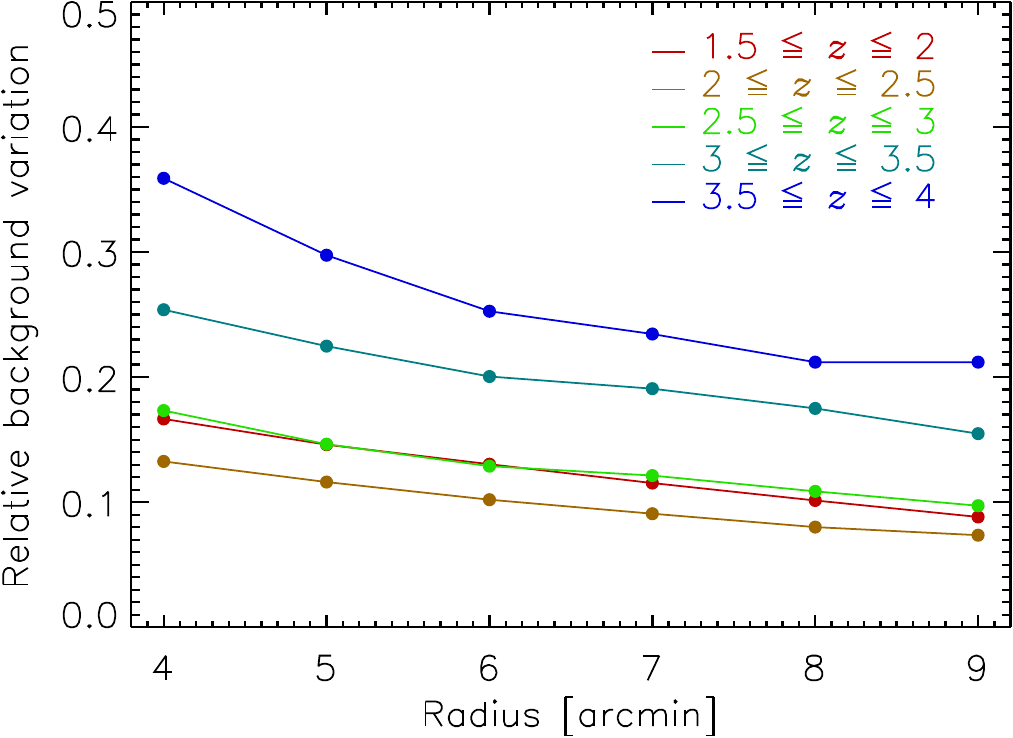}
\caption{Relative (RMS)variations of the projected background galaxy densities (solid lines) with a local assessment as a function of aperture radius for five different redshift intervals.
}\label{fig26b}
\end{figure}

In Fig.~\ref{fig26b} the rms of the density fluctuations is shown as a function of aperture radius and redshift in an equivalent way to the results in Fig.~\ref{fig20}. Thus, the results can be directly compared in these figures, and we note, in most cases, an improvement of almost a factor of 2. 
The improvement is due to the fact that the variations in the projected galaxy density background are caused only by a minor degree of Poisson noise but mostly by large-scale structure, which can, for example, be well described by clustering statistics like the two-point correlation function. 
Therefore we obtain a better estimate of the local background if we take a measurement in its immediate neighbourhood.

\begin{figure}[h]
   \includegraphics[width=\columnwidth]{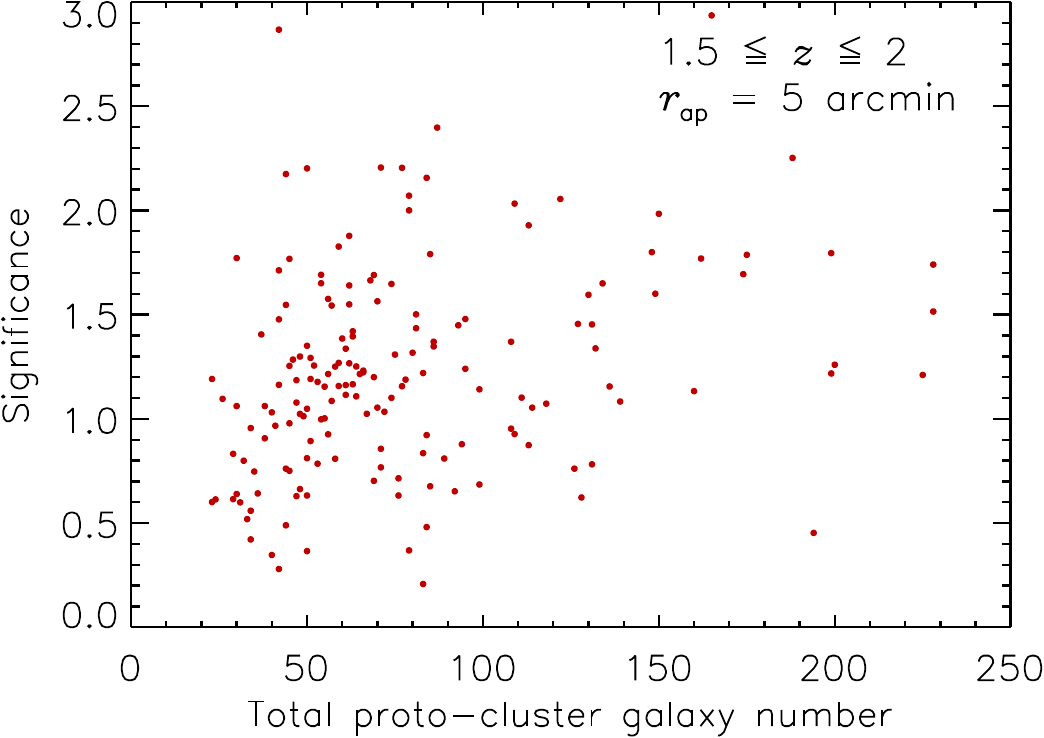}
   \includegraphics[width=\columnwidth]{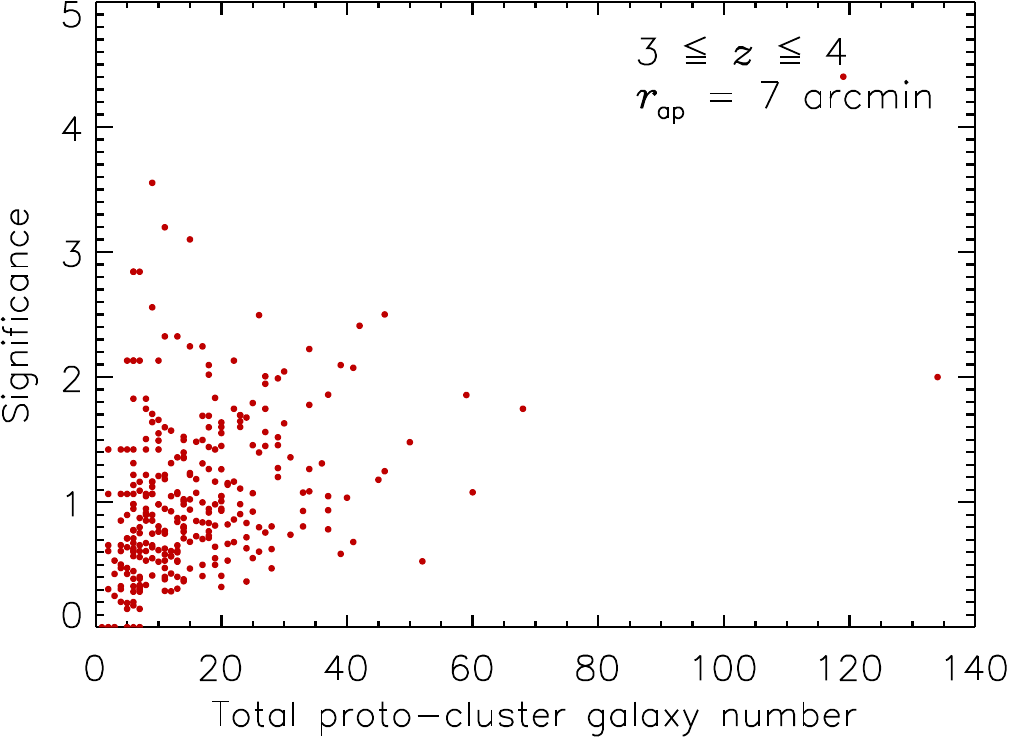}
\caption{{\bf Top:} Detection significance for proto-clusters at $z = 1.5$ to 2 and a detection aperture of 5 arcmin was obtained with a local background assessment.
{\bf Bottom:} Detection significance for a redshift range of $z = 3$ to 4 and an aperture of 7 arcmin with local background.
For the galaxy selection a redshift interval corresponding to $\Delta_{z~(50\%)}$ was used.
}\label{fig26c}
\end{figure}

This has, of course, an effect on the detection significance, which is illustrated in Fig.~\ref{fig26c}. 
We note that now more proto-clusters reach a detection significance of $2\,\sigma$, and a large fraction of detections has significances above $1\,\sigma$, which was not the case in the previous section.

\begin{figure}[h]
   \includegraphics[width=\columnwidth]{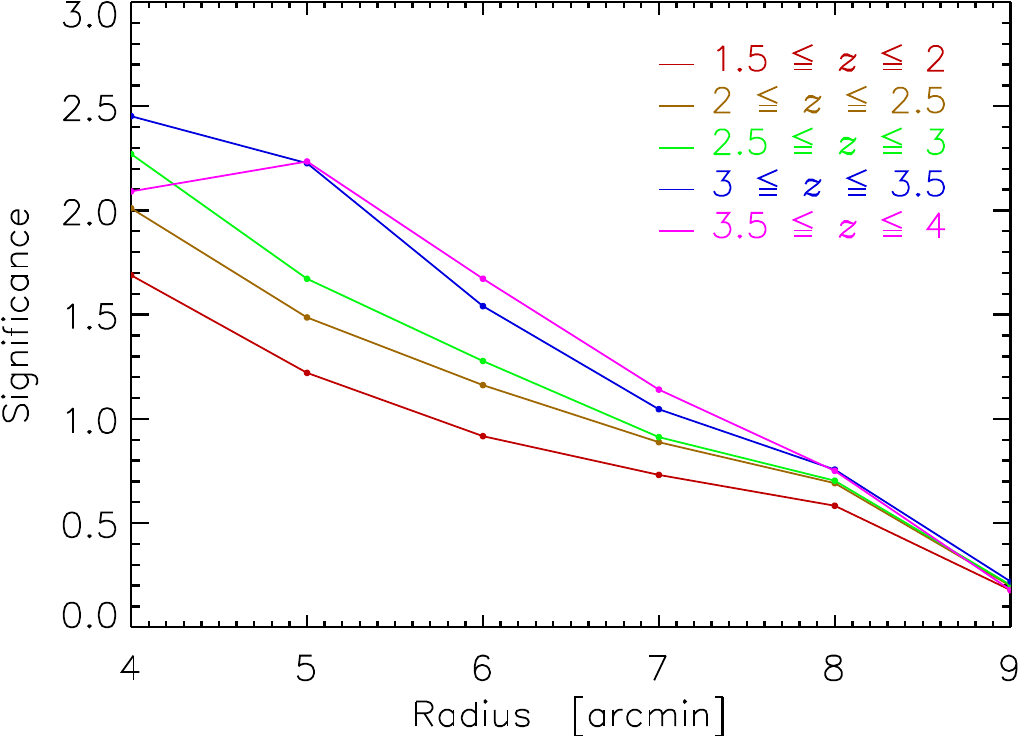}
   \includegraphics[width=\columnwidth]{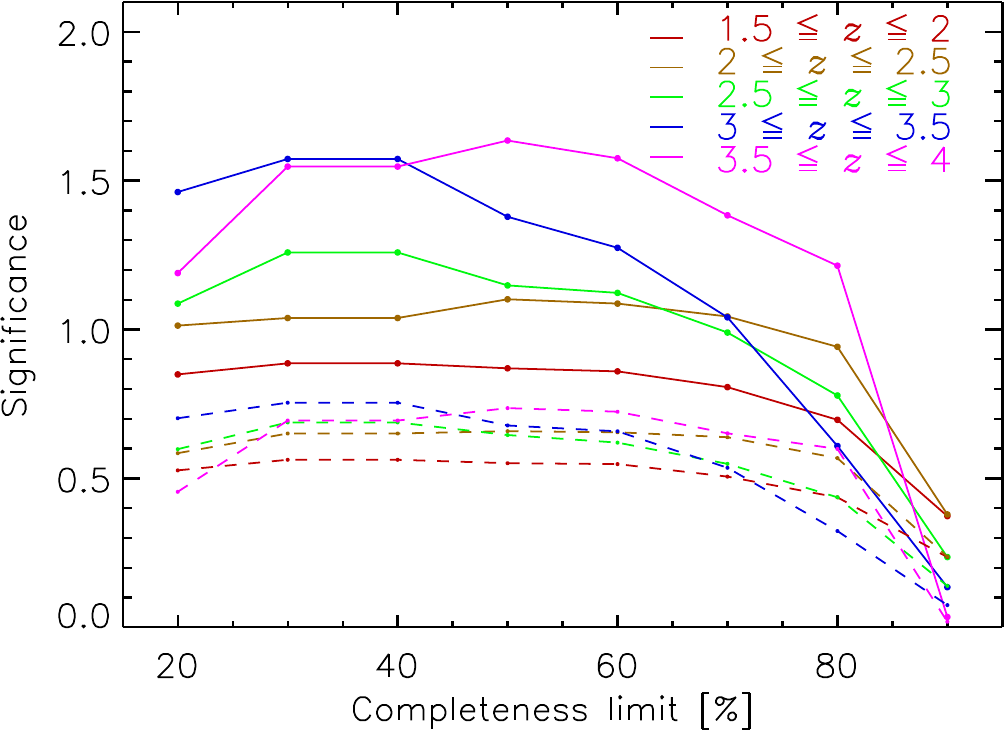}   
\caption{{\bf Top:} Mean detection significance of proto-clusters as a function of aperture radius, as in Fig.~\ref{fig23} but using a local background and aperture radius in arcmin. A redshift interval of $\Delta_{z~(50\%)}$ was used.
{\bf Bottom:} Mean detection significance of proto-clusters as a function of the completeness limit, $\Delta{z~(X\%)}$, in the range 30 to 90\%, as in Fig.~\ref{fig23} but using a local background. The solid lines show the results for an aperture radius of 0.5 $r_{\rm pc}$ and the dashed lines that for 1 $r_{\rm pc}$.
}
\label{fig:detsign_local}
\end{figure}

Figure~\ref{fig:detsign_local} (top) summarises the results on the detection significances. 
This figure can be compared to Fig.~\ref{fig23}, with a change in the radius units to arcmin.
We note that the local background assessment can provide a significant improvement. 
For real observations, the method can be tested by studying the spatial correlation of the galaxy density fluctuations in the field outside the proto-clusters and by using these results to determine the rms of the background fluctuations. 
This can actually be applied in most detection methods, for example, for matched-filter detections. 
In practice, the filter would, for example, include the background ring with a negative weighting.

The bottom panel of Fig.~\ref{fig:detsign_local} shows, analogously to Fig.~\ref{fig23} the mean detection significance as a function of the photometric redshift completeness limit for five different redshift intervals. 
The solid lines are for an aperture radius of $0.5~r_{\rm pc}$ and the dashed lines for $1~r_{\rm pc}$. We note that the results for the redshift intervals $\Delta_{z~(30\%)}$ and  $\Delta_{z~(40\%)}$ have improved in comparison to the other redshift intervals, but $\Delta_{z~(50\%)}$ is still a good choice for the detection.

The practical meaning of the significance of a detection algorithm depends also on the rareness of the objects to be detected. 
If a large sky area has to be inspected to find an object, one needs a high detection significance threshold to keep the detected samples reasonably pure. In our case, we will find below that proto-clusters are quite abundant in projection on the sky. Therefore, we can still obtain a valuable proto-cluster candidate sample with a low significance threshold. The discussion section provides further details on this point.

In summary, we conclude that the best strategy for the detection of proto-clusters is to use an aperture radius smaller than $r_{\rm pc}$ and a local background assessment. Also, a redshift range given by $\Delta_{z~(50\%)}$ is quite optimal, but an exploration of higher completeness limits is often not much worse.
A more sophisticated detection algorithm will, in this respect, anyway, include a probability distributions of photometric redshifts.

\section{Mass-richness relation}

For galaxy clusters, the richness, the number of member galaxies inside a given radius and magnitude limit can be used as a proxy for the cluster mass (e.g. \citealt{And2010,Ryk2014,Cas2016}). Therefore, we test in this section how tight the mass-richness relation is for the proto-clusters in the simulations. Here, we inspect the intrinsic relation, including all known proto-cluster members in the simulations
(with magnitude limits defined in Sect. 3) and not only the ones that would be detected with a certain prescription.

The mass-richness relation was determined for the MAMBO, GAEA H, and GAEA F samples of proto-clusters. 
Here, the mass of the system is that of the descendent cluster at $z = 0$ inside $r_{200}$ since we attribute all the mass of the descendent cluster to the proto-cluster at any redshift.
We binned the proto-clusters into subsamples of redshift bins with a width of $\Delta z = 0.25$ starting at $z = 1.5$ and an extra bin for $z = 1.5$ to $2$. 
This leaves several hundred proto-clusters per bin for the GAEA samples and an order of a hundred for MAMBO.

\begin{figure}[h]
   \includegraphics[width=\columnwidth]{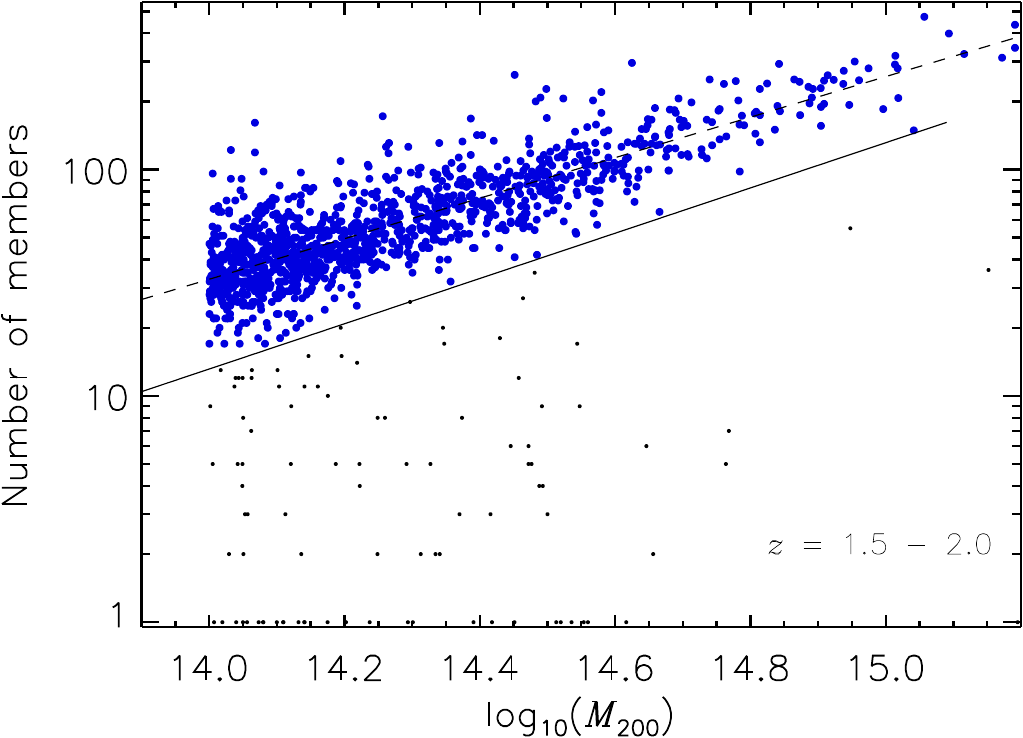}
   \includegraphics[width=\columnwidth]{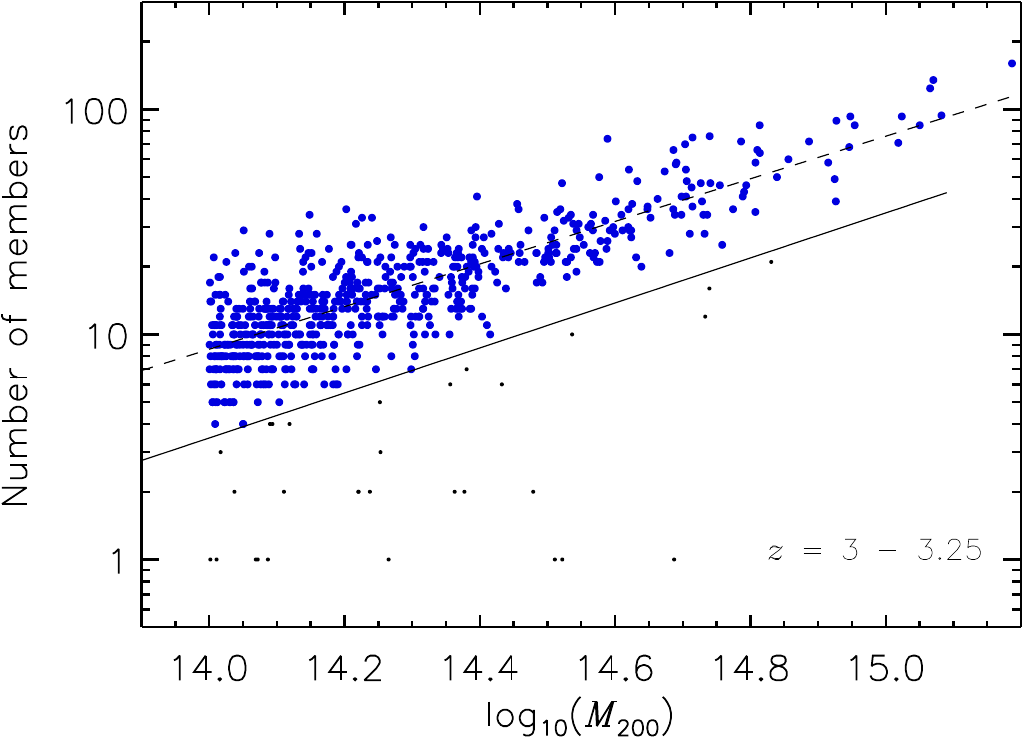}
\caption{{\bf Top:} mass-richness relation for GAEA H at $z = 1.5$ to $2.$ The
small black dots show all proto-clusters; the blue dots are the ones above the cut, shown as a solid black line. 
The linear regression fit to the relation is shown by a dashed line. $M_{200}$ is in units of \si{\solarmass}.
{\bf Bottom:} mass-richness relation for GAEA H at $z = 3$ to $3.25$. The notation is the same as in the top panel.
}\label{fig27}
\end{figure}

The results are shown in Fig.~\ref{fig27} for the GAEA H simulations.
The results from the other lightcones look similar.
We find a few proto-clusters with very low member numbers well outside the variance of the number counts. 
They were identified as artefacts due to some common problems in the production of the lightcones. 
We excluded them from our study by a cut, which removes all cases with a negative 3$\,\sigma$ deviation from the mean relation.The relative scatter that we observe in the relations, as shown for some examples in
Fig.~\ref{fig27}, is typically around 40\% and decreases with richness to about 20\%.
This scatter is distinctly different from Poisson uncertainties und usually significantly larger. We provide some further illustrations of this fact towards the end of this Section. 

\begin{figure}[h]
   \includegraphics[width=\columnwidth]{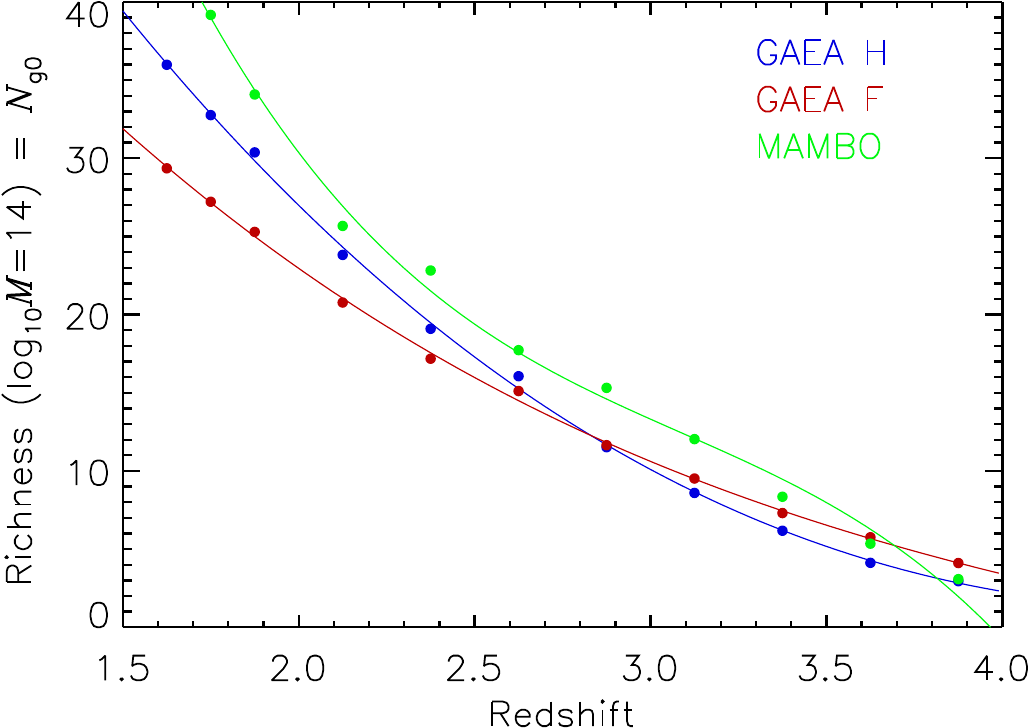}
   \includegraphics[width=\columnwidth]{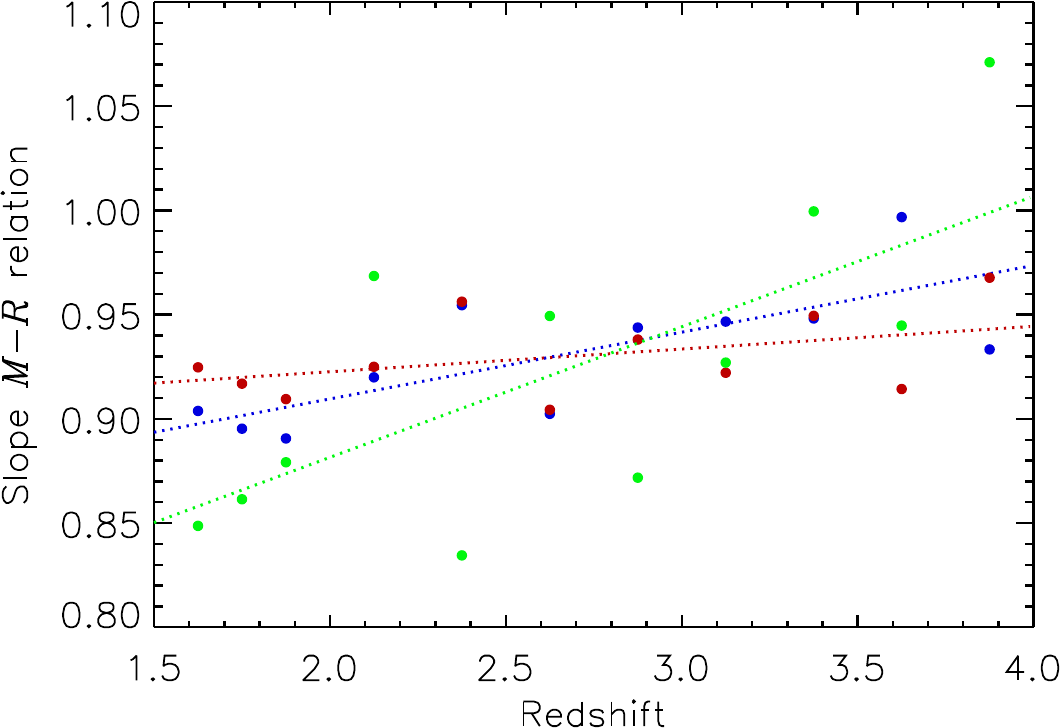}
\caption{{\bf Top:} Normalisation, $N_{\rm g0}$, for the mass-richness relation as a function of redshift. The lines show the third-order polynomial approximation. The normalisation is given for $\log_{10} M = 14$, with $M$ in units of solar mass.
{\bf Bottom:} Slope parameter, $\alpha$, of the mass-richness relation as a function of redshift. The dotted lines show fits of linear relations. The colours have the same meaning as above.
}\label{fig29}
\end{figure}

The distribution of the number counts in Fig.~\ref{fig27} and all other relations studied
is highly suggestive of a linear relation in logarithmic space. 
Therefore, we fitted the distribution by a relation of the form:

\begin{equation}
N_{\rm g}(M,z) = N_{\rm g0}(z) ~ \left({M \over 10^{14} \si{\solarmass}}\right) ^{\alpha} ~~.
\end{equation}

   \begin{table}
      \caption{Fit parameters of the polynomial expression for the normalisation,
$N_{\rm g0}(z) = a + b~ z + c~ z^2 + d~ z^3$, of the mass-richness relation as
a function of redshift for the three proto-cluster samples. We also show the slope,
$\alpha$, averaged over the redshift intervals.
}
      \[
         \begin{array}{lrrrrr}
            \hline
            \noalign{\smallskip}
{\rm sample}&{\rm a}& {\rm b} & {\rm c} & {\rm d} & <\alpha >\\
           \noalign{\smallskip}
 \hline
           \noalign{\smallskip}
{\rm GAEA~H} & 95.512 & -47.695  & 7.339 & -0.3104 & 0.93 \\
{\rm GAEA~F} & 73.268 & -36.778  & 6.577 & -0.4452 & 0.93 \\
{\rm MAMBO}  &206.466 & -168.640 & 51.386& -5.5446 & 0.92 \\
           \noalign{\smallskip}
            \hline
            \noalign{\smallskip}
         \end{array}
      \]
\label{tab7}
   \end{table}

\begin{figure}[h]
   \includegraphics[width=\columnwidth]{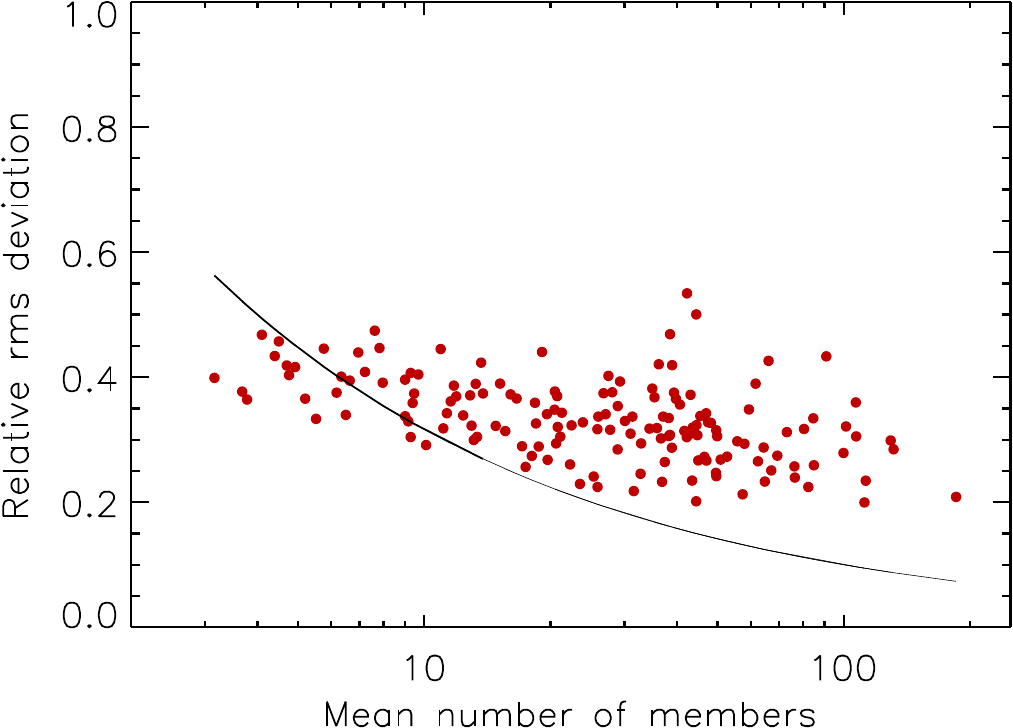}
\caption{Relative RMS deviation of the galaxy number counts as a function of the mean number of galaxies for the proto-clusters in each mass-redshift bin (data point). We also show the expectation for the scatter if it would be governed by Poisson statistics, which is shown by the solid line.
}\label{fig30}
\end{figure}

\begin{figure}[h]
   \includegraphics[width=\columnwidth]{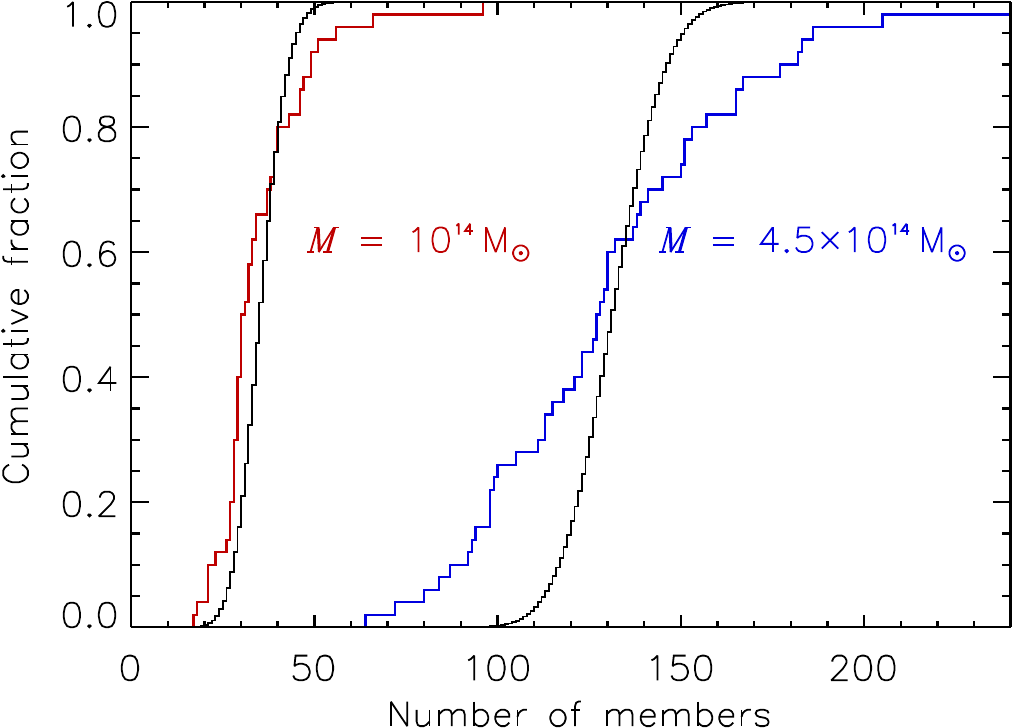}
\caption{Comparison of the distribution of the galaxy membership counts (richness) for given mass to the prediction for Poissionian scatter (black curves). The mean masses in the interval are $10^{14}$ \si{\solarmass} (red curve) and $4.5 \times 10^{14}$ \si{\solarmass} (blue curve), respectively. 
The curves show cumulative histograms normalised to unity. The data are taken from the GAEA F lightcone in the redshift interval $z= 1.5$ to $2$.
}\label{fig31}
\end{figure}

The fits are shown in the figures as dashed lines. The fit results for the normalisation as a function of redshift are shown
in the top panel of Fig.~\ref{fig29} for the three sets of simulations. 
The results are encouragingly similar. We fitted this mass-richness relation normalisation as a function of redshift with a third-order polynomial expression. The lines in Fig.~\ref{fig29} show the fitted functions, and the resulting parameters are given in Table~\ref{tab7}.

The slope of the fitted mass-richness relation for the different redshift shells is shown in the bottom panel of Fig.~\ref{fig29}. 
The values are about 0.9 or a little higher. Table~\ref{tab7} also shows the mean slopes averaged over the redshift intervals.
This is in line with the observational finding that the efficiency of galaxy formation decreases
with halo mass in the mass range above $10^{12}$\,\si{\solarmass} (e.g. \citealt{Beh2013, Kra2018}).
The simulations attempted to reproduce this empirical finding. 
It is also good to observe that the mass-richness relations are similar in all three approaches of painting galaxy evolution onto the cosmological simulations. 
The small deviations are also due to the limited statistics.

We also studied the variance of the richness in the mass-richness relations because we wanted to see if we could approximate this quantity with simple Poisson statistics. 
We used the GAEA sample here for better statistics. We divided the sample into redshift bins with the division described above. 
Each of the eleven redshift bins was subdivided into several mass bins, with at least 50 members per interval. 
This left us with 12 to 25 mass bins per redshift bin. For each subsample, we determined the root mean square (RMS) deviation between the mean number of galaxies in the bin and the actual counts. 
The results are plotted for the GAEA H sample in Fig.~\ref{fig30}. We compare the results for the scatter in the galaxy counts with the expectation for Poisson statistics.
The relative scatter for the galaxy counts in the simulations is fairly constant as a function of the mean galaxy number: it decreases from about 0.4 to 0.25 over the range of galaxy counts from 3 to 200. 
This is distinctly different from the Poisson distribution, where the relative RMS changes by a factor of about 8.
This deviation from Poisson statistics is observed in the simulations.
At least here, the formation of galaxies in the overdense region of the proto-cluster is not a Poisson point process. 
It is not immediately clear if this is purely a result of the process of ’painting galaxies onto the $N$-body simulations’ or if we should expect a
similar behaviour in nature. 
{But this exercise gives at least a warning that we should not use simple Poisson statistics for the uncertainty of the mass-richness relation unless we have shown that it works for observational data.}

To further illustrate this difference between the variance of the number counts for a given mass and a Poisson distribution, we study the distribution of the richness for a given mass in two narrow mass intervals as shown in Fig.~\ref{fig31} for the redshift range $z = 1.5$ to $2$. 
The mean masses in the intervals are $\sim 1 \times 10^{14}\, \si{\solarmass}$ and $\sim 4.5 \times 10^{14}$ \si{\solarmass}. 
To remove the scatter in the number counts due to the width of the mass interval, we normalise the number of members of each proto-cluster
by a correction factor $\bar m / m_{\rm pc}$. If for example the mass of a proto-cluster in the first bin is $ 1.15 \times 10^{14}$ \si{\solarmass}, we multiply the number counts by a factor of 1/1.15.
We note that in both cases, the observed distribution is wider than the Poisson prediction, and this difference is much more pronounced in the higher mass bin.

Overall, we find mass-richness relations that are close to linear with logarithmic slopes close to 0.9. 
Again, we emphasise that these are the relations found in the simulations with known galaxy memberships.

\section{Proto-cluster abundance and sky coverage}

The abundance of proto-clusters is determined by their definition, which in our study is given by the statement that they evolve
into present-day galaxy clusters with masses $M_{200} \ge 10^{14}$\,\si{\solarmass}. 
Practically, we defined proto-clusters above as all galaxies that will end up in the descendent cluster at $z = 0$ inside $r_{200}$. 
Similarly, we associate all matter that will finally be assembled in the zero redshift cluster as belonging to the proto-cluster. 
This is exactly the mass contained inside $r_{\rm pc}$ in the top-hat overdensity model. 
Thus, in this context, the proto-cluster has the same mass as the descendent cluster at all times.
Therefore, the density of proto-clusters for a certain mass limit in comoving coordinates is that of the present-day clusters above that mass limit. The cumulative present-day cluster mass function, $n(> M)$, can be obtained from observations, for example, from the cluster abundance in the REFLEX survey of X-ray luminous galaxy clusters \citep{Boe2013,Boe2014}, one of the best defined and comprehensive galaxy cluster samples in the nearby Universe. 
The cumulative mass function can be approximated by the following function:

\begin{equation}
n(>M) = \alpha \left({M_{\rm 14} \over 2 }\right)^{-\beta}
~ \exp\left({-M_{\rm 14}^{\delta}\over \gamma}\right) ~~~,
\end{equation}

\noindent
where $M_{\rm 14}$ is the proto-cluster mass in units of $10^{14} \si{\solarmass}$  for a fiducial radius  of $r_{200}$, $\alpha = 1.237 \times 10^{-5}~ $ Mpc$^{-3}$, ~$\beta = 0.907$, ~$\gamma = 0.961$, and $\delta = 0.625$
\citep{Boe2017}. 
For the calculation of the comoving volume as a function of redshift, we used the reference cosmology. 
Table~\ref{tab8} shows the results for different lower mass limits for volumes for which proto-cluster counts were integrated over redshift intervals of $\Delta z = 0.5$. The redshifts listed in the Table, give the upper bound of the redshift interval.
The Table also provides values for the proto-cluster radii calculated by means of Eq. (2) in units of arcmin on the sky for proto-clusters with masses of $10^{14}$\,\si{\solarmass} and  $10^{15}$\,\si{\solarmass} and sky surface areas of proto-clusters with $ \ge 10^{14}$\,\si{\solarmass}.  

\begin{figure}[h]
   \includegraphics[width=\columnwidth]{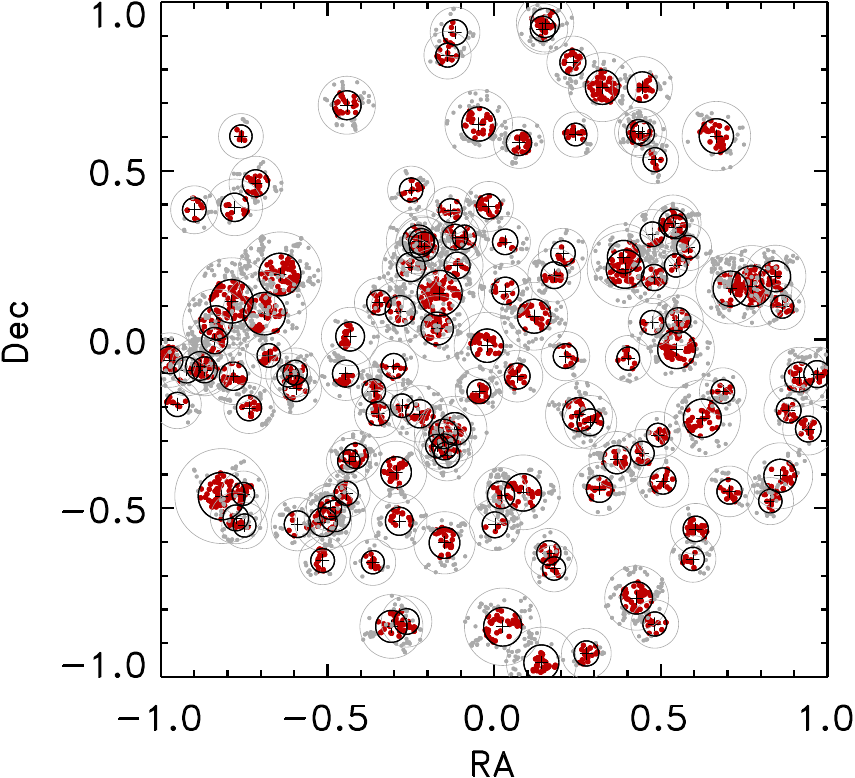}
\caption{Proto-clusters and their galaxies in a redshift slice around $z = 2$ with a redshift range according to the photometric redshift accuracy limits for 50\% completeness and $M_{\rm pc} \ge 10^{14}$ \si{\solarmass}. 
The dense core regions inside $r \le 0.5\,r_{\rm pc}$ are shown by black circles and red dots for the galaxies, while the outer regions are shown in grey.
}\label{fig32}
\end{figure}

\begin{figure}[h]
   \includegraphics[width=\columnwidth]{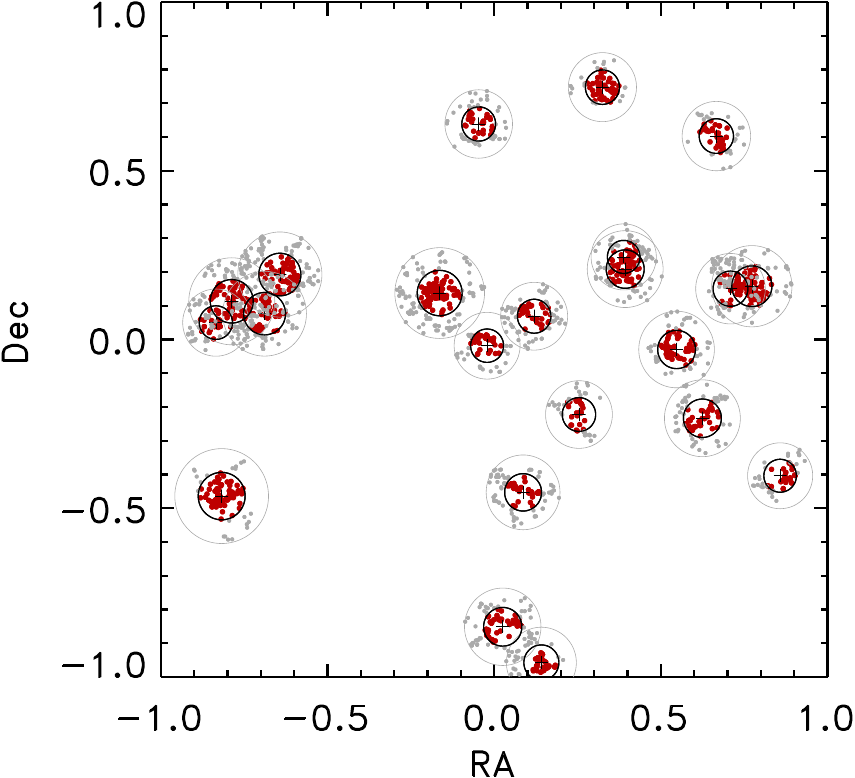}
\caption{Proto-clusters and their galaxies in a redshift slice around $z = 2$ as in Fig.~\ref{fig32} but with a higher mass limit of the proto-clusters of $M_{\rm pc} = 3 \times 10^{14}$\,\si{\solarmass}. 
Symbols and colours have the same meaning as in  Fig.~\ref{fig32}. 
}\label{fig33}
\end{figure}

\begin{figure}[h]
   \includegraphics[width=\columnwidth]{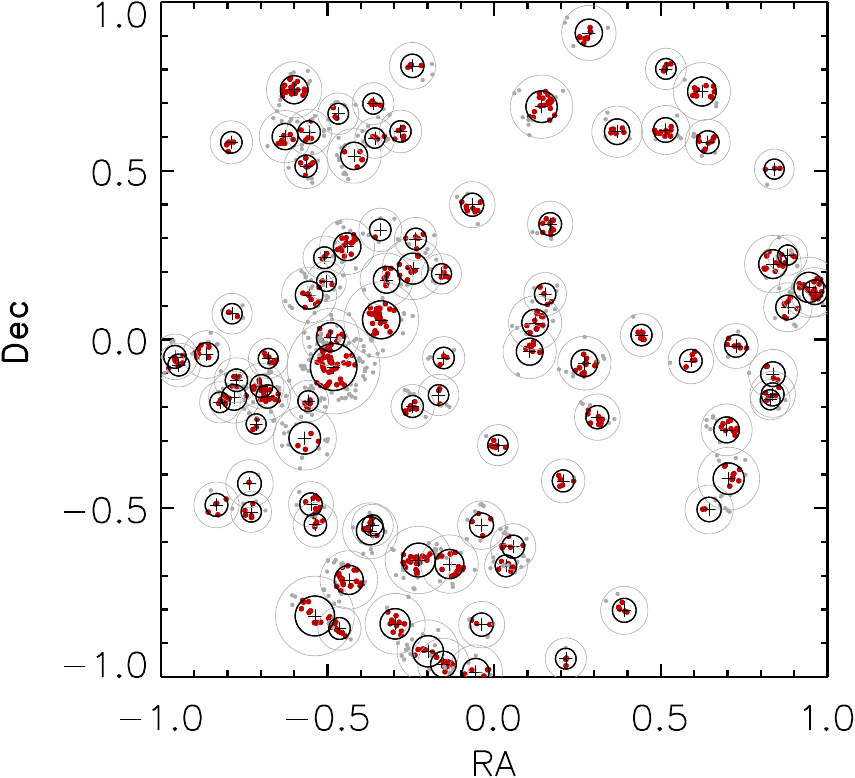}
\caption{Proto-clusters with $M_{\rm pc} \ge 10^{14}$ \si{\solarmass}
and their galaxies in a redshift slice around $z = 3.5$ with a redshift range according to the photometric redshift accuracy limits for 50\% completeness. 
Symbols and colours have the same meaning as in  Fig.~\ref{fig32}. 
}\label{fig34}
\end{figure}

   \begin{table*}
      \caption{Statistics concerning the proto-cluster abundance and sky coverage as a function of redshift,
column (1) gives the upper bound of the redshift interval with width $\Delta z = 0.5$, column (2) the volume per deg$^2$ in units of $10^6$ Mpc$^3$, column (3 to 6) the number of proto-clusters in this volume with mass $> 10^{14}, 2\times 10^{14}, 5\times 10^{14}$ and $10^{15}$\,\si{\solarmass}, respectively, column (7,8) the physical radii of proto-clusters with masses of $10^{14}$ and $10^{15}$\,\si{\solarmass}, and column (9) the sky area covered by a proto-cluster with a mass  $\ge 10^{14}$\,\si{\solarmass}.
}
      \[
         \begin{array}{lrrrrrrrr}
            \hline
            \noalign{\smallskip}
{\rm redshift}&{\rm volume}&N(>M_{14})&N(>2~M_{14})&N(>5~M_{14})& N(> M_{15}) &~~~~ {\rm rad}(M_{14})&~~ {\rm rad}( M_{15}) &~~~~~~~~~~{\rm area}\\
              & {\rm deg}^{-2}&         &           &            &                & {\rm arcmin}     &  {\rm arcmin}       & {\rm deg}^{2}   \\
            \noalign{\smallskip}
            \hline
            \noalign{\smallskip}
  0.50 &    0.68 &    5.47 &    1.64 &    0.21 &    0.02 &  6.528 & 14.064 &  0.0372  \\
  1.00 &    2.98 &   23.82 &    7.15 &    0.89 &    0.11 &  5.079 & 10.943&  0.0225  \\
  1.50 &    4.78 &   38.21 &   11.46 &    1.43 &    0.17 &  4.447 &  9.581 &  0.0173  \\
  2.00 &    5.67 &   45.38 &   13.62 &    1.70 &    0.20 &  4.084 &  8.799 &  0.0146  \\
  2.50 &    5.98 &   47.86 &   14.36 &    1.79 &    0.22 &  3.842 &  8.278 &  0.0129  \\
  3.00 &    5.97 &   47.78 &   14.33 &    1.79 &    0.22 &  3.665 &  7.896 &  0.0117  \\
  3.50 &    5.80 &   46.43 &   13.93 &    1.74 &    0.21 &  3.527 &  7.599 &  0.0109  \\
  4.00 &    5.56 &   44.48 &   13.34 &    1.67 &    0.20 &  3.415 &  7.358 &  0.0102  \\
  4.50 &    5.29 &   42.29 &   12.69 &    1.59 &    0.19 &  3.322 &  7.157 &  0.0096  \\
  5.00 &    5.01 &   40.07 &   12.02 &    1.50 &    0.18 &  3.243 &  6.986 &  0.0092  \\
  5.50 &    4.74 &   37.91 &   11.37 &    1.42 &    0.17 &  3.174 &  6.839 &  0.0088  \\
  6.00 &    4.48 &   35.86 &   10.76 &    1.34 &    0.16 &  3.115 &  6.712 &  0.0085  \\
  6.50 &    4.24 &   33.92 &   10.18 &    1.27 &    0.15 &  3.064 &  6.601 &  0.0082  \\
  7.00 &    4.02 &   32.13 &    9.64 &    1.20 &    0.14 &  3.019 &  6.505 &  0.0080  \\
  7.50 &    3.81 &   30.45 &    9.13 &    1.14 &    0.14 &  2.981 &  6.422 &  0.0078  \\
  8.00 &    3.62 &   28.92 &    8.68 &    1.08 &    0.13 &  2.948 &  6.350 &  0.0076  \\
  8.50 &    3.43 &   27.46 &    8.24 &    1.03 &    0.12 &  2.919 &  6.290 &  0.0074  \\
  9.00 &    3.27 &   26.16 &    7.85 &    0.98 &    0.12 &  2.896 &  6.239 &  0.0073  \\
            \noalign{\smallskip}
            \hline
            \noalign{\smallskip}
         \end{array}
      \]
\label{tab8}
   \end{table*}

Based on the proto-cluster abundances described above, we can study how well we can separate proto-clusters in the sky if we have only photometric redshifts. 
We look at the fraction of the sky covered by proto-clusters within a redshift slice, which corresponds to the uncertainty of photometric redshifts.
For the study, we select all the proto-clusters and their member galaxies from the MAMBO lightcone in redshift intervals given by the photometric redshift accuracy for a 50\% completeness limit and show how they are projected onto the sky. Figure~\ref{fig32} shows proto-clusters and member galaxies selected from redshifts around $z=2$ with a redshift range of $\Delta_{z,(50\%)}$. The core regions ($r \le 0.5\, r_{\rm pc}$), which will stick out due to their higher density contrast, are highlighted in the plot.

In total, there are 132 proto-clusters; in this exercise, we also included proto-clusters that are not completely contained in the field of view of the MAMBO lightcone.

\begin{figure}[h]
   \includegraphics[width=\columnwidth]{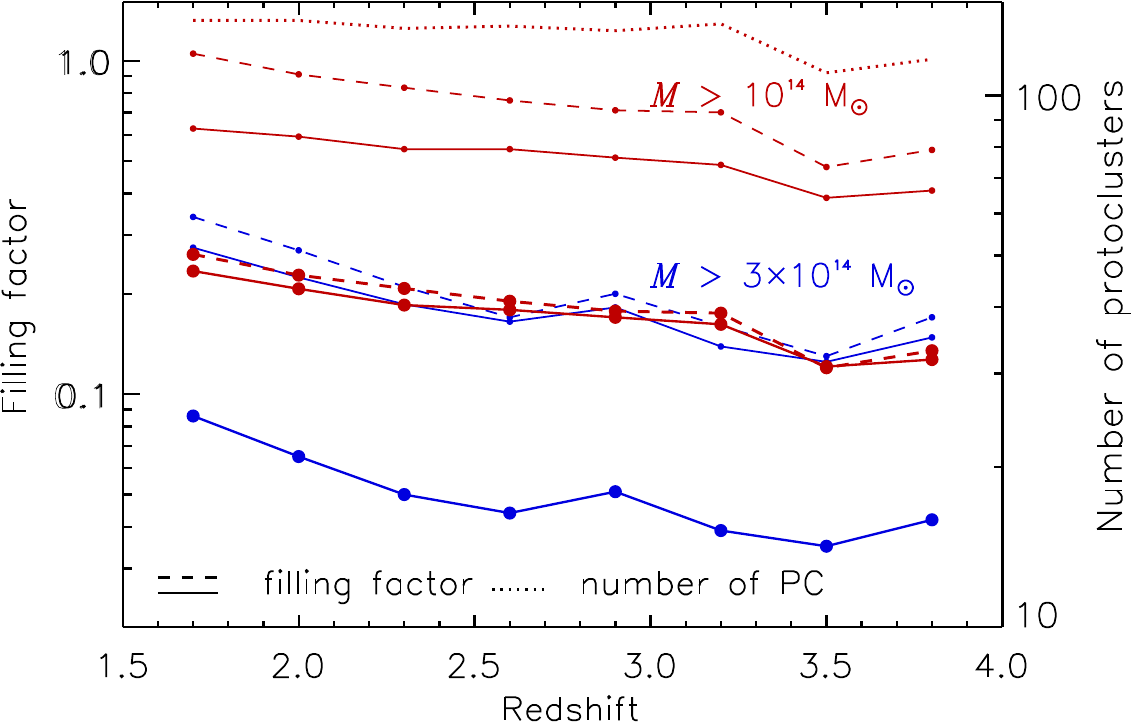}
\caption{Number of MAMBO proto-clusters and sky filling factor in a $\Delta_{z~(50\%)}$ redshift interval as a function of redshift and proto-cluster mass limit. The numbers are only shown for the sample with $M_{\rm pc} \ge 10^{14}$ \si{\solarmass} as a dotted line.
Solid lines show the fraction of the sky covered by the proto-clusters, and the dashed lines display the combined sky area of the proto-clusters. 
The thick lines with larger data points
show the results for the core regions, while the thin lines show those for the region out to $r_{\rm pc}$. For the core regions of the $M_{\rm pc} \ge 3 \times 10^{14}$ \si{\solarmass} proto-clusters, the dashed and solid lines overlap.
}\label{fig34b}
\end{figure}

We note that the sky is indeed densely covered by the proto-clusters. 
By just integrating the surface area of each proto-cluster (inside $r_{\rm pc}$), taking into account that some proto-cluster have some of their areas outside the lightcone, we find that the sum of all proto-cluster areas is 91\%, while the fraction of covered sky is about 59\%, due to overlaps.
We have seen in Sects. 6 and 7, however, that proto-clusters are detected with sufficient significance around $2\,\sigma$ only at radii smaller than the proto-cluster radius and in Sect. 5, we showed that typically half the member galaxies are found inside $0.5~r_{\rm pc}$ where they have a 2.7 times higher projected density than in the annulus around.
Therefore, to give an impression of the sky coverage of the recognisable core regions of the proto-clusters, we also quote the results for proto-cluster radii of $0.5~r_{\rm pc}$, which is a total fractional area of 23\% and a fraction of the sky covered of 21\%. Figure~\ref{fig34b} illustrates the case for both proto-cluster radii. 
We note that in some regions, the proto-cluster cores overlap, which is the result of clustering in large-scale structures.

We can, however, look at this from {a more practical} point of view. If we relax our ambition to detect all proto-clusters with a final mass of $10^{14}$\,\si{\solarmass} and look, for example, only at proto-clusters above a mass limit of $3 \times 10^{14}$\,\si{\solarmass}, the situation looks much better, as shown in Fig.~\ref{fig33}.
If we increase the mass limit even further, for example, to $8 \times 10^{14}$\,\si{\solarmass} we find only none to three proto-clusters per $\Delta_{z~(50\%)}$ in the MAMBO lightcone, with filling factors of a few per cent. 
Only in the highest redshift bin, are there four proto-clusters with a filling factor of 7\%. 
These larger mass objects correspond probably better to the typical proto-clusters that have been identified in optical surveys.
If we had included smaller systems ($< 10^{14}\,\si{\solarmass})$, the picture would have looked worse. 
Thus, the whole problem is that of detecting massive proto-clusters in the presence of large-scale structures in the background, a problem that we have already encountered in section 6, which has already been taken into account in the calculation of the detection significance.
It therefore appears that targeting proto-clusters with a mass limit of $10^{14}$\,\si{\solarmass} may be ambitious.
In Fig.~\ref{fig34} we also show the sky coverage for $M > 10^{14}$\,\si{\solarmass} proto-clusters around redhift $z \sim 3.5$. 
The sky coverage is less dramatic than in the low redshift case. 
But we note in all three pictures the substantial clumping of the proto-clusters. 
This may sometimes cause difficulties in clearly separate structures, which are identified with separate proto-clusters in the simulations.

In Fig.~\ref{fig34b}, we summarise the results for our target redshift range at eight redshift values. 
We show the sky coverage of the proto-clusters in the $\Delta_{z~(50\%)}$ redshift slices and also the sum of their sky area.
The sky coverage factors are decreasing with redshift. 
Since only the centres of the proto-clusters will be detectable and their sky-filling factors never exceed 30\% even for the worst case, proto-clusters should, in general, be separable from detection algorithms, except for the regions of proto-superclusters.

\section{Discussion}

The previous sections have shown that the approximate analytic concept for the prediction of the proto-cluster properties such as size, density contrast, and abundance, developed above, can provide a robust guideline to explore proto-clusters in the Euclid Wide Survey.
In particular, the prediction for the proto-cluster radius gives a good approach to defining a fiducial radius for a proto-cluster.
In addition, we have obtained further useful descriptions, such as the typical proto-cluster profiles and their variations by means of the simulations. 
We note that some of the presented properties are intrinsic in the sense that they are described on the basis of the information given by the simulations. This includes the proto-cluster radii, profiles, abundances, and mass-richness relation. The detection significances are on the contrary observed properties, for which the information from the simulations is used as would be available to an observer.

As an application of the information given in Table~\ref{tab8}, we calculate the expected volumes of proto-clusters and compare them to some observed examples in the literature.
From the radii given in arcmin, we calculate the comoving volumes for proto-clusters with a descendent mass of $10^{15}$ \si{\solarmass} at redshifts 2, 3, 4, and 5 with results of about 9700, 13\,000, 15\,000, and 16\,000 Mpc$^3$,
respectively, and compare them to the volumes assigned to some prominent proto-clusters by \citet{Casey2016}. 
The quoted volumes for COSMOS ($z=2.10$) and ($z=2.47$) of about 15\,000 Mpc$^3$ are larger than what is expected for the most massive proto-clusters, and one might conclude that we are looking at two or more closely neighbouring structures, as we see them in Figs.~\ref{fig32}, \ref{fig33}, and \ref{fig34}. 
The combined volume of three fields in SSA22 ($z=3.09$) with a value of 21\,000 Mpc$^3$ is consistent with the expectation for massive proto-clusters if we take the three fields as different systems. 
The value of 20\,000 Mpc$^3$ for HDF 850.1 ($z = 5.18$) is a bit high but could still be explained by one very massive proto-cluster if the boundary was considered slightly too generous. For MRC1138$-$256 ($z=2.16$) and AzTEC-3 ($z=5.18$) the observed values are comparably small, and either 
the volumes assigned concern only part of the proto-cluster, or these proto-clusters have descendants with a mass much lower than $10^{15}$
\si{\solarmass}. 
In a recent CO survey around MRC1138$-$256 \citep{Jin2021}, a large overdensity of CO emitters was found as part of a filamentary structure
with an extent of about 120 comoving Mpc. The extent of 38 Mpc in physical scale is too large for one proto-cluster and a structure of this size should break up to form several virialised units. But within this filament, MRC1138$-$256 could indeed be a larger proto-cluster than mentioned above.

With these tools, we can prepare the strategy for the detection of proto-clusters in the Euclid Survey and the assessment of their
properties. 
They could, in particular, provide a good orientation for the parameter selection for more sophisticated detection algorithms than the aperture counts used here. 

In Sects. 6 and 7, we reported the moderate detection significance for proto-clusters, which can be achieved with the Euclid Survey with the given photometric redshift uncertainties based on broadband photometry. This will make it difficult, in general, to characterise proto-clusters quantitatively without further follow-up observations. On the other hand, many proto-clusters will be detectable with sufficient significance ($\sim 2\,\sigma$) to obtain a large number of highly likely candidates. Since the proto-clusters have a large sky-filling factor, as shown in Sect. 10, one can actually afford a low detection significance threshold. 
The following example may best illustrate this.
If we study a contiguous sky region of 200 times the area of a typical proto-cluster and the region contains 100 such systems, corresponding to a filling factor of 50\%, a 2$\,\sigma$ detection threshold would lead to an average false detection rate of 5 contaminating events (since only the positive deviations of the distribution count), a false detection rate of 5\%, which could be tolerable for statistical studies.

What we have described here is a worst-case scenario using a very simple detection aperture. Several sophisticated detection algorithms have been developed for the detection of proto-clusters in the Euclid Survey. Using a particular aperture and redshift interval exploits only part of the information given by the observations. A good detection algorithm will make use of the complete available information, including galaxies at all relevant radii and the complete probability distributions of the photometric redshifts with optimal weights. 
Therefore, we can easily expect a significant increase in the detection efficiency with the proper use of dedicated algorithms. 
One has to be careful, however, when proto-cluster shapes are assumed in the algorithms about the possible introduction of selection effects. 
The examples of unusual proto-cluster profiles found in Sect. 5 are, in this respect, interesting test cases.

We can illustrate the unique opportunity offered by \Euclid, also with the following example. 
We focus on the most massive proto-clusters with $M \ge 8 \times 10^{14}$ \si{\solarmass} mentioned above with a typical sky-filling factor of $\sim 5\%$. 
A dedicated algorithm is expected to improve the detection significance from $\sim 2$ to at least $\sim$ 3$\,\sigma$. 
This would correspond to false detections with a sky-filling factor of 0.15\%. 
Thus, we would obtain a ratio of about 34 true to one false detection. 
In total, we can expect of the order of 40\,000 proto-clusters with a mass limit of $M \ge 8 \times 10^{14}$ \si{\solarmass} in the complete Euclid Wide Survey in the redshift range $z = 1.5$ to 4. 
This will definitely provide a relatively pure, interesting, and unique proto-cluster candidate sample.

A sweet spot for the detection of proto-clusters in the considered redshift range is $z \sim 3$, where the density contrast turns
out to be the highest. This is an interesting region where  Ly-break galaxy surveys revealed the first proto-clusters.
With a further boost of significance by means of dedicated detection algorithms, the data should be sufficient to estimate selection functions and
perform population statistics. What will be particularly interesting is the unprecedentedly large survey volume of the Euclid Wide Survey, which will turn up the most extreme and rare objects that have not been observed in the available much smaller survey areas.    

The present study also leads to the question: how can the contrast above the background be improved? Narrowband surveys will, of course, lead to more precise photometric redshifts and thus provide an improvement.
We can take a look at two examples. The survey of \citet{Yam2012} in the SA22 field at $z \sim 3$ with a narrowband filter $\Delta \lambda = 77 \AA$ provides an accuracy of about $\Delta z = 0.063$. For the COSMOS survey \citet{Chi2014} obtained $\Delta z = 0.025~ ( 1 + z)$. 
Compared to the numbers for the \Euclid data given above, this is an improvement of about a factor of two, which is, of course, very helpful, but it does not improve the situation so much that it allows a precise assessment of the proto-cluster properties. 
Thus, spectroscopic follow-up observations will be mandatory for an accurate study of the most interesting proto-clusters.

\section{Summary and Conclusions}

We provided an overview of how galaxy proto-clusters in the redshift range $z = 1.5$ to $4$
are expected to appear in the Euclid Wide Survey to assist in the preparation of the survey analysis. 
The paper provides studies on the following proto-cluster properties.

\begin{itemize}
\item[$\bullet$] A practical estimate of the proto-cluster radius was obtained on the basis of an analytical model for the evolution of a homogeneous top-hat overdensity. Comparison with cosmological simulations shows that about 80\% of the members of the descendent cluster (inside $r_{200}$) are contained within this radius with contamination of about five to ten per cent. 
An analytic approximation of the evolution of the proto-cluster radius with redshift was provided (Sect. 4).
\smallskip

\item[$\bullet$]
The mean radial galaxy distribution of proto-clusters can be well described with a cored profile with a specified inner and outer logarithmic slope. We provide best-fit formulae for the mean three-dimensional and projected profiles (Sect. 5). Individual proto-clusters show, however, a large variety. About three-quarters of the proto-clusters have a pronounced dense core. However, proto-clusters with multiple substructure components can have profiles with a density maximum at larger radii.  
\smallskip

\item[$\bullet$] 
We investigated with which density contrast in the galaxy distribution proto-clusters can be observed against the galaxy background. We assumed that photometric data from the Euclid Survey and broad band ground-based auxiliary data (at the time of the third data release) are available for this study to estimate photometric redshifts. We show that the small overdensity of proto-clusters, the large redshift depth that has to be sampled, and the large-scale structure variance of the background lead to a moderate detection significance for simple detections in circular apertures. With a local background assessment and apertures focused on the inner regions of the proto-clusters, significances of the order of 2$\,\sigma$ can be reached. Because proto-clusters are not very rare, useful samples of highly likely proto-cluster candidates can be obtained with such detection thresholds.
\smallskip

\item[$\bullet$] 
The cluster richness, the number of galaxies being members of the proto-cluster, was found to be tightly correlated with the mass of the descendent cluster. We found a relation with a logarithmic slope around 0.9, which may reflect the decreasing galaxy formation efficiency with increasing halo mass for the most massive halos. We provided analytical approximations to the mass-richness relation as a function of mass and redshift for the GAEA and MAMBO simulations (Sect. 8). 
\smallskip

\item[$\bullet$] 
The abundance of proto-clusters is determined by their definition as being the precursors of galaxy clusters at redshift zero with masses, $M_{200} \ge 10^{14}$ \si{\solarmass}. The matter fraction in such clusters was, e.g., determined in \citet{Boe2017} to be about 4\% and also, the comoving number density can be obtained from this result. Knowing the objects' abundance and sizes, we can calculate how densely the proto-clusters fill the sky for a redshift range comparable to the uncertainties of the photometric redshifts. We found that the area filling is quite large, and proto-clusters with the quoted mass limit start to overlap significantly.
\end{itemize}

There are several efforts in the Euclid Consortium to devise optimised detection algorithms for the search of proto-clusters in the Euclid Survey. They will use more sophisticated techniques, which also involve probability distributions for the photometric galaxy redshifts. This will provide the leverage to improve the detection efficiency over what has been described here. With the large sky area explored by \Euclid, an enormously large database of likely proto-clusters and their properties will be provided, which can carry proto-cluster studies and follow-up observations to a new level. 

\begin{acknowledgements} 
HB thanks the Munich Excellence Cluster 'Origins' for support. 
GC acknowledges support by the Deutsches Luft- und Raumfahrt Zentrum through grant no. 50 OR 2204. 
HD acknowledges financial support from the Agencia Estatal de Investigaci\'on del Ministerio de Ciencia e Innovaci\'on (AEI-MCINN) under grant (La evoluci\'on de los c\'umulos de galaxias desde el amanecer hasta el mediod\'ia c\'osmico) with reference (PID2019-105776GB-I00/DOI:10.13039/501100011033) and del Ministerio de Ciencia, Innovaci\'on y Universidades (MCIU/AEI) under grant (Construcci\'on de c\'umulos de galaxias en formaci\'on a trav\'es de la formaci\'on estelar ocurecida por el polvo) and the European Regional Development Fund (ERDF) with reference (PID2022-143243NB-I00/DOI:10.13039/501100011033).
LM acknowledges the financial contribution from the grant PRIN-MUR 2022 20227RNLY3 ``The concordance cosmological model: stress-tests with galaxy clusters'' supported by Next Generation EU and from the grants ASI n.2018-23-HH.0 and n. 2024-10-HH.0 ``Attività scientifiche per la missione Euclid – fase E''
\AckEC\\
\end{acknowledgements}

%
%

\bibliography{Euclid}

\appendix
\section{Morphological classification of the proto-cluster profile}

\begin{figure}[htpb!]
\centering
\includegraphics[width=.95\columnwidth]{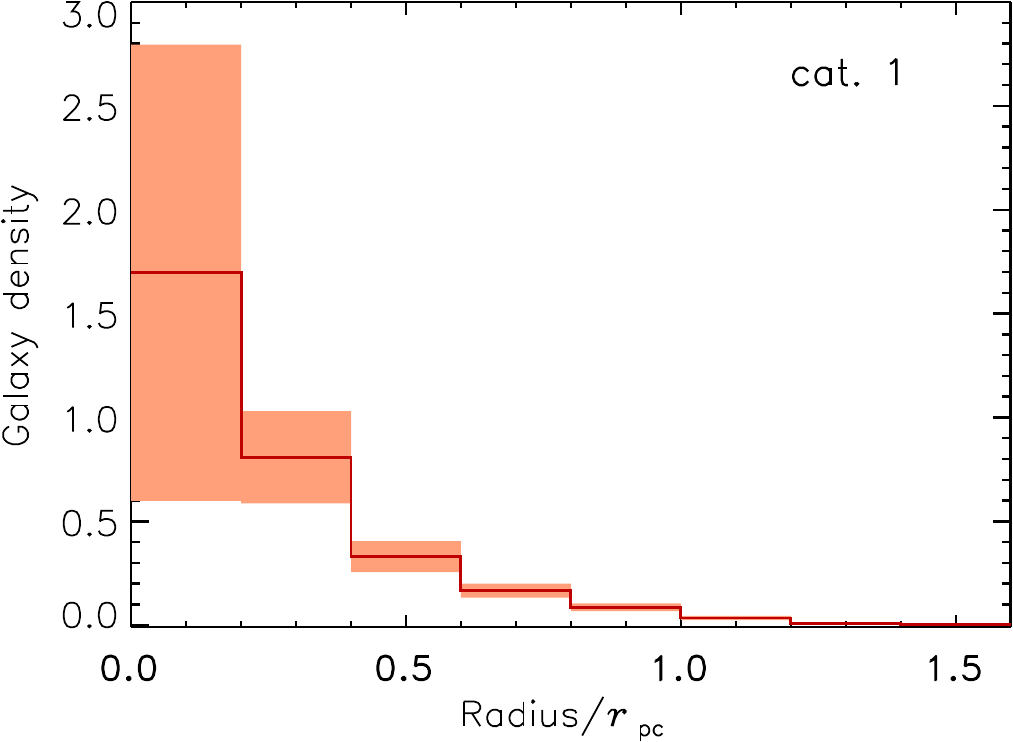}
\includegraphics[width=.95\columnwidth]{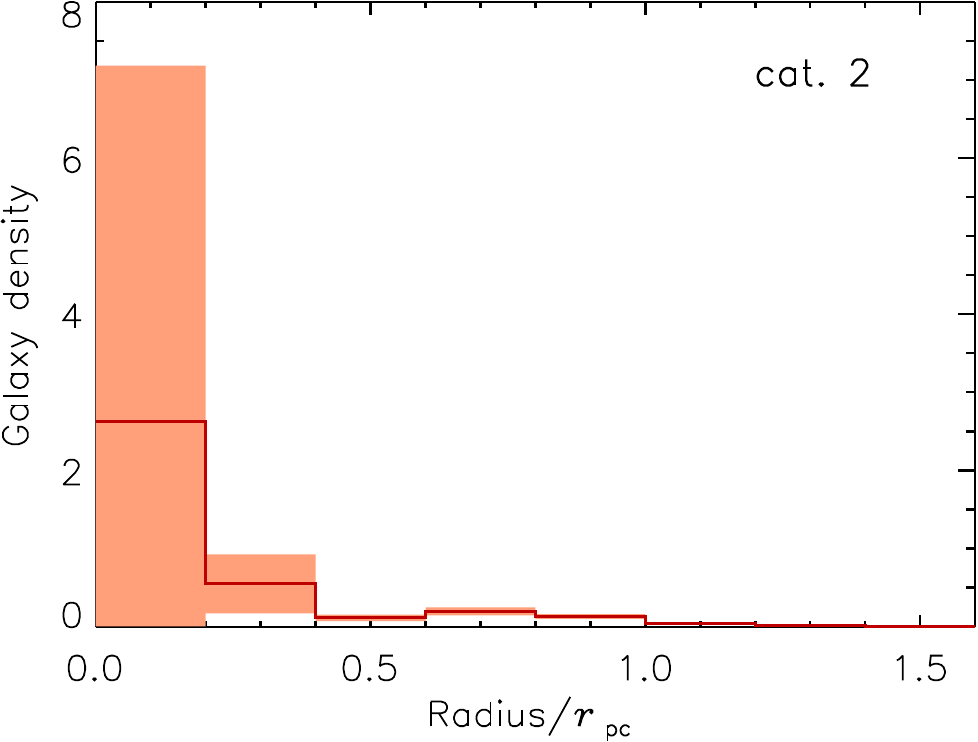}

\caption{Mean profiles of the proto-clusters for two of the five structural categories defined in the text, where the categories are labelled in the panels. The shaded regions in the histograms indicate the rms variance of the density.
}
\label{fig50}
\end{figure}

\begin{figure}[htpb!]
\centering
\includegraphics[width=.95\columnwidth]{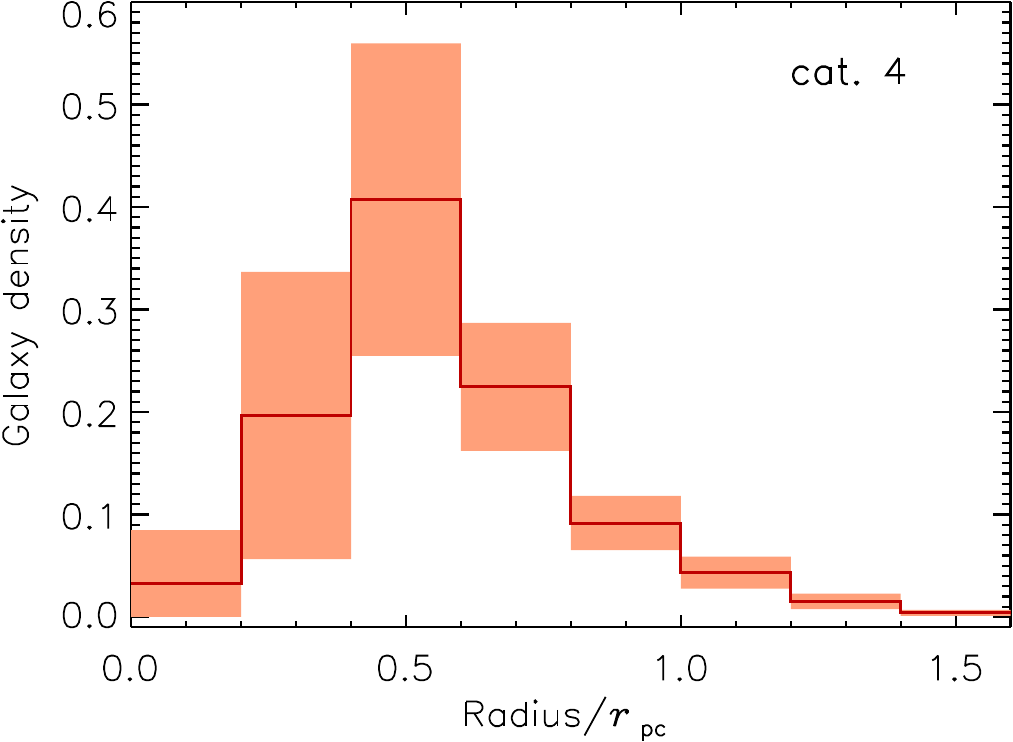}
\includegraphics[width=.95\columnwidth]{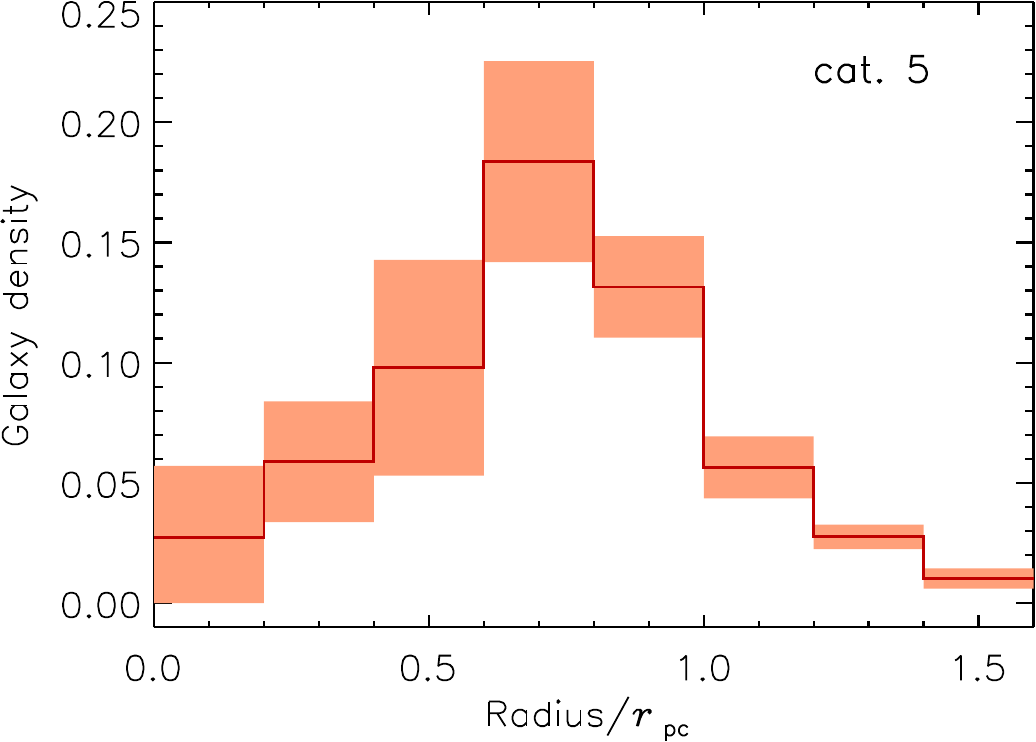}
\includegraphics[width=.95\columnwidth]{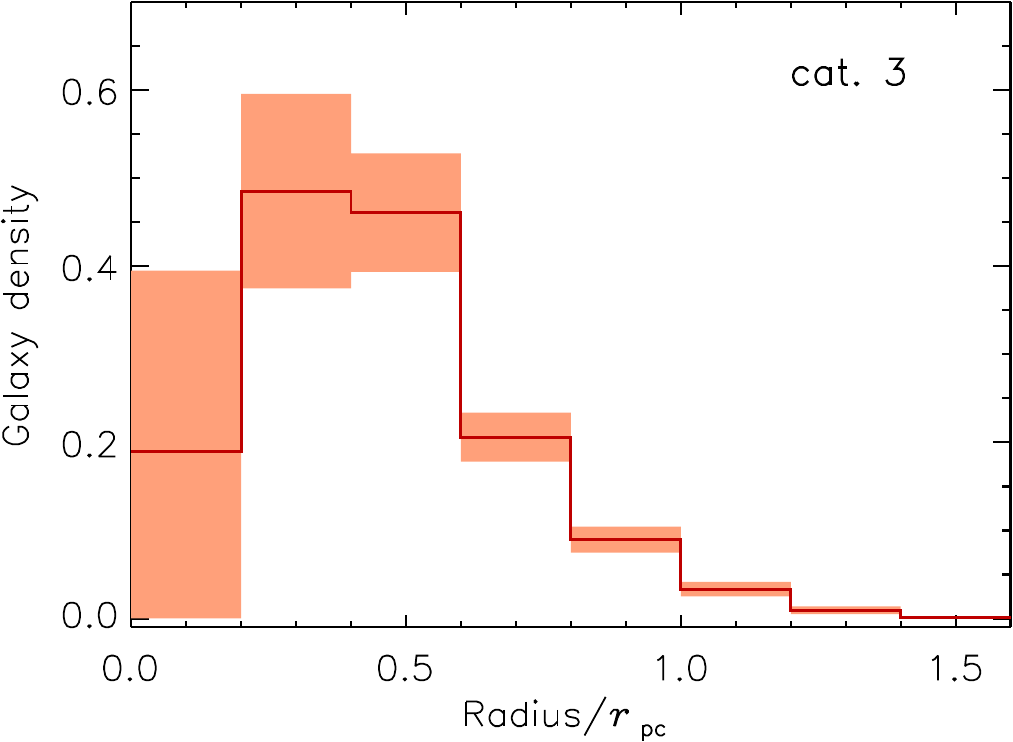}
\caption{Mean profiles of the proto-clusters for the remaining three of the five structural categories defined in the text, where the categories are labelled in the panels. 
}
\label{fig51}
\end{figure}

\begin{figure}[htpb!]
\centering

\includegraphics[width=.90\columnwidth]{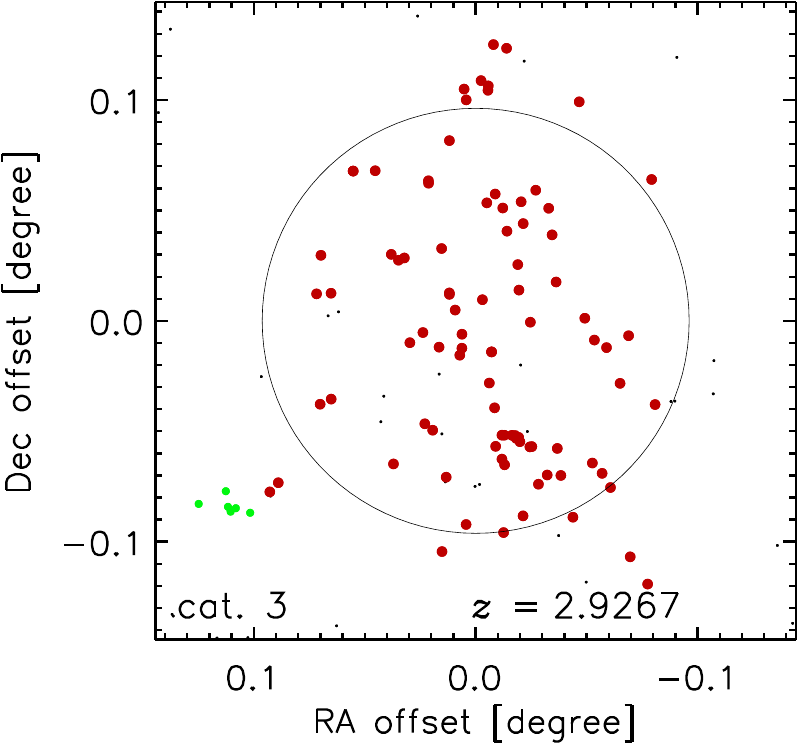}
\includegraphics[width=.90\columnwidth]{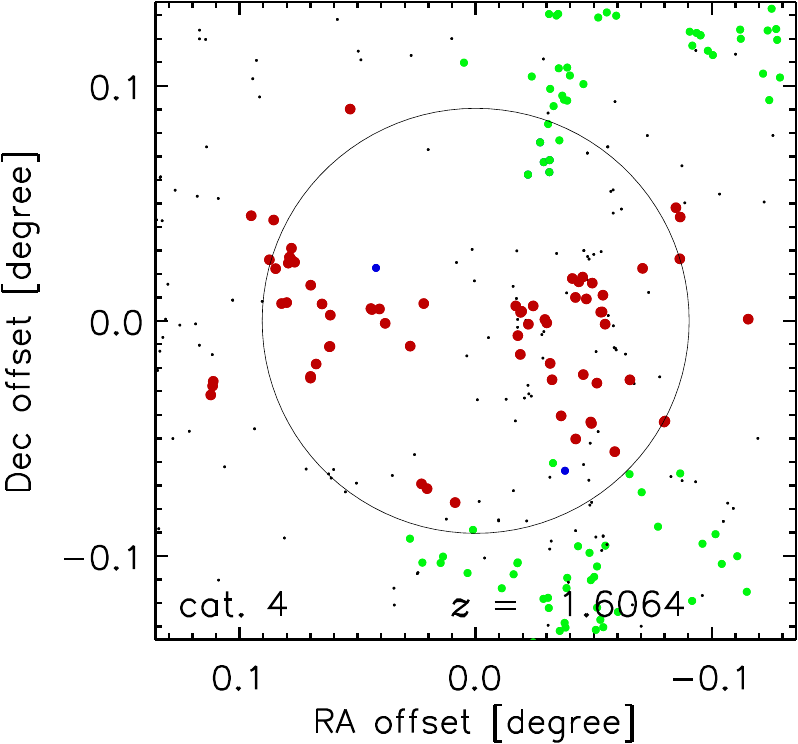}
\includegraphics[width=.90\columnwidth]{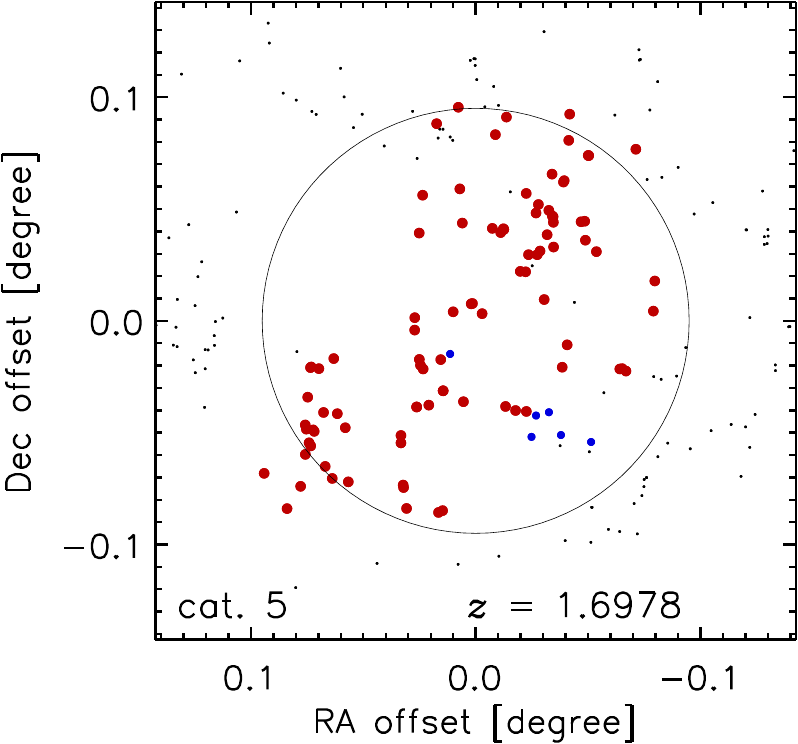}

\caption{
Examples of three proto-clusters for the categories 3 to 5. The category and the redshift of the proto-clusters is given in the legend of the Figure.
}
\label{fig52}
\end{figure}

In Section 5, we have explored the variation of proto-cluster profiles in the simulation lightcones. 
Here, we present figures for the mean profiles for each of the five categories in Figs.\,\ref{fig50} and \ref{fig51}. The figures also show the standard deviations of the profiles in each category. Fig.\,\ref{fig52} shows examples of proto-clusters in categories 3 to 5. 
All these proto-clusters show large substructures or even multi-modality.

\end{document}